\def\version{March 21, 2014}

\documentclass[12pt]{article}
\def\macrosPb{}
\ifdefined\macrosPa
  \usepackage[textwidth=465pt,textheight=650pt,centering]{geometry} 
\fi
\ifdefined\macrosPb
  \usepackage[textwidth=500pt,textheight=650pt,centering]{geometry} 
\fi

\usepackage[psamsfonts]{amsfonts}
\usepackage{amsmath,amssymb,amsthm}
\usepackage[dvips]{graphicx}
\usepackage{appendix}
\usepackage{bbm} 
\usepackage{amsbsy}
\usepackage{enumerate}
\usepackage{cite}

\ifdefined\macrosS


  \usepackage{mathptmx}
  \DeclareMathAlphabet{\mathcal}{OMS}{cmsy}{m}{n}
\fi

\usepackage[usenames]{color}

\newcommand{\LT}{{\rm Loc}  }

\newcommand{\DV}{\Dcal}
\newcommand{\DVa}{\alpha}

\def\UseSection{
        \numberwithin{equation}{section}
	\theoremstyle{plain}
        \newtheorem{theorem}    {Theorem}[section]
        \DefineTheorems 
}

\def\DefineTheorems{
	
	\newtheorem{lemma}      [theorem] {Lemma}
	
	\newtheorem{prop}       [theorem] {Proposition}
	
	\newtheorem{cor}        [theorem] {Corollary}

	\theoremstyle{definition}
	\newtheorem{defn}       [theorem] {Definition}

	\theoremstyle{definition}

}

\newcommand{\bt}   {\begin{theorem}}
\newcommand{\et}   {\end  {theorem}}
\newcommand{\bl}   {\begin{lemma}}
\newcommand{\el}   {\end  {lemma}}
\newcommand{\bp}   {\begin{prop}}
\newcommand{\ep}   {\end  {prop}}
\newcommand{\bc}   {\begin{cor}}
\newcommand{\ec}   {\end  {cor}}
\newcommand{\bd}   {\begin{defn}}
\newcommand{\ed}   {\end  {defn}}

\newcommand{\ba}   {\begin{array}}
\newcommand{\ea}   {\end  {array}}
\newcommand{\be}   {\begin{enumerate}}
\newcommand{\ee}   {\end  {enumerate}}
\newcommand{\bi}   {\begin{itemize}}
\newcommand{\ei}   {\end  {itemize}}

\def\eq#1\en{\begin{equation}#1\end{equation}}  
\def\eqsplit#1\ensplit{
	\begin{equation}\begin{split}#1\end{split}\end{equation}
	}
\def\eqalign#1\enalign{
	\begin{align}#1\end{align}
	}
\def\eqmul#1\enmul{
	\begin{multline}#1\end{multline}
	}
\newcommand{\eqarrstar} {\begin{eqnarray*}} 
\newcommand{\enarrstar} {\end{eqnarray*}} 
\newcommand{\eqarray}   {\begin{eqnarray}} 
\newcommand{\enarray}   {\end{eqnarray}} 
\newcommand{\nnb}	{\nonumber \\} 

\newcommand{\lbeq}[1]  {\label{e:#1}}
\newcommand{\refeq}[1] {\eqref{e:#1}}    
\newcommand{\lbfg}[1]  {\label{fg: #1}}

\makeatletter
\newcommand{\labelcounter}[2]{{%
	\stepcounter{#1}
	\protected@write\@auxout{}%
	{\string\newlabel{#2}{{\csname the#1\endcsname}{\thepage}}}%
	{\ref{#2}}
	}}
\makeatother
\newcommand{\Nbold} {{\mathbb N}}

\newcommand{\Rbold} {{\mathbb R}}

\newcommand{\Zbold} {{\mathbb Z}}

\newcommand{\Bcal}   {\mathcal{B}} 
 
\newcommand{\Dcal}   {\mathcal{D}} 
\newcommand{\Ecal}   {\mathcal{E}}

\newcommand{\Kcal}   {\mathcal{K}} 
\newcommand{\Lcal}   {\mathcal{L}} 
 
\newcommand{\Ncal}   {\mathcal{N}} 
 
\newcommand{\Pcal}   {\mathcal{P}}

\newcommand{\Scal}   {\mathcal{S}} 
 
\newcommand{\Ucal}   {\mathcal{U}} 
\newcommand{\Vcal}   {\mathcal{V}} 
\newcommand{\Wcal}   {\mathcal{W}}

\newcommand{\Zd}    {{ {\Zbold}^d }}

\newcommand{\spose}[1] {{\hbox to 0pt{#1\hss}} }
\newcommand{\ltapprox} {\mathrel{\spose{\lower 3pt\hbox{$\mathchar"218$}}
 \raise 2.0pt\hbox{$\mathchar"13C$}}}
\newcommand{\gtapprox} {\mathrel{\spose{\lower 3pt\hbox{$\mathchar"218$}}
 \raise 2.0pt\hbox{$\mathchar"13E$}}}

\newcommand{\dist} {{  \rm dist }}

\UseSection   
\renewcommand{\to} {\rightarrow}

\newcommand{\shift}   {\!\!\!\!}

\newcommand{\R}{\Rbold}
\newcommand{\Z}{\Zbold}

\newcommand{\N}{\Nbold}

\newcommand{\volume}{\mathbb{V}}

\newcommand{\1}{\mathbbm{1}}

\newcommand{\FDel}{\lambda}

\newcommand{\jm}{j_\Omega}

\newcommand{\Ex}{\mathbb{E}}

\newcommand{\chicCov}{{\chi}}

\newcommand{\bubble}{{\sf B}}

\newcommand{\pair}[1]{\langle #1 \rangle}

\newcommand{\Phipol}{\Pi}

\newcommand{\diam}[1]{\textrm{diam}(#1)}

\newcommand{\pt}{{\rm pt}}

\newcommand{\Upt}{U_{\rm pt}}
\newcommand{\Vpt}{V_{\rm pt}}
\newcommand{\gpt}{g_{\mathrm{pt}}}
\newcommand{\nupt}{\nu_{\mathrm{pt}}}

\newcommand{\zpt}{z_{\mathrm{pt}}}
\newcommand{\ypt}{y_{\mathrm{pt}}}

\newcommand{\Kch}{\check{K}}
\newcommand{\Vch}{\check{V}}
\newcommand{\Rch}{\check{R}}
\newcommand{\gch}{\check{g}}
\newcommand{\zch}{\check{z}}
\newcommand{\much}{\check{\mu}}
\newcommand{\nuch}{\check{\nu}}

\newcommand{\gbar}{\bar{g}}

\newcommand{\ggen}{\tilde{g}}
\newcommand{\sgen}{\tilde{s}}

\newcommand{\mgen}{\tilde{m}}
\newcommand{\Iint}{\mathbb{I}}
\newcommand{\Igen}{\tilde{\mathbb{I}}}

\newcommand{\zbar}{\bar{z}}
\newcommand{\mubar}{\bar{\mu}}

\newcommand{\domRG}{\mathbb{D}}

\newcommand{\half}{\textstyle{\frac 12}}

\newcommand{\ddp}[2]{\frac{\partial #1}{\partial #2}}

\newcommand{\epV}{\epsilon_{V}}

\newcommand{\Kspace}{\Kcal}

\DeclareMathOperator{\Loc}{Loc} 

\usepackage{xr}
\title {
  Scaling limits and critical behaviour of the \\
  $4$-dimensional $n$-component $|\varphi|^4$ spin model
}

\author{
  Roland Bauerschmidt\thanks{School of Mathematics,
    Institute for Advanced Study,
    Princeton, NJ 08540 USA.
    E-mail: {\tt brt@math.ias.edu}.},\;
  David C.\ Brydges\thanks{Department of Mathematics,
    University of British Columbia,
    Vancouver, BC, Canada V6T 1Z2.
    E-mail: {\tt db5d@math.ubc.ca}, {\tt slade@math.ubc.ca}.}\;
  and Gordon Slade$^\dagger$}

\date\version

\newcommand{\T}{\mathbb{T}}
\renewcommand{\chicCov}{{\vartheta}}
\newcommand{\chicCovgen}{{\tilde\vartheta}}

\begin{document}

\maketitle

\begin{abstract}
  We consider the $n$-component $|\varphi|^4$ spin model on $\Z^4$,
  for all $n \ge 1$, with small coupling constant.  We prove that the
  susceptibility has a logarithmic correction to mean field scaling,
  with exponent $\frac{n+2}{n+8}$ for the logarithm.
  We also analyse the asymptotic behaviour of the pressure as the
  critical point is approached, and prove that the specific heat has
  fractional logarithmic scaling for $n =1,2,3$; double logarithmic
  scaling for $n=4$; and is bounded when $n>4$.  In addition, for the
  model defined on the $4$-dimensional discrete torus, we prove that
  the scaling limit as the critical point is approached is a multiple
  of a Gaussian free field on the continuum torus, whereas, in the
  subcritical regime, the scaling limit is Gaussian white noise with
  intensity given by the susceptibility.
  The proofs are based on a rigorous renormalisation group method in
  the spirit of Wilson, developed in a companion series of papers to
  study the 4-dimensional weakly self-avoiding walk, and adapted here
  to the $|\varphi|^4$ model.
\end{abstract}

\section{Introduction and main results}

\subsection{Introduction}
\label{sec:intro}

The renormalisation group has become the principal theoretical approach to a wide
variety of problems in physics involving infinitely many degrees of
freedom cooperating over all length scales,
including quantum field theory, condensed matter physics, and
the theory of critical phenomena and phase transitions.  The renormalisation
group approach simultaneously supplies both the means to carry out otherwise intractable
calculations of physically relevant quantities such as critical exponents, as well
as the conceptual understanding that universality classes arise
as different domains of attraction of fixed points in
a dynamical system in a space of Hamiltonians.  While the concept of the renormalisation
group approach predates the work of Kenneth Wilson, it is Wilson's formulation of
the method in the 1970s that fully revealed its power and became
the dominant approach \cite{WK74}.
In particular, he showed that the evolution of the dynamical system is typically
dominated by a finite-dimensional part (the relevant and marginal directions),
with an infinite dimensional irrelevant part.
A fascinating historical perspective on the role of the
 renormalisation group in critical phenomena can be found in \cite{Domb96}.
In this paper, we discuss a mathematically rigorous
implementation of Wilson's renormalisation
group approach for the $n$-component $|\varphi|^4$ spin system.

Despite the successes of the early 1970s, in 1975 Wilson was careful to emphasise
that the renormalisation group is not a panacea.  He wrote in \cite{Wils75}:
\begin{quote}
``It is at present an approach of last resort, to be used only when all other
approaches have been tried and discarded.
The reason for this is that it is rather difficult to formulate
renormalization group methods for new problems; in fact, the renormalization
group approach generally seems as hopeless as any other approach
until someone succeeds in solving the problem by the renormalization group
approach. Where the renormalization group approach has been successful
a lot of ingenuity has been required: one cannot write a renormalization
group cookbook. [\,\dots]
It
will probably require several years of stagnation in elementary particle
theory before theorists will accept the inevitability of the renormalization
group approach despite its difficulties.''
\end{quote}
Four decades later, this quote reads partly as overly pessimistic.
For a myriad of problems in theoretical physics it is now the
standard approach rather than one of last resort, with a well-developed
calculus established for its use \cite{Amit84,Card96,Zinn07}.

On the other hand, concerning mathematically rigorous studies, the
characterisation as a method of last resort is hard to dispute.  A
mathematically rigorous implementation involves proving that a flow of
finitely many parameters contains all the information about critical
exponents. Wilson's contribution was to provide the reason behind this
fact, but converting his insight about irrelevant terms into a proof
remains a major challenge.  Nevertheless, there have been
many successes which demonstrate that the renormalisation group can be
harnessed in a mathematically rigorous manner to solve difficult
problems of physical significance.  Recent examples in
statistical mechanics include: the rigorous construction of an
interacting Fermi liquid at temperature zero in two space dimensions
\cite{FKT04I}, universality of the conductivity in graphene
\cite{GMP12}, the analysis of the two-dimensional Coulomb gas near its
Kosterlitz--Thouless transition \cite{Falc12,Falc13}, the analysis of
gradient interface models \cite{AKM12}.  Books and major reviews have
been written, including \cite{BG95,Bryd09,FKT02,Mast08,Riva91,Salm99}.
However the mathematical difficulties are severe enough to tempt us to
recast the last sentence in Wilson's 1975 quote to apply instead to
mathematicians.  Certainly a lot of ingenuity is required, and basic
principles are sometimes difficult to distill from the sea of
technicalities.  After examination of the mathematical literature on
the renormalisation group method, one can only agree with Wilson that
there is no cookbook.

Recently we have developed a recipe for a mathematically rigorous
implementation of the renormalisation group approach
\cite{BBS-rg-flow,BBS-rg-pt,BS-rg-norm,BS-rg-loc,BS-rg-IE,BS-rg-step},
which we used to analyse the critical behaviour of the 4-dimensional
continuous-time weakly self-avoiding walk \cite{BBS-saw4-log,BBS-saw4}.
In particular, we proved that its critical two-point function has
$|x|^{-2}$ decay and that its susceptibility diverges with logarithmic
correction $\varepsilon^{-1}(\log \varepsilon^{-1})^{1/4}$, as was
predicted forty years earlier in the physics literature but not
previously proved rigorously.  In the present paper, we apply the
recipe to the $n$-component $|\varphi|^4$ model, for all $n \ge 1$.
The $|\varphi|^4$ model, which is also called the
\emph{Landau--Ginzburg--Wilson model}, is a continuous-spin analogue
of the Ising model, and is the simplest example of an interacting
Euclidean boson quantum field.  For $n=1$, it is predicted to be in
the same universality class as the Ising model.
Our analysis of the $|\varphi|^4$ model is simpler
than that of the weakly self-avoiding walk
in \cite{BBS-saw4-log,BBS-saw4}, since the
latter requires a fermion field which is now absent.

The $|\varphi|^4$ model and related spin systems
have been studied in depth by mathematicians \cite{GJ87,FFS92}.
In the early 1980s, it was proved without using renormalisation group
ideas that its upper critical dimension
is $4$, with Gaussian scaling in dimensions $d>4$ \cite{Aize82,Froh82}.
In particular, the susceptibility has mean field divergence of the
form $\varepsilon^{-1}$ for $d>4$, where $\varepsilon \downarrow 0$
corresponds to the approach
to the critical point.
For $d=4$ and $n=1$, the divergence of the
susceptibility at the critical point was proven to lie between $\varepsilon^{-1}$
and $\varepsilon^{-1}(\log \varepsilon^{-1})$  \cite{AG83}
(some related ideas appear in \cite{ACF83}).
Similarly, it was shown that there is at most logarithmic deviation from
mean field behaviour for the specific heat \cite{Soka79} and the magnetisation \cite{AF86}.
The methods used to show these facts were, however, insufficient to
identify the logarithmic correction with precision.  These results
rely on differential inequalities obtained from stochastic geometric
representations of correlation functions, on correlation inequalities,
and on the infrared bound of \cite{FSS76}.  The correlation
inequalities usually rely on restricted values of $n$, such as $n
=1,2$, while the infrared bound is proved for general $n$.  On the
other hand, the infrared bound is proved using reflection positivity,
which requires special geometry.  For example, reflection positivity
breaks down in the presence of next-nearest neighbour spin
interactions, or for the self-avoiding walk, despite the fact that in
both cases the infrared bound is predicted to remain valid in all
dimensions.

Lace expansion methods have been used to obtain detailed information
about critical behaviour in dimensions $d>4$ for Ising and $\varphi^4$
models \cite{Saka07,Saka14}, as well as for other models above their
upper critical dimensions, including self-avoiding walk, lattice
trees, lattice animals, percolation, and the contact process
\cite{Slad06}.  The lace expansion is a robust method that does not
rely on reflection positivity (in fact, when the lace expansion
applies, it proves the infrared bound directly).  However, the lace
expansion is not effective at or below the upper critical dimension,
and for this the use of the renormalisation group approach appears to
be almost inevitable.  We write ``almost,'' because for 2-dimensional
models there have been major successes which bypass renormalisation
group methods that had often made prior predictions, including various
exact solutions \cite{Baxt82}, as well as the invention of ${\rm SLE}$
with its exploitation of conformal invariance (see, e.g.,
\cite{CS12}).

The $|\varphi|^4$ model is a natural testing ground for the
renormalisation group method.  Wilson applied his renormalisation
group approach to the $|\varphi|^4$ model \cite{Wils71II} and computed
its critical exponents in dimensions $d=4-\epsilon$ for small positive
$\epsilon$ \cite{WF72,WK74}.  Logarithmic corrections to scaling that
had been identified earlier in the critical dimension $d=4$
\cite{LK69} were computed using the Wilson approach in
\cite{BGZ73,WR73}.  In the mid-1980s, mathematically rigorous
implementations of the renormalisation group method were successfully
applied to the $n=1$ model in dimension $d=4$, for the case of small
coupling constant.  The first results were proofs of $|x|^{-2}$ decay
for the critical two-point function, independently by Gaw\c{e}dzki and
Kupiainen \cite{GK85,GK86} and by Feldman et al \cite{FMRS87}.
Logarithmic corrections to scaling were then proved by Hara and Tasaki
\cite{Hara87,HT87}, including
$\varepsilon^{-1}(\log\varepsilon^{-1})^{1/3}$ divergence of the
susceptibility as the critical point is approached.

In the present paper, we revisit the 4-dimensional $|\varphi|^4$
model, again for small coupling constant but now for general $n \geq
1$, and prove the following.  Precise statements are given in
Section~\ref{sec:mr}.
\begin{enumerate}
\item
The susceptibility diverges in the subcritical approach to the critical point $\varepsilon = \nu-\nu_c \downarrow 0$
as
\[
    \varepsilon^{-1}(\log\varepsilon^{-1})^{(n+2)/(n+8)}.
\]
This is consistent with the exponent $\frac 13$ mentioned above for $n=1$,
and with the exponent $\frac 14$ for the self-avoiding walk for $n=0$
(that $n=0$ corresponds formally to self-avoiding walk was pointed out in
\cite{Genn72}, but the results of \cite{BBS-saw4-log,BBS-saw4} are obtained without
assuming any formal correspondence).
\item
At and above the critical point, the pressure is approximately equal to that of
the Gaussian free field with
renormalised parameters.
 \item
 As $\varepsilon = \nu -\nu_c \downarrow 0$, the
 singular behaviour of the specific heat is given by
 \begin{alignat*}{2}
   &(\log \varepsilon^{-1})^{(4-n)/(n+8)}  &\quad&  (n=1,2, 3)\\
   &\log \log \varepsilon^{-1}  && (n=4)\\
   &1  && (n>4).
\end{alignat*}
\item
When the critical point is suitably approached, the scaling limit of the $|\varphi|^4$ field is
a multiple of a massive Gaussian free field on the continuum torus, while for
$\nu>\nu_c$, the scaling limit is white noise with intensity given by the susceptibility.
The convergence to white noise is
under standard central limit theorem rescaling, but the convergence to the
free field requires anomalous scaling, which is a manifestation of strong correlations.
\end{enumerate}
Our renormalisation group approach, which we describe in more detail in
Section~\ref{sec:method} below, is substantially different from the block spin approach
used in \cite{GK85,GK86} or the phase space expansion method of \cite{FMRS87}.
We plan to treat critical correlation functions, including the two-point function,
 in a future publication
\cite{BST-phi4}.

Our analysis applies at the critical point and in the high temperature
phase, but is restricted to small coupling constant $g>0$.  The
analysis of the low temperature phase is out of the reach of our
current methods.
Indeed, the development of robust techniques for the study of
the low temperature phase of models with continuous symmetry breaking
($n>1$) is an outstanding challenge.
The most promising way forward has been found by Ba{\l}aban who has
developed a renormalisation group scheme that is able to prove results
in the low temperature phase away from the critical point (see e.g.,
\cite{Bala83,BO99}, and \cite{Dimo13} for an overview
in a simpler context).

Although our results are proved only for isotropic nearest-neighbour spin coupling,
unlike with methods using reflection-positivity
this is not essential for our method of proof, and we expect that it would be
possible to obtain similar results for
more general non-isotropic models with finite-range interaction (see \cite{AKM12}
for related work without isotropy).
We do not pursue these potential generalisations here.

For quantum fields, the case of $d=4$ space-time dimensions is of
greatest physical interest, whereas for critical phenomena such as the
ferromagnetic phase transition it is $d=3$ that is most important.
Wilson and Fisher had the idea to study $d=3$ via a perturbation
around $d=4$, giving rise to the so-called $\epsilon$ expansion
\cite{WF72}.  In the $\epsilon$ expansion, the analysis is performed
in dimension $d=4-\epsilon$.  Rigorous analysis in non-integer
dimensions may seem problematic initially, but there is a way to mimic
fractional dimensions by use of long-range couplings.  It was pointed
out in \cite{FMN72} that a spin system in dimension $d$ whose
microscopic couplings are long range with decay $1/r^{d+\alpha}$ has
upper critical dimension $d_c=d_c(\alpha)=\min\{ 2\alpha,4\}$.  The
long range case corresponds to $\alpha \in (0,2)$, and critical
exponents were calculated for small positive $\epsilon = d_c -d$ in
\cite{FMN72}.  Rigorous results establishing mean field behaviour for
$d>d_c(\alpha)$ are given in \cite{AF88} (for Ising and $\varphi^4$
spins), in \cite{Saka14} (for $\varphi^4$), and in \cite{Heyd11,HHS08}
(for Ising spins, self-avoiding walk, percolation and emphasising the
connection with $\alpha$-stable processes for $\alpha<2$).  From this
perspective, it is possible to mimic a short range model in
$d=4-\epsilon$ dimensions by instead considering a long range model in
$d=3$ dimensions with $\alpha = \frac{3+\epsilon}{2}$, so that
$3=d_c-\epsilon$.  A rigorous version of this for $\varphi^4$ was
initiated in \cite{BMS03}, where for small $\epsilon$ a
non-perturbative construction of a non-Gaussian fixed point of the
renormalisation group map was achieved.  In \cite{Abde07}, the
renormalisation group trajectory from the Gaussian to the non-Gaussian
fixed point was fully constructed in this setting.  The related
construction of the non-Gaussian fixed point for the weakly
self-avoiding walk was carried out in \cite{MS08}. In
\cite{ACG13}, an analogous fractional dimension is tested in a
continuum hierarchical model context, that is for quantum fields in a
a p-adic continuum.  In particular, the anomalous dimension of the
composite operator $\phi^{2}$ is studied. These references provide a
first step towards the computation of critical exponents in dimension
$d_c-\epsilon$, and it would be of considerable interest to carry out
the remaining steps to obtain a mathematically rigorous version of the
$\epsilon$ expansion.

\subsection{Definition of model}
\label{sec:model}

We now give the precise definition of the $n$-component $|\varphi|^4$
model.  As usual in statistical mechanics, it is necessary to define
the model first in finite volume.  Let $n > 0$ and $d > 0$ be
integers.  Let $L >1$ be an integer (eventually chosen large), and let
$\Lambda_N$ be the $d$-dimensional discrete torus $\Z^d/L^N\Z^d$ of
side length $L^N$.  Ultimately we are interested in the thermodynamic
limit $N \to \infty$.

We view fields $\varphi$ either as functions $\varphi : \Lambda_N \to
\R^n$, or equivalently as vectors $\varphi \in (\R^n)^{\Lambda_N}$.
We use subscripts to index $x\in \Lambda$ and superscripts for the
component $i=1,\dots, n$.  We write $|v|$ for the Euclidean norm
$|v|^2 = \sum_{i=1}^n (v^i)^2$ and $v \cdot w = vw= \sum_{i=1}^n v^i
w^i$ for the Euclidean inner product on $\R^n$.  For $e\in\Z^d$ with
$|e|_1 = \sum_{i=1}^d {|e_i|} = 1$, we define the discrete gradient by
$(\nabla^e \varphi)_x =\varphi_{x+e}-\varphi_x$.  The discrete
Laplacian is defined by
$\Delta = -\frac{1}{2}\sum_{e\in\Z^d:|e|_1=1}\nabla^{-e} \nabla^{e}$, and we
write $\varphi_x(-\Delta \varphi)_x = \sum_{i=1}^n \varphi_x^i(-\Delta\varphi^i)_x$.
We use same symbol for the Laplacian $\Delta_{\Lambda}$ regarded as an
operator on scalar functions and on vector-valued functions, acting
diagonally in each component, i.e., if $h: \Lambda \to \R^n$ then
$(\Delta h)^i = \Delta h^i$.  We also identify $\Delta_{\Lambda}$ with
a matrix $(\Delta_\Lambda)_{x,y}, \, (x,y \in \Lambda)$; thus, as a
matrix, $\Delta_{\Lambda}$ is viewed as the scalar version of the
Laplacian.

Given $g>0, \nu \in \R$, we define a function $V_{g,\nu,N}$ of the fields by
\begin{equation} \label{e:Vdef1}
  V_{g,\nu,N}(\varphi)
  = \sum_{x\in\Lambda}
  \Big(\tfrac{1}{4} g |\varphi_x|^4 + \half \nu |\varphi_x|^2
  + \half \varphi_x(-\Delta \varphi_x)  \Big)
  .
\end{equation}
Then we define $P_{g,\nu,\Lambda}$ to be the probability measure on
$(\R^n)^{\Lambda_N}$ given by
\begin{equation}
  \lbeq{Pdef}
  P_{g,\nu,N}(d\varphi) = \frac{1}{Z_{g,\nu,N}} e^{-V_{g,\nu,N}(\varphi)} d\varphi,
\end{equation}
where $Z_{g,\nu,N}$ is a normalisation constant (the \emph{partition
  function}), and $d\varphi$ is the Lebesgue measure on
$(\R^n)^{\Lambda_N}$.  Expectation with respect to $P_{g,\nu,N}$ is
denoted $\pair{\,\cdot\,}_{g,\nu,N}$.  Then $\varphi$ is a field of
classical continuous $n$-component spins.
The measure $P_{g,\nu,N}$ defines the finite volume $|\varphi|^4$
model on the torus $\Lambda_N$, i.e., with periodic boundary
conditions.

The measures \refeq{Pdef} are related to other standard models in
statistical mechanics.  With an appropriate negative and $g$-dependent
choice of $\nu$, \refeq{Vdef1} corresponds to a nearest-neighbour
ferromagnetic interaction with single spin density
$e^{-g(|\varphi|^2-1)^2}$.  In the limit $g\to\infty$, these measures
converge to those of the Ising model ($n=1$), the rotor model ($n=2$),
and the classical Heisenberg model ($n=3$).  Conversely, for $n=1$,
the $\varphi^4$ model can be realised as a limit of Ising models
\cite{SG73}.

Two fundamental quantities are the \emph{pressure} and the \emph{susceptibility},
defined as the limits
\begin{align}
  \label{e:pressuredef}
  p(g,\nu) &= \lim_{N \to \infty} \frac{1}{|\Lambda_N|} \log Z_{g,\nu,N},
  \\
  \label{e:susceptdef}
  \chi(g, \nu)
  &= \lim_{N \to \infty} \sum_{x\in\Lambda_N}
  \pair{\varphi_0^1\varphi_x^1}_{g,\nu,N}
  = n^{-1} \lim_{N\to\infty} \sum_{x\in\Lambda_N}
  \pair{\varphi_0 \cdot \varphi_x}_{g,\nu,N}
  .
\end{align}
Existence of the limit defining the pressure has been proved under
quite general assumptions, including the $n$-component $|\varphi|^4$
model for any $d>0$, any $n \ge 1$, and any $g>0$ and $\nu\in \R$
\cite{LP76}.  Moreover the pressure is independent of the boundary
conditions, and in particular is identical to the infinite volume
limit of the pressure under free boundary conditions \cite{LP76}.  For
$n=1,2$, standard correlation inequalities \cite{FFS92} imply that the
pressure is convex, and hence also continuous, in $\nu$, and that for
the case of free boundary conditions the limit defining the
susceptibility exists (possibly infinite) and is monotone
non-increasing in $\nu$.  Proofs are lacking for $n>2$ due to a lack
of correlation inequalities in this case (as is discussed, e.g., in
\cite{FFS92}), but one expects that these facts known for $n\le 2$ are
true also for $n>2$.  In our theorems below, we prove the existence of
the infinite volume limit with periodic boundary conditions directly
in the situations covered by the theorems, without application of any
correlation inequalities.

\bigskip\noindent \emph{Asymptotic notation.}
Throughout the paper, we write $p \sim q$ to denote $\lim p/q =1$.
We also use the usual big-$O$ notation, in which all implied constants
are allowed to depend on the number of components $n \in \N$,
but are uniform in $g\in(0,\delta)$,$\nu\in (\nu_c,\nu_c+\delta)$,
$\varepsilon \in (0,\delta)$, for some small $\delta>0$.
Constants are also uniform in the scale parameter $j \in \N_0$
and the mass parameter $m^2\in(0,\delta)$ that begin to play important
roles in Section~\ref{sec:method}.

\subsection{Main results}
\label{sec:mr}

Our results concern the critical behaviour of the $n$-component
$|\varphi|^4$ model in dimension $d=4$ and for any $n \ge 1$, and
scaling limits of the measures $P_{g,\nu,N}$ with $\nu \downarrow \nu_c$
and with $\nu \to \nu_c+\varepsilon$, as $N \to \infty$.
The proofs are restricted to small $g>0$, but presumably our
conclusions should remain valid for all $g>0$.
Moreover, throughout this paper, we tacitly
assume that $L$ is chosen sufficiently large. This is needed,
in particular, for the results of \cite{BS-rg-IE,BS-rg-step}, on which our
results rely.

\subsubsection{Susceptibility}
\label{sec:mrsuscept}

In the first theorem, we identify a \emph{critical} value
$\nu_c=\nu_c(g,n)$ such that $\chi(g,\nu) \uparrow \infty$ as
$\nu \downarrow \nu_c$ with $\nu_c \sim -{\sf a}g$ for some
${\sf a}={\sf a}(n)>0$, and we identify the precise asymptotic
form of the divergence of $\chi$.

\begin{theorem} \label{thm:suscept}
  For $d=4$, $n \ge 1$, and for $g>0$ sufficiently small,
    there exist $\nu_c=\nu_c(g,n)< 0$ and $A=A(g,n)>0$ such that, as $\varepsilon \downarrow 0$,
  \begin{equation} \label{e:suscept4}
    \chi(g,\nu_c+\varepsilon) \sim A \varepsilon^{-1} (\log \varepsilon^{-1})^{(n+2)/(n+8)}.
  \end{equation}
  As $g \downarrow 0$,
  \begin{equation} \label{e:Anuc}
    A(g,n) = ({\sf b}g)^{(n+2)/(n+8)} (1+O(g)),
    \quad
    \nu_c(g,n) = -{\sf a} g + O(g^2),
  \end{equation}
  with ${\sf b}= (n+8)/(16\pi^2)$ and ${\sf a}=(n+2)(-\Delta_{\Z^4}^{-1})_{0,0} > 0$.
\end{theorem}

Some related bounds on critical values are obtained in \cite{FSS76}
using the infrared bound; our method does not use the infrared bound.
For dimensions $d>4$ and $n=1$,
it is shown in \cite{Saka14} that the formula
for $\nu_c$ in \refeq{Anuc} holds with ${\sf a} = 3(-\Delta_{\Zd}^{-1})_{0,0}$.
The exponent $\frac{n+2}{n+8}$ was predicted in the physics literature, see e.g.,
\cite[(A2.7)]{LK69}, \cite[(4.18)]{WR73}, \cite[(D16)]{BGZ73}.
For $n=1$, in \cite{Hara87,HT87}
the block spin renormalisation group approach of \cite{GK85,GK86} was adapted and
extended to confirm the exponent $\frac 13$ rigorously, and also to obtain
results for the correlation length and the renormalised coupling constant.

Throughout the remainder of the paper, we denote the exponent $\frac{n+2}{n+8}$ by
\begin{equation}
  \lbeq{gammadef}
  \gamma = \gamma(n) = \frac{n+2}{n+8}
  .
\end{equation}
The proof of Theorem~\ref{thm:suscept} is an extension of the proof of
an analogous statement with exponent $\gamma = \frac{1}{4}$ for the
$4$-dimensional weakly self-avoiding walk, given in
\cite{BBS-saw4-log}.  This is consistent with the well-known
interpretation of the self-avoiding walk as the $n \to 0$ limit
\cite{Genn72}, but our analysis of the weakly self-avoiding walk in
\cite{BBS-saw4-log} does not use any formal limit.

The constant ${\sf b}$ and the logarithm (but not its exponent) in
Theorem~\ref{thm:suscept} both arise in our proof from the free bubble
diagram, which is defined as follows.  First, the lattice Green
function is defined, for $m^2 > 0$, by
\begin{equation}
\label{e:lGf}
    C_{m^2}(x)
    = (-\Delta_{\Zd}+m^2)^{-1}_{0x}
    \qquad
    (x \in \Zd)
    .
\end{equation}
The inverse is bounded in $l^2(\Z^d)$-sense for $m^2>0$,
and the limit $m^2 \downarrow 0$ exists (pointwise in $x$) if $d>2$.
The free \emph{bubble diagram} is the squared
$\ell^2$ norm $B_{m^2}=\sum_{x \in \Zd} C_{m^2}(x)^2$.
Let $\T^d=[0,1]^d$ denote the unit torus, which is the Fourier dual space to $\Zd$.
We denote the Fourier multiplier of $-\Delta_\Zd$ by
  \begin{equation}
  \lbeq{FDel}
    \FDel(k)
    =
    4 \sum_{j=1}^{d} \sin^2 (\pi k_j)
    \qquad
    (k \in \T^d)
    .
  \end{equation}
By Parseval's formula and elementary calculus, for $d=4$ and
as $m^2 \downarrow 0$, the bubble diagram can be expressed as an
integral over the torus $\T^4=[0,1]^4$ as
\begin{equation}
  \label{e:freebubble}
  B_{m^2}
  =
  \int_{\T^4}
  \frac{1}{(\FDel(k) +m^2)^2}
  dk
  \sim
  \frac{\log m^{-2}}{16\pi^2} .
\end{equation}
It is the divergence of $B_{0}$ that makes $d=4$ more difficult than $d>4$,
for which differential inequalities or lace expansion methods have
been used to prove mean field behaviour $\varepsilon^{-1}$
(for $n=1,2$) instead of \refeq{suscept4}
\cite{Aize82,Froh82,Saka14}.
The constant ${\sf a}$ in \eqref{e:Anuc} can also be expressed in terms
of \refeq{FDel}, as
\begin{equation}
  {\sf a} = (n+2)(-\Delta_{\Z^4}^{-1})_{0,0} = (n+2)\int_{\T^4} \frac{1}{\FDel(k)}  dk.
\end{equation}

\subsubsection{Pressure and its derivatives}
\label{sec:mrpressure}

The next theorem compares the pressure for small positive $g$ with the pressure
for $g=0$ and renormalised parameter, and studies the singular behaviour of
its second derivative with respect to $\nu$.
We use the notation $\pair{A;B} = \pair{AB}-\pair{A}\pair{B}$ to denote
the \emph{covariance} or \emph{truncated expectation} of $A$ and $B$.
Assuming the derivatives exist and commute with the infinite volume limit,
the derivative
\begin{equation}
  -
  \ddp{}{\nu} p(g,\nu) = \lim_{N\to\infty} \half \langle |\varphi_x|^2 \rangle_{g,\nu,N}
\end{equation}
is half of the mean of $|\varphi_x|^2$ in the infinite volume limit, and
\begin{equation}
  \ddp{^2p}{\nu^2}(g,\nu)
  = \lim_{N\to\infty}
  \tfrac 14 \sum_{y\in\Lambda_N}
  \langle |\varphi_x|^2 ; |\varphi_y|^2 \rangle_{g,\nu,N}
\end{equation}
expresses the second derivative as a quarter of the covariance of $|\varphi_x|^2$
with its sum.
We call the second derivative the \emph{specific heat}, and write
\begin{equation} \label{e:cHdef}
  c_H(g,\nu) =  \ddp{^2p}{\nu^2}(g,\nu).
\end{equation}
This is not the usual definition of the specific
heat, but we expect it to have the same singular behaviour as the standard definition
involving differentiation of the pressure
with respect to temperature.

\begin{figure}\label{fig:pressure}
  \begin{center}
    \input{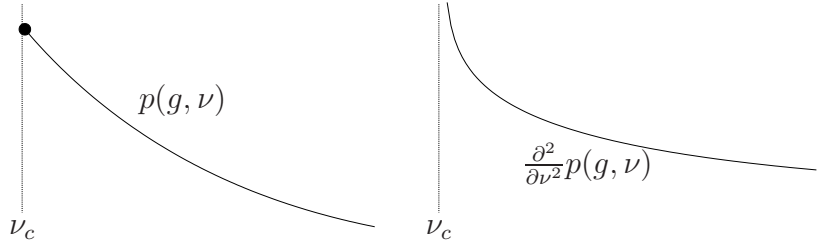}
  \end{center}
  \caption{Schematic plot of the pressure $p(g,\nu)$ and its second
    derivative (the specific heat) for fixed $g$.  For $n\leq 4$, the
    specific heat diverges slowly as $\nu \downarrow \nu_c$,
    while for $n>4$, it converges to a finite value.}
\end{figure}

The finite volume pressure is given by
\begin{equation}
\lbeq{pNdef}
  p_N(g,\nu) = |\Lambda_N|^{-1} \log Z_{g,\nu,N}.
\end{equation}
For $g=0$ and $m^2>0$,
exact evaluation of the Gaussian integral shows
that
\begin{equation}
  p_N(0,m^2) =
  -
  n|\Lambda_N|^{-1} \log \det(-\Delta_{\Lambda_N}+m^2)
  ,
\end{equation}
where $\Delta_{\Lambda_N}$ is the matrix corresponding to the Laplace
operator acting on scalar functions, as discussed in Section~\ref{sec:model}.
The eigenvalues of $-\Delta_{\Lambda_N}$ are given by
\refeq{FDel} for $k \in \T_N^d = \T^d \cap L^{-N}\Zd$,
and $|\T_N^d|=|\Lambda_N|$, so
\begin{equation}
  \lbeq{psum}
  p_N(0,m^2)
  = -
  n
  |\T_N^d|^{-1} \sum_{k \in\T_N^d}
  \log\left(
  \FDel(k)
  + m^2\right).
\end{equation}
The right-hand side is a Riemann sum, so for $m^2>0$ we obtain
\begin{equation}
\lbeq{pressure0}
  p(0,m^2)
  = \lim_{N\to\infty} p_N(0,m^2)
  =
  -
  n \int_{\T^d} \log\left(\FDel(k) + m^2\right)  dk.
\end{equation}
The integral on the right-hand side is absolutely convergent when $m^2=0$, and
we use its value to define also $p(0,0)$.
Differentiation of \eqref{e:pressure0} shows that the specific heat of the
non-interacting field is given by
\begin{equation}
\lbeq{cH0}
  c_H(0,m^2) = \ddp{}{m^2} p(0,m^2) = nB_{m^2} \sim \frac{n \log m^{-2}}{16 \pi^2},
\end{equation}
and thus diverges logarithmically as $m^2 \downarrow 0$.

For $g>0$ and $n=1$, it is known rigorously that the specific heat
diverges at most logarithmically in dimension $d=4$ \cite{Soka79}.
The following theorem shows that $c_H$ has the interesting
$n$-dependent asymptotic behaviour first predicted for the specific
heat in \cite[(A2.7)]{LK69} (see also \cite[(4.17)]{WR73}). In its
proof, we show that the derivatives indeed commute with the infinite
volume limit.  The theorem also shows that the pressure and its
derivative with respect to $\nu$ are close to the corresponding free
quantities with an effective mass given by the reciprocal of the
susceptibility.

\begin{theorem} \label{thm:pressure}
  Let $d=4$, $n \ge 1$, and let $\delta >0$ be sufficiently small.
    Let $g,\varepsilon \in (0, \delta]$ and $\nu =\nu_c+\varepsilon$.

  \smallskip\noindent
  (i) The limit \eqref{e:pressuredef} exists and is given by
  \begin{equation}
    p(g,\nu)
    =
    p(0,1/\chi)(1+ O(g))
    ,
  \end{equation}
  and $\lim_{\nu \downarrow \nu_c} p(g,\nu) = p(0,0)(1+  O(g))$.
  Here $\chi=\chi(g,\nu)$ is the susceptibility \refeq{susceptdef}.

  \smallskip\noindent
  (ii) The derivative of the pressure exists and satisfies
  \begin{equation}
    \ddp{}{\nu}p(g,\nu)
    =
    \half \lim_{N\to\infty} \langle |\varphi_0|^2 \rangle_{g,\nu,N}
    =
    \half C_{1/\chi}(0) (1+O(g))
    ,
  \end{equation}
  and $\lim_{\nu \downarrow \nu_c} \ddp{}{\nu} p(g,\nu) = C_0(0)(1+O(g))$.
  Here $C_{m^2}$ is the Green function \refeq{lGf}.

  \smallskip\noindent
  (iii) The specific heat \eqref{e:cHdef} exists, and there exists
  $D(g,n)>0$ such that, as $\varepsilon \downarrow 0$,
  \begin{equation}
  \lbeq{cHasy}
    c_H(g,\nu_c+\varepsilon)
    \sim D(g,n)
    \begin{cases}
      (\log \varepsilon^{-1})^{(4-n)/(n+8)} & (n=1,2, 3)\\
      \log \log \varepsilon^{-1} & (n=4)\\
      1 & (n>4).
    \end{cases}
  \end{equation}
\end{theorem}

The proof shows that $D(g,n)$ satisfies, as $g\downarrow 0$,
\begin{equation}
  \lbeq{Dgn}
  D(g,n) = (1+O(g))
  \begin{cases}
    \frac{n}{2(4-n)}
    \left( \frac{n+8}{16\pi^2} \right)^{(4-n)/(n+8)}
    \shift g^{-(2n+4)/(n+8)}
    & (n=1,2,3) \\
    \frac{1}{6} g^{-1} & (n=4)\\
    \frac{n}{2(n-4)}g^{-1} & (n>4).
  \end{cases}
\end{equation}
In particular, $D(g,n) \to \infty$ as $g\downarrow 0$, compatible with
the fact that $c_H$ diverges more rapidly as $\nu \downarrow \nu_c$
than the behaviour for $g=0$ given in \refeq{cH0}.

As discussed below \eqref{e:susceptdef}, for $n=1,2$, it is known that
$p(g,\nu)$ is a convex function of $\nu$. In particular, $p(g,\nu)$ is
continuous in $\nu$ in this case, and Theorem~\ref{thm:pressure}(i)
implies $p(g,\nu_c) = \lim_{\nu\downarrow \nu_c} p(g,\nu) = p(0,1/\chi)(1+O(g))$.
To show right-continuity of $p(g,\nu)$ at $\nu=\nu_c$ also for $n>2$ would
require further development of our methods.

The logarithmic divergence proved in \eqref{e:cHasy} is difficult to
observe numerically, even in the most divergent case
of $n=1$ \cite{LM09}.  For $n=4$, numerical identification of
$\log\log\varepsilon^{-1}$ would be impossible.

\subsubsection{Scaling limits}
\label{sec:mrclt}

Next, we consider scaling limits of the field $\varphi$,
and prove a central limit theorem, valid for small $g>0$ and
for any $n\ge 1$.
Its more elementary statement applies in the subcritical regime
$\nu >\nu_c$, and shows that the
scaling limit of the field $\varphi$ is equal to white noise on the continuum
torus $\T^d$, with intensity equal to the susceptibility.
Its deeper statement concerns the scaling limit in the vicinity of the critical point,
and shows that under an anomalous scaling the scaling limit is the \emph{massive}
Gaussian free field (GFF) on the continuum torus.

The GFF was much studied
in the 1970s and 1980s
as the point of departure for the rigorous construction of Euclidean quantum
field theories \cite{Simo79,GJ87}.  Recently it has received renewed interest
from mathematicians, notably due to its connections with ${\rm SLE}$ in
dimension $d=2$ \cite{Shef07}.  The massless GFF on the torus $\T^d$ is the Gaussian measure
with covariance $(-\Delta_{\T^d})^{-1}$, suitably defined to omit the
zero mode.
It is supported on distributions (continuous
linear functionals on the space $C^\infty(\T^d,\R^n)$, not
\emph{probability} distributions!).
We require the $n$-component \emph{massive} GFF, which we parametrise to have
covariance $(-m^{-2}\Delta_{\T^d}+1)^{-1}$
(as an operator on $C^\infty(\T^d,\R^n)$), and we write
${\rm GFF}_{m^2}$ for its measure.
Also,
for $a>0$ we write $\dot W_a$ for the $n$-component white noise with intensity $a$.
Its covariance is given by $a$ times the identity operator on $C^\infty(\T^d,\R^n)$.
With our parametrisation, ${\rm GFF}_{m^2}$
converges to $\dot W_1$ as $m^2 \to \infty$.

We identify the discrete torus $\Lambda=\Lambda_N$
of side length $L^N$ with the subset $\T_N^d$ of the continuum unit torus
$\T^d = \R^d/\Z^d$ with lattice spacing (mesh size) $L^{-N}$.
Note that here $\T_N^d$ arises as position space, as opposed to Fourier
space as in \refeq{psum}.
The scaling limit is formulated in terms of
continuum limits on the discrete torus $\T_N^d$, as the lattice spacing
goes to zero.
We write $\varphi^{(N)}$ for the field on $\T_N^d$ induced by $\varphi$, i.e.,
$\varphi^{(N)}(x)=\varphi_{L^{-N}x}$ for $x \in \T_N^d$.
Given a test function $h: \T^d \to \R^n$, we write
\begin{equation}
    \varphi^{(N)}(h) =
    \sum_{x\in\T^d_N} h(x) \cdot \varphi^{(N)}(x),
\end{equation}
and in the continuous case we write $\varphi(h)$ for the pairing
of the distribution $\varphi$ with $h$.
Then $\dot W_a$ and ${\rm GFF}_{m^2}$ are characterised by their Laplace transforms, which are
\begin{align}
\lbeq{covs}
    \dot W_a( e^{ \varphi(h)} )
    & =
    e^{\frac a2 (h,h)}
    \\
    {\rm GFF}_{m^2} ( e^{\varphi(h)})
    &=
    e^{\frac 12 (h,(-m^{-2}\Delta_{\T^d}+1)^{-1}h)}
    ,
\end{align}
where $(\cdot,\cdot)$ is the inner product on $L^2(\T^d, \R^n)$.

It is natural to rescale the random variable $\varphi^{(N)}(h)$ by its standard
deviation.  We do this in an $h$-independent manner, by using the
$n$-component constant test function
$\1=(1,0,\dots,0)$ in the rescaling.
The variance of $\varphi^{(N)}(\1)$ is
\begin{equation}
\lbeq{sigN}
    \pair{ (\varphi^{(N)}(\1))^2}_{g,\nu,N} =  L^{dN} \chi_N
    \quad\text{with}\quad
    \chi_N =  \sum_{x\in \Lambda_N}\pair{\varphi_0^1 \varphi_x^1}_{g,\nu,N}.
\end{equation}
It is convenient in the following theorem to use the infinite volume
susceptibility instead of $\chi_N$, in the rescaling.

\begin{theorem} \label{thm:clt}
  Let $d = 4$, $n \ge 1$, and let $g > 0$ be sufficiently small.
  Let $m^2 > 0$ and $\varepsilon_N \in (0,\delta)$ with small $\delta$,
  let $\chi^{(N)}=\chi(g, \nu_c+\varepsilon_N)$,
  and set $\sigma_N^2= L^{dN}\chi^{(N)}$.
  There exists $\rho>0$ such that, for any $h \in C^\infty(\T^d,\R^n)$,
\begin{equation}
\lbeq{clt}
    \lim_{N\to\infty}
    \left\langle
    e^{  \varphi^{(N)}(h)/\sigma_N }
    \right\rangle_{g,\nu_c+\varepsilon_N,N}
    =
    \begin{cases}
    \dot W_1 ( e^{ \varphi(h)} )
    & \text{if $\chi^{(N)} L^{-2N} \to 0$}
    \\
    {\rm GFF}_{m^2} ( e^{\varphi(h)} )
    & \text{if $\chi^{(N)} L^{-2N} \to \rho m^{-2}>0$}
    .
    \end{cases}
\end{equation}
In particular:\\
\smallskip\noindent
(i) If $\varepsilon_N \to\varepsilon$ for some $\varepsilon>0$,
then $\sigma_N \sim L^{2N}\sqrt{\chi(g,\nu_c+\varepsilon)}$ and
  \begin{equation}
   \label{e:cltm}
    \lim_{N\to\infty}
    \left\langle e^{ \varphi^{(N)}(h)/\sigma_N } \right\rangle_{g,\nu_c+\varepsilon_N,N}
    =
    \dot W_1 ( e^{ \varphi(h)} ).
  \end{equation}
(ii) There exists $\alpha>0$ such that if
$\varepsilon_N \sim \alpha m^2 L^{-2N} (\log L^N)^{(n+2)/(n+8)}$
with $m^2>0$,
then $\sigma_N \sim \rho^{1/2}m^{-1} L^{3N}$ and
  \begin{equation}
  \label{e:cltm2}
    \lim_{N\to\infty}
    \left\langle e^{ \varphi^{(N)}(h)/\sigma_N } \right\rangle_{g,\nu_c+\varepsilon_N,N}
    =
    {\rm GFF}_{m^2} ( e^{\varphi(h)} ).
  \end{equation}
\end{theorem}

Theorem~\ref{thm:clt} is a statement of convergence of Laplace transforms,
and as such, implies convergence of moments. 
In particular,
  \begin{alignat}{2}
    \sigma_N^{-1} \varphi^{(N)}(h)
    &\stackrel{D}{\Rightarrow}
    \mathrm{N}\left(0, 
      (h,h)\right)
    &&\qquad \text{in case~(i)},
    \\
    \sigma_N^{-1} \varphi^{(N)}(h)
    &\stackrel{D}{\Rightarrow}
    \mathrm{N}\left(0, (h, (-m^{-2}\Delta+1)^{-1}h) \right)
    &&\qquad \text{in case~(ii)},
  \end{alignat}
where $\mathrm{N}(0,\sigma^2)$ denotes
a normal random variable with mean $0$ and variance $\sigma^2$,
and the convergence is in distribution.
Note that for fixed $\varepsilon \in (0,\delta)$, \refeq{clt} gives a statement of
convergence of $L^{-2N}\varphi^{(N)}$ to Gaussian white noise with intensity equal
to the susceptibility $\chi(g,\nu_c+\varepsilon)$.

The scaling for $\nu >\nu_c$ in Theorem~\ref{thm:clt}(ii) is standard
central limit scaling, since $L^{-2N}\varphi^{(N)}(h) = |\T_N^4|^{-1/2} \sum_{x\in \T_N^4}
h(x) \cdot \varphi^{(N)}(x)$.
The critical case in Theorem~\ref{thm:clt}(iii) has anomalous scaling, and this
is a manifestation of the strong correlations of the
random variables $\varphi_x$ as $\nu \downarrow \nu_c$.
The convergence to white noise in the subcritical case was
proved much earlier for $n=1$, for any $d\geq 2$ and any $g>0$,
as a consequence of the FKG inequality \cite{Newm80}.  Our proof is different.
Theorem~\ref{thm:suscept} provides the precise asymptotics for the
divergence of the intensity
of the white noise as $\nu \downarrow \nu_c$.

Our method would require further development in order to prove that
the correlation length, defined as
$\pair{\varphi_0\varphi_x}_{\nu} \approx e^{-|x|/\xi(\nu)}$,
has the predicted behaviour
$\xi(\nu_c+\varepsilon) \approx \varepsilon^{-1/2} (\log \varepsilon^{-1})^{\gamma/2}$
with $\gamma$ given by \refeq{gammadef}.
This behaviour is proved for $n=1$ in \cite{HT87}.
However, Theorem~\ref{thm:clt}(iii) is
consistent with the predicted behaviour, in the following sense.
Since the ${\rm GFF}_{m^2}$ field is correlated over distances of order $1$,
we expect
the discrete model to have correlation length of
the same order $L^N$ as the side of the torus, and
solving for $\xi \approx L^N$ in the formula
for $\varepsilon_N$ gives the predicted behaviour
 $\xi \approx (\varepsilon^{-1} N^{\gamma})^{1/2}
\approx \varepsilon^{-1/2} (\log \varepsilon^{-1})^{\gamma/2}$.

Theorem~\ref{thm:clt} considers the \emph{scaling limit} in which the lattice spacing
approaches $0$
while the original (``bare'')
coupling constants
remain fixed, as is natural in the context of statistical mechanics.
This limit is different from the continuum limit of interest in quantum field theory,
in which the coupling constants are adjusted as the lattice spacing approaches $0$ in such a way
as to obtain a non-trivial limit. In different terminology, we address the \emph{infrared problem}
rather than the \emph{ultraviolet problem}.

The condition that $\delta$ be small should
not be necessary for the white noise limit in Theorem~\ref{thm:clt},
but we have made no effort to remove it. It
arises because some ingredients of the proof of Theorem~\ref{thm:clt} were written with the
primary goal to study the more difficult $\varepsilon \downarrow 0$ limit,
as in Theorem~\ref{thm:clt}(ii) and Theorems~\ref{thm:suscept}--\ref{thm:pressure}.

\subsection{Wilson's approach}
\label{sec:wilson}

In this section, we briefly recall Wilson's general strategy, and mention some of
the ingredients in our implementation of that strategy.  These ingredients are
discussed at much greater length in Sections~\ref{sec:method}--\ref{sec:rgmap} below.

The general problem of understanding the macroscopic behaviour
of a system consisting of an infinite number of interacting degrees of freedom
remains the great challenge in statistical physics, today as much as 40
(or 140) years ago.
The phenomena of greatest interest, both
physically and mathematically, often arise at or near critical points
where the correlation length is infinite and interactions over all length scales remain important.

Wilson's renormalisation group approach
provides a general quantitative strategy to ``thin'' degrees of freedom:
the reduction of irrelevant microscopic details to reveal macroscopic properties.
For the Ising model, a proposal to thin degrees of freedom via block spins
was introduced by Kadanoff \cite{Kada66}.
Wilson was inspired by Kadanoff's block spin approach \cite{Kada66},
but emphasised in \cite{Wils71I} that it should not be taken
too literally, stating in particular:
\begin{quote}
``In short the Kadanoff block picture, although
absurd, will be the basis for generalizations which
are not absurd.''
\end{quote}
Four years later, in \cite{Wils75b}, after comparison of the Kadanoff picture
with the result of detailed computations, he revised this view and wrote:
\begin{quote}
``Thus the old idea of Kadanoff that there would be
effective nearest-neighbor Ising models for block spins is
very close to the truth.''
\end{quote}
The detailed computations showed that the nearest-neighbour
and the next nearest-neighbour block spin
couplings were dominant, with longer range couplings nonzero but
qualitatively of lesser importance.
In a mathematically rigorous analysis, the qualitative must be made
quantitative, and estimates are required to prove that long range couplings
are truly dominated by short range couplings.

Rather than the \emph{real-space}
renormalisation of block spins, in
\cite{Wils71II} Wilson used a \emph{momentum-space} (Fourier transform) renormalisation,
in which he computed a partition function by successive integration over
momenta shells $2^{-(j+1)}\le |k| \le 2^{-j}$.  This is essentially a covariance
decomposition in momentum space.
The first rigorous implementation
of the renormalisation group used a real-space covariance decomposition \cite{BCGNOPS78}.
Our analysis also uses a real-space covariance decomposition, but a different one
discussed below.
Block spins have been used with success in rigorous analysis, including
\cite{GK85,GK86,Hara87,HT87} for the 4-dimensional
$\varphi^4$ model; an earlier example is the application to
the $(\nabla \phi)^4$ model for dimensions $d \ge 2$ in \cite{GK80}.

For simplicity, we restrict to $n=1$ in the following discussion;
the generalisation to $n\ge 1$ is straightforward.
The density in \eqref{e:Pdef} can be rewritten as
\begin{equation} \label{e:eVG}
    e^{-V_{g,\nu,N}(\varphi)}
    =
    e^{-\frac 12 (\varphi,(-\Delta +m^2)\varphi)}
    e^{-\sum_x ( \frac 14 g|\varphi_x|^4+\frac 12(\nu-m^2)|\varphi_x|^2)}.
\end{equation}
The first factor on the right-hand side is proportional to the density of
the Gaussian measure with covariance $(\Delta +m^2)^{-1}$, and the second factor
provides a perturbation to the Gaussian measure.
The parameter $m^2 >0$ is required for the Gaussian measure to be
non-degenerate (normalisable) in finite volume,
but temporarily we are not careful about this point.
Moments of Gaussian measures can be evaluated explicitly in
terms of the covariance, via Wick's theorem \cite{GJ87}.
In a naive attempt to understand
the measures \eqref{e:Pdef}, one might try to expand the exponential in the perturbation
as
\begin{equation}
    e^{-\sum_x ( \frac 14 g|\varphi_x|^4+ \frac 12  \nu|\varphi_x|^2)}
    = 1 - \sum_x (\tfrac 14 g |\varphi_x|^4 + \half \nu|\varphi_x|^2) + \cdots
\end{equation}
and evaluate all moments that arise in this way individually by Wick's theorem.
However, there are serious difficulties with this approach.
One is that the covariance matrix $(-\Delta+m^2)^{-1}$ becomes long range
as $m^2\downarrow 0$, and as a consequence many of the
moments encountered will diverge as $m^2\downarrow 0$. An example is the moment
$\sum_{x,y} E(\varphi_x^4\varphi_y^4)$.  In its calculation, the
bubble diagram \eqref{e:freebubble} appears,
with its logarithmic divergence as $m^2 \downarrow 0$ when $d=4$.
This difficulty is more severe than lack of convergence of an infinite series:
there are divergences in individual terms.

Wilson's solution to this problem involves splitting the Gaussian field with covariance $(-\Delta+m^2)^{-1}$ into
two parts, one corresponding to small distances (high momenta)
and one corresponding to large distances (low momenta),
and then first integrating the short distance part \cite{Wils71II,WK74}.
A convenient formulation in probabilistic terms was given in \cite{BCGNOPS78}.
To describe this, we first recall the elementary fact from probability theory that if $X_1$ and $X_2$ are Gaussian
random vectors with covariances $C_1$ and $C_2$, then $X_1+X_2$ is also Gaussian and its covariance is $C_1+C_2$.
Given a covariance matrix $C$,
let $P_C$ denote the Gaussian probability measure with covariance $C$,
and let $\Ex_C$ denote the corresponding expectation.
We write $\Ex_C\theta F$  for the convolution of $F$ with $P_C$,
i.e., given $F \in L^1(P_C)$,
\begin{equation}
\lbeq{thetadef}
  (\Ex_C\theta F)(\varphi) = \Ex_C F(\varphi+\zeta),
\end{equation}
where the expectation $\Ex_C$ acts on $\zeta$ and leaves $\varphi$ fixed.
It is thus a conditional expectation.
Wilson's approach amounts to splitting the Gaussian field in distribution as $\varphi = \varphi_1 + \zeta_1$,
or equivalently the covariance as $(-\Delta)^{-1} = C_{1} + C_1'$, where the
\emph{fluctuation field}  $\zeta_1$ of covariance $C_1$ is the rapidly varying
(short range) part of
$\varphi$.
Let $V_0 = \sum_x (\frac{1}{4}g|\varphi_x|^4+ \half \nu|\varphi_x|^2)$.
Then the Gaussian integral with covariance $C=C_1+C_1'$ can be performed progressively, as
\begin{equation}
  \lbeq{CC1}
  \Ex_C e^{-V_0} = \Ex_{C_1'+C_1} e^{-V_0} = \Ex_{C_1'} \Ex_{C_1}\theta e^{-V_0}.
\end{equation}
The integral over $\zeta_1$ can be performed without
difficulty---at least on the formal level. By the cumulant expansion, formally,
\begin{align} \label{e:cumexpan}
  \Ex_{C_1} \theta e^{-V_0}
  & \approx
  e^{ - \Ex_{C_1}\theta V_0 + \frac 12 \Ex_{C_1}\theta ( V_0;  V_0)  + \cdots}
  ,
\end{align}
where the second term in the exponent on the right-hand side of \eqref{e:cumexpan} is
the \emph{truncated expectation} (or \emph{variance})
\begin{equation}
  \label{e:Etrunc}
  \Ex_{C_1} \theta ( V_0;  V_0)
  =
  \Ex_{C_1} \theta  V_0^2
  -
  (\Ex_{C_1} \theta V_0)^2.
\end{equation}

To paraphrase Wilson, one argues now that the right-hand side of \refeq{cumexpan}
can be effectively approximated in terms of
a renormalised polynomial $V_1$ of the form
\begin{equation}
\lbeq{V1}
    V_1(\varphi) = \sum_x \left(
    \tfrac{1}{4} g_1|\varphi_x|^4 + \half\nu_1 |\varphi_x|^2 + \half z_1 \varphi_x(\Delta\varphi)_x + u_1
    \right)
\end{equation}
as
\begin{equation}
\label{e:C1int}
    \Ex_{C_1} \theta e^{-V_0}
    \approx
    e^{ - \Ex_{C_1}\theta V_0 + \frac 12 \Ex_{C_1}\theta ( V_0;  V_0)  + \cdots}
    \approx
    e^{-V_1(\varphi_1) + \cdots}.
\end{equation}
The expectation generates the $z_1$ and $u_1$ terms that were not present initially in $V_0$.
The field $\varphi_1$ is then rescaled as
$\varphi_1(x) \mapsto L^{(d-2)/2}\varphi_1(Lx)$, so that the rescaled field will
resemble the original field.
This procedure of conditional expectation and rescaling can be repeated, with the original potential changing slightly in each step.
The renormalised coupling constants can be computed to any order in the formal expansion
\eqref{e:cumexpan}.  The renormalisation group transformation is the map that expresses
the new coupling constants in terms of the old ones.
Now, the analogue of the bubble diagram $B_{m^2}$ is a bubble diagram
$\beta_j$ defined in terms of the partial covariance $C_j$ rather than the original
full covariance, and this partial bubble diagram is finite and bounded in $j$.
For $d=4$, this leads in particular to an evolution
\begin{align}
  g_{j+1} &=  g_j - \beta_j g_j^2 + \cdots 
  ,
\end{align}
with well-behaved coefficients $\beta_j$.
In our particular implementation,
$\sum_{j} \beta_j$ is a multiple of the bubble diagram.
Wilson's crucial observation is that of the terms generated by this
\emph{renormalisation group transformation}, only finitely many are
important, here $g_j,\nu_j,z_j,u_j$ as in \refeq{V1}. After
rescaling, the other terms \emph{contract} under the renormalisation
group transformation and no detailed information is required.
The study of the flow of the important (\emph{relevant} and \emph{marginal}) terms
then provides accurate information about the macroscopic properties of the system.

Critical theories are scale invariant,
and are characterised as fixed points of the renormalisation
group transformation. In our example, the fixed point $V_0=0$ corresponds to the
massless Gaussian free field.
Initial conditions $V_0$ such that $V_j \to 0$ appropriately are said to belong
to the same universality class, and the vanishing of the interaction in the limit
goes by the name of \emph{infrared asymptotic freedom}.
Thus Wilson's renormalisation group relates the
long-distance behaviour of spins, or their universality classes,
to fixed points and their domains of attractions
of the renormalisation group transformation $V_j \mapsto V_{j+1}$.
It is predicted that in dimensions $d\geq 4$ there are exactly two renormalisation
group fixed points, the massless free field and white noise.
Our focus is on the critical dimension $d=4$.
As $d$ decreases, the number of fixed points increases.
Wilson and Fisher argued that in dimension $d=4-\epsilon$, with small $\epsilon>0$,
a non-Gaussian fixed point arises \cite{WF72}.
A rigorous version of this is given in \cite{BMS03,Abde07}.
In dimension $d=2$, there are infinitely many fixed points (conformal
field theories).

There are serious mathematical difficulties in the rigorous implementation
of this procedure.
Our approach follows Wilson's general strategy, and develops general mathematical methods
for its implementation.
The decomposition of the field is achieved via a decomposition of the covariance of
the free field on $\Lambda$, in a particularly convenient finite-range manner \cite{BGM04,Baue13a}.
 To control approximations, suitable function spaces and norms
are developed in \cite{BS-rg-norm}.  In the second approximate equality of \refeq{C1int},
a nonlocal functional of the fields is approximated by the local polynomial $V_1$,
which should capture all the relevant and marginal components.
In \cite{BS-rg-loc}, we develop a general approach to such approximations, which can
be achieved via an operator we call $\LT$.  Our method for performing perturbative
calculations such as \refeq{cumexpan} is laid out in \cite{BBS-rg-pt}.
This involves replacement of the formal expression \eqref{e:cumexpan}
by an exact second order version
with remainder term, in the spirit of Taylor's formula.
To obtain control uniformly in the volume, we represent error terms using a
\emph{polymer gas}, a widely successful concept in statistical mechanics that provides
a generalised notion of locality in which Wilson's strategy can
be put to work in a non-perturbative manner.  This is the technically most demanding
part of our approach, and is carried out in \cite{BS-rg-IE,BS-rg-step}.
Inclusion of all error terms in the renormalisation group transformation leads to
an infinite dimensional dynamical system, which we study in \cite{BBS-rg-flow}.
In Sections~\ref{sec:method}--\ref{sec:rgmap}, we describe some of these ingredients
in greater detail.
In Section~\ref{sec:pf}, we assemble the ingredients to prove
Theorems~\ref{thm:suscept}--\ref{thm:clt}.

\section{Elements of renormalisation group approach}
\label{sec:method}

We now summarise some basic elements of
our rigorous implementation of the renormalisation
group approach.

\subsection{Approximation by renormalised free field}
\label{sec:approxgff}

An adjustment is needed to implement the notion of infrared asymptotic freedom
mentioned in Section~\ref{sec:wilson}.
The vanishing of $V_j$ as $j\to\infty$ leaves only the
Gaussian, which in the limits of infinite volume and $m^2 \downarrow 0$ has
covariance exactly $-\Delta^{-1}$, the Gaussian free field.  It cannot be
expected that the GFF  exactly describes the large distance behaviour of the critical
$|\varphi|^4$ model in dimension $4$.
Rather, there should be a renormalised scaling factor, such as $\rho\neq 1$ in
Theorem~\ref{thm:clt}(i).  We anticipate this fact by making an initial change of variables.
This change of variables is our implementation of the \emph{wave function renormalisation}
in the physics literature.

For this, it is convenient to generalise \eqref{e:Pdef} and define
\begin{equation}
  V_{g,\nu,z}(\varphi) = \tfrac{1}{4} g |\varphi_x|^4 + \half\nu|\varphi_x|^2 + \half z \varphi_x(-\Delta\varphi)_x .
\end{equation}
By definition, for any $m^2 > 0$ and for any $z_0 > -1$,
\begin{equation} \label{e:Vsplit}
  V_{g,\nu,1}(\varphi_x)
  = V_{0,m^2,1}((1+z_0)^{-1/2}\varphi_x)
  + V_{g_0,\nu_0,z_0}((1+z_0)^{-1/2}\varphi_x)
  ,
\end{equation}
where we have set
\begin{equation} \label{e:g0gnu0nu}
  g_0 = g(1+z_0)^2, \quad \nu_0 = (1+z_0)\nu-m^2
  .
\end{equation}
We define, for $X \subset \Lambda$,
\begin{equation}
\lbeq{V0Z0}
    V_0(X) = V_0(\varphi,X) = \sum_{x\in X} V_{g_0,\nu_0,z_0}(\varphi_x), \quad
  Z_0(\varphi) = e^{-V_{0}(\varphi,\Lambda_N)}
  .
\end{equation}
Let $C=(-\Delta_{\Lambda_N}+m^2)^{-1}$.
By making the change of variables $\varphi_x \mapsto (1+z_0)^{1/2}\varphi_x$,
and writing $\tilde F (\varphi) = F((1+z_0)^{1/2}\varphi)$, we obtain
\begin{equation}
\lbeq{ExF}
    \pair{F}_{g,\nu,N}
    = \frac{\Ex_C \tilde F Z_0}{\Ex_C Z_0} .
\end{equation}
We evaluate such expectations by separate evaluation of the numerator
and denominator on the right-hand side, and this becomes our principal task.
We define a generalisation of the denominator by
\begin{equation}
\label{e:ZNdef}
  Z_N(\varphi) = \Ex_{C}\theta Z_0 = \Ex_C Z_0(\varphi +\zeta),
\end{equation}
with $\theta$ the convolution operator of \refeq{thetadef} (i.e., the
expectation on the right-hand side of the last equality acts on $\zeta$).
Then $Z_N(0)=\Ex_CZ_0$.

\subsection{Decomposition of free field}
\label{sec:decomp}

We use decompositions of both of the covariances
$(-\Delta_\Zd + m^2)^{-1}$ and $(-\Delta_{\Lambda_N} +m^2)^{-1}$, where $\Lambda_N$ is
the torus of side length $L^N$.
For $\Zd$, this Green function exists for $d>2$ for all
$m^2 \ge 0$, but for finite $\Lambda$ we restrict to $m^2>0$.
In \cite[Section~\ref{pt-sec:Cdecomp}]{BBS-rg-pt},
we use results from \cite{Baue13a,BGM04} to
define a sequence
$(C_j)_{1 \le j < \infty}$ (depending on $m^2 \ge 0$)
of positive definite covariances on $\Zd$
such that
\begin{equation}
\lbeq{ZdCj}
    (\Delta_\Zd + m^2)^{-1} = \sum_{j=1}^\infty C_j
    \quad
    \quad
    (m^2 \ge 0).
\end{equation}
The covariances $C_j$ are translation invariant,
and have the \emph{finite-range} property
\begin{equation}
  \label{e:frp}
  C_{j;x,y} = 0 \quad \text{if \; $|x-y| \geq \frac{1}{2} L^j$}
  .
\end{equation}
For $j<N$, the covariances $C_j$ can therefore be identified with
covariances on $\Lambda$, and we use both interpretations.
For $m^2>0$, there is also a covariance $C_{N,N}$ on $\Lambda$ such that
\begin{equation}
\lbeq{NCj}
    (-\Delta_\Lambda + m^2)^{-1} = \sum_{j=1}^{N-1} C_j + C_{N,N}
    .
\end{equation}
Thus the finite volume decomposition agrees with the infinite volume
decomposition except for the last term in the finite volume decomposition.

The covariances $C_j$ satisfy a number of estimates which are important
for the analysis.
We write
$\nabla_x^\alpha=\nabla_{x_1}^{\alpha_1} \dotsb \nabla_{x_d}^{\alpha_d}$ for a
multi-index $\alpha=(\alpha_1,\dotsc,\alpha_d)$, where
$\nabla_{x_k}$ denotes the
finite-difference operator $\nabla_{x_k}f(x,y)=f(x+e_k,y)-f(x,y)$.
The number
\begin{equation}
\lbeq{dimphi}
    [\varphi]=\frac {d-2}{2}
\end{equation}
is referred to as the \emph{scaling dimension} or
\emph{engineering dimension} of the field, or, more briefly, simply as
the field's \emph{dimension}.
It is shown in \cite[Proposition~\ref{pt-prop:Cdecomp}]{BBS-rg-pt}
that for multi-indices $\alpha,\beta$ with
$\ell^1$ norms $|\alpha|_1,|\beta|_1$ at most
some fixed value $p$, for $j <N$, and for any $k \in \N$,
\begin{equation}
  \label{e:scaling-estimate}
  |\nabla_x^\alpha \nabla_y^\beta C_{j;x,y}|
  \leq c(1+m^2L^{2(j-1)})^{-k}
  L^{-(j-1)(2[\varphi]+(|\alpha|_1+|\beta|_1))},
\end{equation}
where $c=c(k)$ depends on $k$ but is independent of $j$.
The same bound holds for $C_{N,N}$ if
$m^2L^{2(N-1)} \ge \delta$ for some $\delta >0$,
with $c$ depending on $\delta$ but not on $N$.
We thus consider $\delta>0$ as a fixed parameter, and consider the
covariances in \eqref{e:NCj} as fixed functions
of $m^2$ in the interval $\Iint_j$ defined by
\begin{equation}
\lbeq{massint}
    \Iint_j = \begin{cases}
    [0,\delta) & (j<N)
    \\
    [\delta L^{-2(N-1)},\delta) & (j=N).
    \end{cases}
\end{equation}
With $m^2 \in \Iint_j$,
\eqref{e:scaling-estimate} holds for $j\leq N$, including the covariance $C_{N,N}$.
Constants are permitted to depend on $\delta$.

We define the \emph{mass scale} $j_m$ by
\begin{equation}
     j_m =
     \begin{cases}
     \lfloor \log_L m^{-1}\rfloor & (m>0)
     \\
     \infty & (m=0).
    \end{cases}
\end{equation}
Thus $j_m$ is the largest integer $j$ such that $mL^{-j} \le 1$,
and $\lim_{m \downarrow 0}j_m=\infty$.
For fixed $\Omega >1$ ($\Omega=2$ is a good choice), the factor
$(1+m^2L^{2(j-1)})^{-k}$ is bounded above by a multiple of
\begin{equation}
\lbeq{chicCovdef}
    \chicCov_j = \Omega^{-(j-j_m)_+},
\end{equation}
where $x_+=\max\{x,0\}$.
In our estimates, we find it convenient to capture the fast decay of \refeq{scaling-estimate}
beyond the mass scale using $\chicCov_j$.
(The left-hand side of \refeq{chicCovdef} is comparable to
$\chi_j=\Omega^{-(j-\jm)_+}$ used in \cite{BBS-rg-pt,BS-rg-IE,BS-rg-step},
by \cite[Proposition~\ref{pt-prop:rg-pt-flow}]{BBS-rg-pt};
we reserve the Greek letter $\chi$
for the susceptibility in the present paper, and employ $\chicCov_j$
rather than $\chi_j$.)

The finite-range property of the covariance decomposition can be contrasted with
the \emph{block spin} method used in \cite{GK85,GK86},
in which the fluctuation covariances $C_j$ are chosen such the fields
$\varphi_j$ are constant over blocks of side length $L^j$.
Block spin covariances decay exponentially, but do not have the finite-range property
\eqref{e:frp}.
In our setup, fields are only \emph{approximately} constant over blocks by \eqref{e:scaling-estimate},
but this is compensated by an independence property that
allows for an effective construction of a renormalisation group map,
using independence rather than cluster expansion.
To state the independence property, we make the following definitions.
First, we let
\begin{equation}
\lbeq{Ncaldef}
    \Ncal = \Ncal(\Lambda) = C^{p_\Ncal}((\R^n)^\Lambda,\R)
\end{equation}
denote the space of real-valued functions of the fields having at least
$p_\Ncal$ continuous derivatives, where $p_\Ncal$ is a fixed integer.
More generally, for $X \subset \Lambda$, we set
$\Ncal(X)=C^{p_\Ncal}((\R^n)^X,\R)$.
Then for $F \in \Ncal(X)$
and $G \in \Ncal(Y)$ such that $\dist(X,Y) > \frac 12 L^j$,
\begin{equation}
\label{e:Exfac}
  \Ex_{C_j}\theta(FG) = (\Ex_{C_j}\theta F)(\Ex_{C_j}\theta G),
\end{equation}
since uncorrelated Gaussian random
variables are independent.

In addition, the following extension of \refeq{CC1} holds:
\begin{equation}
    \label{e:progressive}
    \Ex_{C}\theta F
    =
    \big( \Ex_{C_{N,N}}\theta \circ \Ex_{C_{N-1}}\theta \circ \cdots
    \circ \Ex_{C_{1}}\theta\big) F
    .
\end{equation}
This expresses the expectation on the left-hand side as a progressive
integration.
To compute the expectations on the right-hand side of \eqref{e:ExF}, we use
\refeq{progressive}
to evaluate it progressively.  Namely, if we define
\begin{equation}
\label{e:Z0def}
  Z_{j+1} = \Ex_{C_{j+1}}\theta Z_j \quad\quad
  (j<N),
\end{equation}
with $Z_0 = e^{-V_0(\Lambda)}$ as in \refeq{V0Z0},
then, consistent with \eqref{e:ZNdef},
\begin{equation}
\label{e:ZN}
    Z_N = \Ex_C\theta Z_0.
\end{equation}
Thus we are led to study the recursion $Z_j \mapsto Z_{j+1}$.
To simplify the notation, we use the short-hand notation $\Ex_{j} = \Ex_{C_j}$,
and leave implicit the dependence of the covariance $C_j$
on the mass $m$.

\subsection{Marginal and relevant directions}
\label{sec:margrel}

A field functional $M_x$ is said to be a \emph{local field monomial} (located at $x$) if
\begin{equation}
\lbeq{Mx}
  M_x = \prod_{k=1}^m \nabla^{\alpha_k} \varphi_{x}^{i_k}
\end{equation}
for some integer $m$, where $\alpha_i$ are multi-indices.
The \emph{dimension} of $M_x$
is defined to be
$[M_x] = \sum_k ([\varphi] + |\alpha_k|)$, with $[\varphi]$ given by \refeq{dimphi}.
A local field monomial is said to be
\begin{alignat*}{2}
  \text{\emph{relevant}} &\quad \text{if $[M_x] < d$,}\\
  \text{\emph{marginal}} &\quad \text{if $[M_x] = d$,}\\
  \text{\emph{irrelevant}} &\quad \text{if $[M_x] > d$.}
\end{alignat*}
We include the degenerate case of the empty product in \refeq{Mx}, which defines the constant
monomial $1$, of dimension zero.
These definitions are motivated by the fact that, roughly,
assuming that $\nabla^{\alpha_k} \varphi_x^{i_k}$  typically has size
$|\nabla^{\alpha_k} C_{j;x,x}^{1/2}| \approx L^{-j([\varphi]+|\alpha_k|)}$
as in \refeq{scaling-estimate},
\begin{equation}
  \sum_{x\in B} \Ex_{C_j} M_x \approx L^{(d-[M_x])j},
\end{equation}
where $B \subset \Lambda$ is a cube of side length $L^j$.
Examples of relevant monomials for $d=4$ are $1$, $\varphi$, $\varphi(\nabla\varphi)$,
$\varphi^2$, and $\varphi^3$, but in practice the set of relevant monomials
is limited by symmetry requirements.
A \emph{local polynomial} is a finite sum of local monomials with
constant coefficients.
The following local polynomials
play a central role in our analysis:
\begin{align}
  &
  1, \quad \tau_x = \half |\varphi_x|^2, \quad \tau_x^2 = \tfrac14 |\varphi_x|^4
  ,
  \nnb &
  \tau_{\Delta,x} = \half \varphi_x \cdot (-\Delta \varphi)_x
  , \quad
  \tau_{\nabla\nabla,x} =
  \tfrac14 \sum_{e\in\Z^d:|e|_1=1} \nabla^e\varphi_x \cdot \nabla^e \varphi_x
  .
\end{align}

Two important symmetries are Euclidean and $O(n)$ invariance, discussed next.

\smallskip \noindent
\emph{Euclidean symmetry:}
Let $\Ecal$ denote the set of
lattice automorphisms $E:\Lambda\rightarrow \Lambda$;
these are bijections that preserve nearest neighbours.
An automorphism $E$ induces an action
on $\Ncal$ via  $(EF)(\varphi)=F(E\varphi)$,
where
$(E\varphi)_{x} = \varphi_{Ex}$.  A local monomial $M$ is \emph{Euclidean
invariant} if automorphisms that fix $x$ also fix $M_{x}$.
For example, $\nabla_{e}\varphi$ is not Euclidean
invariant because there is a reflection that changes $\varphi_{x+e}$
into $\varphi_{x-e}$ so that $(\nabla_{e}\varphi)_{x} \mapsto
(\nabla_{-e}\varphi)_{x}$.  The monomials
$\tau_{\Delta}$ and $\tau_{\nabla\nabla}$ are Euclidean invariant.

\smallskip \noindent
\emph{$O(n)$ symmetry:}
Let $M(n)$ denote the set of $n\times n$ real matrices, and let $O(n)$ be the
group
of orthogonal matrices.
We write the matrix elements of $T\in M(n)$ as $T_{ij}$.
The action of $T$ on $\Ncal$ is induced by the action
$\varphi \mapsto T\varphi$, defined by the matrix multiplication $(T\varphi)^i
=\sum_{j=1}^n T_{ij}\varphi^j$, via $(TF)(\varphi)=F(T\varphi)$.
We say that $F\in \Ncal$ is
\emph{$O(n)$ invariant} if
$AF=F$
for every $A \in O(n)$.
For example, $F(\varphi)=|\varphi_x|^2$ is $O(n)$ invariant, but $F(\varphi)=\varphi_x^1$
is not.

\smallskip
In dimension $d=4$, the general local polynomial which is
Euclidean and $O(n)$ invariant and consists only of relevant local field monomials
has the form $\nu\tau+ u1 = \nu \tau +u$ (we will omit $1$ for the constant
monomial henceforth), whereas for marginal monomials it is
$g\tau^2+z\tau_\Delta+ y\tau_{\nabla\nabla}$.
Thus the general Euclidean and $O(n)$ invariant local
polynomial consisting of relevant and marginal monomials in $d=4$ has the form
\begin{equation}
    \label{e:Vdef-bis}
    U = g\tau^2 + \nu\tau + z\tau_{\Delta} + y\tau_{\nabla\nabla} + u
    ,
\end{equation}
and we define the 5-dimensional linear space $\Ucal$ to consist
of the \emph{local polynomials}
of the form \refeq{Vdef-bis}.  For $U \in \Ucal$, we write
\begin{equation}
  \label{e:Vdef-bisbis}
  U(X) = \sum_{x \in X}U_x
  .
\end{equation}
Typically we write $V$ for elements of $\Ucal$ for which $y=u=0$,
and we write $\Vcal \subset \Ucal$ for the subspace of such elements.

The notion of marginal, relevant, and irrelevant directions is reflected in our
analysis via the application of norms which are defined as follows.
Let $\bar\Lambda = \Lambda \times \{1,\ldots,n\}$, and let
$\Lambda^*$ denote the set of finite sequences of elements of $\bar\Lambda$.
Given $x=(x_1,i_1,\ldots, x_p,i_p)\in \Lambda^*$, we write $x!=p!$ and
\begin{equation}
    F_{x}(\varphi)
    = \frac{\partial^p }{\partial \varphi_{x_p}^{i_p} \cdots  \partial \varphi_{x_1}^{i_1}}
    F(\varphi).
\end{equation}
Functions $f:\Lambda^* \rightarrow \R$ are called
\emph{test functions}.  We define a pairing
between elements of $\Ncal$ (field functionals)
and the set of test functions as follows:
for $F\in\Ncal$,
for a test function $f$, and for $\varphi \in (\R^n)^{\Lambda}$, let
\begin{equation}
    \label{e:pairing}
    \pair{F, f}_\varphi
    = \sum_{x\in \Lambda^*}
    \frac{1}{x!}F_{x}(\varphi) f_{x}.
\end{equation}

We define a normed space of test functions $\Phi_j(\ell_j)$ as follows
(see \cite[Section~\ref{IE-sec:reg}]{BS-rg-IE} and
\cite[Section~\ref{norm-sec:tf}]{BS-rg-norm}).
We set $\ell_j = \ell_0 L^{-j[\varphi]}$ for an appropriate constant $\ell_0$
(it turns out convenient to take it large and $L$-dependent),
fix an integer $p_\Phi \ge 0$ and write
$p(x)$ for the number of components of $x\in \Lambda^*$.
The $\Phi_j(\ell_j)$-norm of a test function $f$ is then defined by
\begin{equation} \label{e:Phinormdef}
  \|f\|_{\Phi_j(\ell_j)} = \sup_{x\in \Lambda^*: \; p(x) \leq p_\Ncal}
  \sup_{\alpha:|\alpha|_1 \leq p_\Phi}
  \ell_j^{-p(x)}   L^{j|\alpha|}  |\nabla^\alpha f_x|
  .
\end{equation}
Given $\varphi \in (\R^n)^\Lambda$, the $T_\varphi=T_{\varphi,j}$ semi-norm is defined by
\begin{equation} \label{e:Tphidef}
  \|F\|_{T_{\varphi,j}} = \sup_{f  : \|f\|_{\Phi_j}=1} |\langle F, f\rangle_\varphi|
  .
\end{equation}
This semi-norm is called the $T_\varphi(\ell_j)$ semi-norm in \cite{BS-rg-IE,BS-rg-step},
where a $T_\varphi(h_j)$ semi-norm is also required with a different parameter $h_j$.
Properties of the $T_\varphi$ semi-norm are systematically developed in \cite{BS-rg-norm}.
In particular, it has the product property
$\|FG\|_{T_\varphi} \le \|F\|_{T_\varphi} \|G\|_{T_\varphi}$.

A particularly important instance of the $T_\varphi$ semi-norm is the $T_0$ semi-norm obtained by setting $\varphi=0$,
as a measure of the size of $F$ when the field is \emph{small}.
In particular, direct computation shows that for $d=4$ (with $\ell_0$-dependent constants),
\begin{gather}
\lbeq{relmon}
  L^{dj} \|1\|_{T_{0,j}} \asymp L^{4j},
  \quad
  L^{dj} \|\tau_x\|_{T_{0,j}} \asymp L^{2j},
  \quad \text{(relevant)}
  \\
\lbeq{marmon}
  L^{dj} \|\tau_x^2\|_{T_{0,j}} \asymp 1,
  \quad
  L^{dj} \|\tau_{\Delta,x}\|_{T_{0,j}} \asymp 1,
  \quad
  L^{dj} \|\tau_{\nabla\nabla,x}\|_{T_{0,j}} \asymp 1,
  \quad \text{(marginal)}
  \\
\lbeq{irrmon}
  L^{dj} \|\tau_x^3\|_{T_{0,j}} \asymp L^{-2j},
  \quad \text{(irrelevant)}
\end{gather}
where we write $a \asymp b$ to denote the existence of $c>0$
such that $c^{-1} a \leq b \leq c a$.
The scaling in \refeq{Phinormdef} is designed to make relevance, marginality, and irrelevance
visible from the size of the norm, as above.  Thus,
in our setup, scaling takes place in the norms and we do not rescale the field.
With this in mind, we define a
norm on $\Ucal$ by
\begin{equation}
\lbeq{Vnorm}
  \|U\|_{\Ucal_j} = \max\{|g|,L^{2j}|\nu|,|z|,|y|,L^{dj}|u|\}.
\end{equation}
For $V \in \Vcal \subset \Ucal$, we also write
$\|V\|_{\Vcal_j} = \|V\|_{\Ucal_j} = \max \{|g|,L^{2j}|\nu|,|z|\}$.

\subsection{Localisation}
\label{sec:loc}

To extract the relevant and marginal parts of an arbitrary, possibly nonlocal element
of $\Ncal$,
we use the projection $\LT$ defined and studied in \cite{BS-rg-loc}.
The operator $\LT$ projects Euclidean- and $O(n)$-invariant
functionals of the field onto
the space of field polynomials spanned by the relevant and marginal
local field monomials of the form \eqref{e:Vdef-bis}, as follows.

The pairing \eqref{e:pairing} provides an interpretation of
$F \in \Ncal$ as a linear
function $f \mapsto \pair{F,f}_{0}$ on test functions, where the
subscript means $\varphi =0$. Given a set $\Phipol$ of test
functions, two elements $F_{1},F_{2} \in\Ncal$ are called \emph{equivalent}
if they define the same linear function on $\Phipol$, and otherwise they
are \emph{separated}.
We are interested in polynomial test functions, but these cannot be defined on
the entire torus $\Lambda$. To avoid this issue,
we restrict to a subset $\Lambda' \subset \Lambda$ whose diameter is
strictly less than the period of $\Lambda$.
This permits $\Lambda'$ to be identified with a subset of $\Zd$,
and therefore permits polynomial test functions to be defined on $\Lambda'$.
Note that two different types of
``polynomial'' are in use: a test function can be a polynomial in
$x\in \Lambda'$, while local polynomials are polynomial in fields.
We define $\Phipol$ in such a way that it is a
minimal space of test functions that separates
all relevant and marginal monomials of the form \refeq{Mx}.
Namely, we define $\Phipol$ to be the set of polynomial test functions $f$
such that $f_x$ is nonzero only if its polynomial degree plus the number of components of
$x$ is at most $d=4$.
Let $\Scal$ be the vector space of local polynomials
that are separated by $\Phipol$ and, for $X\subset \Lambda'$, let
$\Scal(X) = \{P (X):P\in \Scal \}$. The following proposition
associates to any $F \in \Ncal(\Lambda')$ an equivalent local polynomial in
$\Scal (X)$.

\begin{prop}
\label{prop:9LTdef} For
nonempty $X \subset \Lambda'$, there exists a unique linear map
$\LT_{X}: \Ncal(\Lambda') \rightarrow \Scal(X)$ such that
\begin{align}
\label{e:Locpair}
    &\quad
    \pair{\LT_{X} F, f}_0 = \pair{F,f}_0
    \quad
    \text{for $F \in
    \Ncal(\Lambda')$,
    $f \in \Phipol$}.
\end{align}
This map obeys
\begin{align}
\label{e:LocXX}
    &\quad
    (\LT_{X}\circ \LT_{X'}) F= \LT_{X} F \quad
    \text{for $F \in \Ncal(\Lambda')$,
    $X,X' \subset \Lambda'$},
    \\
\label{e:LocE}
    &\quad
    E\big(\LT_{X} F\big) = \LT_{EX} (EF) \quad
    \text{for $F \in \Ncal(\Lambda')$,
    $E\in \Ecal$},
    \\
\label{e:LocOn}
    &\quad
    T\big(\LT_{X} F\big) = \LT_{X} (TF) \quad
    \text{for $\Ncal(\Lambda')$,
    $T\in M(n)$}
    .
\end{align}
\end{prop}

\begin{proof}
The existence and uniqueness of $\LT$ obeying \refeq{Locpair} is
established in \cite[Definition~\ref{loc-def:LTsym}]{BS-rg-loc}.
The fact that $\LT$ obeys \refeq{LocXX}--\refeq{LocE} is proven in
\cite[Propositions~\ref{loc-prop:LT2}, \ref{loc-prop:9LTdef}]{BS-rg-loc}.
The case of \refeq{LocOn} is not discussed in \cite{BS-rg-loc}, so we
sketch the proof here.  First we define $T^{t\otimes} : \Phipol\to\Phipol$
by
\begin{equation}
    (T^{t\otimes} f)_{x_1,i_1,\ldots,x_p,i_p}
    =
    \sum_{j_1,\ldots,j_p=1}^n T_{j_1,i_1}\cdots T_{j_p,i_p}
    f_{x_1,j_1,\ldots,x_p,j_p}.
\end{equation}
This has the property that for any $F\in\Ncal(\Lambda)$,
for any test function $f$, and for any $T \in M(n)$,
\begin{equation}
    \pair{TF,f}_0=\pair{F,T^{t\otimes}f}_0.
\end{equation}
With \refeq{Locpair}, this gives
\begin{equation}
    \pair{T(\LT_XF),f}_0 = \pair{\LT_XF,T^{t\otimes}f}_0= \pair{F,T^{t\otimes}f}_0
    = \pair{TF,f}_0.
\end{equation}
The uniqueness in \refeq{Locpair} then implies that
$T(\LT_XF)=\LT_X(TF)$, and the proof is complete.
\end{proof}

It is a consequence of \refeq{LocE}--\refeq{LocOn} that
when restricted to Euclidean and $O(n)$ invariant elements of $\Ncal$, the
range of $\LT_X$ reduces to the space $\Ucal(X)$ of
polynomials of the form of $U$ in \eqref{e:Vdef-bis}.
Proposition~\ref{prop:9LTdef} asserts the existence of $\Loc_X F$, but it
does not provide an explicit formula.  Nevertheless it is not difficult in
practice to compute $\Loc_X F$ explicitly
when needed.
For example,
\begin{equation}
    \LT_{X}
    |\varphi_y|^4
    = |X|^{-1} \sum_{x\in X}|\varphi_{x}|^4,
    \qquad
    \LT_X
    |\varphi_x|^6
    = 0,
\end{equation}
and, more generally, monomials of degree higher than $4$ are annihilated by $\LT$.
Less trivially,
suppose that $q: \Lambda \to \R$ vanishes if $|x|> \half \diam{\Lambda}$
and that it satisfies, for some $q^{(**)} \in \R$,
\begin{equation}
  \label{e:qprop1}
  \sum_{x \in \Lambda} q_x x_{i} = 0,
  \quad\quad
  \sum_{x \in \Lambda} q_x x_{i}x_{j}
  = q^{(**)} \delta_{i,j},
  \quad\quad\quad
  i,j \in \{1,2,\dotsc ,d \}.
\end{equation}
Then, as in \cite[Section~\ref{loc-sec:locex}]{BS-rg-loc},
\begin{align}
  \label{e:LTF3}
  \LT_{x}
  \left[
    \sum_{y \in \Lambda} q_{x-y} \tau_{y}
  \right]
  &=
  q^{(1)}\tau_{x}
  +
  q^{(**)} (\tau_{\nabla\nabla,x}-\tau_{\Delta,x})
\end{align}

We need to know and take advantage of the fact
that $1-\LT_X$ projects onto irrelevant polynomials.
Our main tool to show this is
\cite[Proposition~\ref{loc-prop:LTKbound}]{BS-rg-loc}, which requires
that we choose the parameter $p_\Phi$ in the definition of the $\Phi$ norm
to obey $p_\Phi \ge\frac{1}{2}d+2$.
A specific example is given in
\cite[\eqref{loc-e:1-LTex}]{BS-rg-loc}, which asserts that
if $F \in \Ncal(X)$ and $Y \subset X \subset \Lambda'$, then
\begin{align}
  \lbeq{1-LTex}
  \|
  (1-\LT_Y) F
  \|_{T_{0,j+1}}
  &\le
  O(L^{-d-1})
  \|F\|_{T_{0,j}}
  ,
\end{align}
with a constant that depends only on $L^{-j}\diam{X}$.
In this sense,
$(1-\LT_X)F$ measured on the scale $j+1$ is significantly smaller than $F$ measured
on scale $j$. Although
\cite[Proposition~\ref{loc-prop:LTKbound}]{BS-rg-loc}
does not play an explicit role
in the present paper, it is crucial in the main result of \cite{BS-rg-step},
and our results here depend on \cite{BS-rg-step}.

\subsection{Blocks, polymers, and circle product}
\label{sec:IK}

To prepare for a multiscale analysis,
we partition $\Lambda = \Z^d/L^N\Z^d$ into a
disjoint union of $L^{d(N-j)}$ scale-$j$ \emph{blocks} of side length $L^j$,
for $j=0,1,\ldots,N$, and denote the
set of all such blocks by $\Bcal_j(\Lambda)$.
A scale-$j$ \emph{polymer} is a union of
scale-$j$ blocks,
and we write
$\Pcal_j=\Pcal_j(\Lambda)$ for the set of scale-$j$ polymers.
The empty set $\varnothing$ is a polymer, as is $\Lambda$.
A polymer $X$ is \emph{connected} if for any $x,y\in X$ there exists a path
$x_0=x,x_1,\ldots,x_{n-1},x_n=y$ with $\|x_{i+1}-x_i\|_\infty = 1$ for all $i$.
Every polymer can be partitioned into connected components, and we denote
set of connected components of $X$ by ${\rm Comp}_j(X)$.
The set of blocks in a polymer $X$ is denoted $\Bcal_j(X)$, and similarly $\Pcal_j(X)$
is the set of polymers formed from blocks in $\Bcal_j(X)$.

\begin{figure}
\begin{center}
\includegraphics[scale = 0.4]{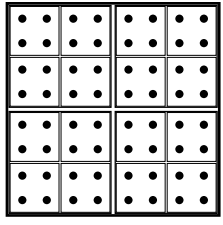}
\end{center}
\caption{\lbfg{reblock}
Illustration of ${\cal B}_j(\Lambda)$ for
$j=0,1,2,3$ when $d=2$, $N=3$, $L=2$.}
\end{figure}

We work with maps $F:\Pcal_j(\Lambda) \to \Ncal(\Lambda)$.
The Euclidean and $O(n)$ symmetries defined in Section~\ref{sec:margrel} extend
to such maps, as follows.
For $E \in \Ecal$, we define $EF:\Pcal_j \to \Ncal$ by $(EF)(X,\varphi)=F(EX,E\varphi)$,
and we say that $F$ is Euclidean invariant if
$EF=F$ for all $E \in \Ecal$.  Also, we say that
$F$ is $O(n)$ invariant if $F(X)$ is $O(n)$ invariant for all $X\in\Pcal_j$,
i.e., if $A(F(X))=F(X)$ for all $X \in \Pcal_j$ and $A\in O(n)$.

For maps $F,G: \Pcal_j(\Lambda) \to \Ncal(\Lambda)$, we define the \emph{circle product}
$F\circ G: \Pcal_j(\Lambda) \to \Ncal(\Lambda)$ by
\begin{equation}
\lbeq{circdef}
  (F \circ G)(X) = \sum_{Y \in \Pcal_j(X)} F(X \setminus Y) G(Y)
  \qquad
  (X \in \Pcal_j(\Lambda)).
\end{equation}
We assume that $F(\varnothing)=1$ for every function
$\Pcal_j(\Lambda) \to \Ncal(\Lambda)$.
The circle product is commutative and associative, with unit element $\1_\varnothing$
defined by $\1_\varnothing(X) = 1$ if $X=\varnothing$ and otherwise $\1_\varnothing(X)=0$.

We define
\begin{equation} \label{e:I0K0}
  I_0(X) = e^{-V_0(X)}, \quad K_0(X) = \1_{\varnothing}(X).
\end{equation}
With this notation, we can write $Z_0=e^{-V_0(\Lambda)}$ as
\begin{equation} \label{e:Z0I0K0}
  Z_0 = I_0(\Lambda) = (I_0 \circ K_0)(\Lambda).
\end{equation}
We wish to maintain the form
of \eqref{e:Z0I0K0} after each expectation in the progressive expectation
\refeq{progressive}.
Namely, we seek to define polynomials $V_j$, interaction
functionals
$I_j=I_j(V_j): \Pcal_j(\Lambda) \to \Ncal_j(\Lambda)$,
remainders  $K_j : \Pcal_j(\Lambda) \to \Ncal_j(\Lambda)$, and constants $u_j \in \R$
such that
$Z_j$ of \refeq{Z0def} is given by
\begin{equation}
\lbeq{IcircKnew}
    Z_j = e^{-u_j|\Lambda|}(I_j\circ K_j)(\Lambda).
\end{equation}
If we set  $\delta u_{j+1} = u_{j+1}-u_j$,
then \refeq{Z0def} can equivalently be written as
\begin{equation} \label{e:IcircKdu}
  \Ex_{j+1}\theta(I_j \circ K_j)(\Lambda)
  =
  e^{-\delta u_{j+1}|\Lambda|}(I_{j+1} \circ K_{j+1})(\Lambda)
  .
\end{equation}
The formula \refeq{IcircKnew} is our exact and well-defined replacement for \eqref{e:C1int}.
In addition, we desire factorisation properties of the form
\begin{equation}
  I_j(X) = \prod_{B \in \Bcal_j(X)} I_j(B),
  \quad
  K_j(X) = \prod_{U \in {\rm Comp}_j(X)} K_j(U).
\end{equation}

The \emph{interaction}
$I_j$ captures the relevant and marginal directions, and is a function of
a local polynomial
\begin{equation}
\lbeq{V}
    V_j = \tfrac{1}{4} g_j|\varphi|^4 + \half \nu_j |\varphi|^2 + \half z_j \varphi(-\Delta\varphi)
\end{equation}
in the fields, whereas $K_j$ is a remainder.
As in Section~\ref{sec:margrel}, we denote the space of polynomials of
the form \refeq{V} by $\Vcal$.
The main difficulty lies in making these definitions
in such a way that it can be
proved that the important behaviour is encapsulated in
$I_j$, with estimates that guarantee that
the \emph{non-perturbative coordinate} $K_j$ is an error term.
In view of \refeq{IcircKnew}, and since $I_j$ is to be determined by $V_j$, we are led
to study the \emph{renormalisation group map}
\begin{equation}
\lbeq{RGu0}
    (V_j,K_j) \mapsto (\delta u_{j+1},V_{j+1},K_{j+1}).
\end{equation}

\section{Renormalisation group map}
\label{sec:rgmap}

We now discuss the definition of the renormalisation group map \refeq{RGu0}
and its important properties.
These properties are used in Section~\ref{sec:pf} to prove Theorems~\ref{thm:suscept}--\ref{thm:clt}.

\subsection{Perturbative coordinate}
\label{sec:pt}

Our choice of the map $(V_j,K_j) \mapsto (\delta u_{j+1},V_{j+1})$ is explicit
and is a small modification of the choice developed in detail for the weakly
self-avoiding walk in \cite{BBS-rg-pt,BS-rg-step}.
It is convenient to unite the coordinates $\delta u$ and $V$ within the larger class
$\Ucal$ of local polynomials defined in \refeq{Vdef-bis}, and
we identify $\Ucal \cong \R^5$ via $U \cong (g,\nu,z,y,\delta u)$.

The map $(V_j,K_j) \mapsto (\delta u_{j+1},V_{j+1})$ is defined in terms of
a simpler map $V \mapsto \Upt\in\Ucal$,
that corresponds to the case $K_j=0$.  The generalisation to $K_j \neq 0$
will be discussed later.
To prepare for the definition of the map $\Upt$, we first introduce some notation
related to Gaussian integration. We define
\begin{equation}
\label{e:LapC}
    \Lcal_C =
    \sum_{i=1}^n
    \sum_{u,v \in \Lambda}
    C_{u,v}
    \frac{\partial}{\partial \varphi_{u}^i}
    \frac{\partial}{\partial \varphi_{v}^i}
\end{equation}
and, for polynomials $A=A(\varphi)$, $B=B(\varphi)$, we define
\begin{equation}
  \label{e:FCAB}
    F_{C}(A,B)
    = e^{\Lcal_C}
    \big(e^{-\Lcal_C}A\big)
    \big(e^{-\Lcal_C}B\big) - AB
    .
\end{equation}
The well-known connection
between Gaussian integration and the heat equation leads to the fact
(see \cite[Lemma~\ref{norm-lem:*heat-eq}]{BS-rg-norm} for a proof) that
for a polynomial $A=A(\varphi)$,
\begin{equation}
  \Ex_C \theta A = e^{\Lcal_C}A.
\end{equation}
The operator $e^{-\Lcal_C}$ is equivalent to Wick ordering \cite{GJ87}, namely
$e^{-\Lcal_C}A = :\!\!A\!\!:_C$, so $F_C$ is related to
truncated expectation (or \emph{covariance}) by
\begin{equation}
  F_C(:\!\!A\!\!:_C, :\!\!B\!\!:_C)
  =
  \Ex_C \theta(A;B)
  = \Ex_C\theta (AB) - (\Ex_C\theta A)(\Ex_C\theta B).
\end{equation}
We find it convenient to work with $F_C$ rather than using Wick ordering.

Given a finite range decomposition $C= \sum_{j=1}^N C_j$ as in Section~\ref{sec:decomp},
we define $w_j=\sum_{i=1}^j C_i$ and $w_0=0$.
For $V \in \Vcal$ and $X \in \Pcal_j(\Lambda_N)$, we then set
\begin{equation}
  \label{e:WLTF}
  W_j(V,X) = \frac 12 \sum_{x\in X} (1-\LT_{x}) F_{w_j}(V_x,V(\Lambda)).
\end{equation}
The range of $w_j$ is the same as that of $C_j$, namely $\frac 12 L^j$, so
by \refeq{Exfac}, $W_j(V,B) \in \Ncal(B^+)$, where $B^+$ denotes the union of
the block $B$ with all other blocks $B'$ such that $B \cup B'$ is connected.
(The definition \refeq{WLTF} cannot be applied when $j=N$, since the diameter
condition of Proposition~\ref{prop:9LTdef} is then violated; an appropriate
alternate definition for the final scale is
provided in \cite[Section~\ref{IE-sec:finalscale}]{BS-rg-IE}.)
Then, for $X \in \Pcal_j$,
we set
\begin{equation}
  \lbeq{Idef}
  I_j(V,X) = e^{-V_j(X)}\prod_{B \in \Bcal_j(X)}(1+W_j(V,B)).
\end{equation}
For the degenerate case $j=0$, for which $w_0=0$, we interpret the above
as $I_0(V,X)=e^{-V(X)}$.
This definition has the following properties:
\begin{itemize}
\item
  Field locality: $I_j(B) \in \Ncal(B^+)$
  for each block $B \in \Bcal_j$;
\item
  Symmetry: $I_j$ is $O(n)$ invariant and Euclidean invariant;
\item
  Block factorisation: $I_j(X) = \prod_{B \in \Bcal_j(X)} I_j(B)$ for $X \in \Pcal_j$.
\end{itemize}

The map $V \mapsto \Upt\in\Ucal$ is defined by
\begin{equation}
  \lbeq{Vptdef}
  \Upt = e^{\Lcal_{j+1}} V - P_j(V)
  ,
\end{equation}
where
\begin{align}
\label{e:PdefF}
    P_j(V,X)
    &= \sum_{x \in X}
    \LT_x \left(
    e^{\Lcal_{C}} W_j(V,x)
    + \frac 12
    F_{C}
    (e^{\Lcal_{C}} V_x,e^{\Lcal_{C}} V(\Lambda))
    \right)
    ,
\end{align}
and $P_j(V)$ is identified with an element in $\Ucal$ in \eqref{e:Vptdef}.
This is possible because of
the appearance of $\LT$ in \eqref{e:PdefF} which ensures that $P_j \in \Ucal$,
i.e., that $P_j(V,X)$ arises as $Q(X)$ for some $Q \in \Ucal$ that is independent of $X$.
The definition of $\Upt$ in \refeq{Vptdef} is subtle but is motivated in
\cite{BBS-rg-pt}, where it is shown that it has been designed so as to
have the desirable property that
\begin{equation} \label{e:Iex}
  \Ex_{j+1}I_j(V,\Lambda) = I_{j+1}(\Upt, \Lambda) + O(V^3),
\end{equation}
where the equality is as formal power series in the coupling constants, with error terms
containing a product of at least three coupling constants.
(The presence of the constant term in $\Upt$ here, absent in \cite{BBS-rg-pt},
does not affect the applicability of \cite{BBS-rg-pt}.)
Equation~\refeq{Iex} shows that, to second order, $I$ enjoys a form of stability under expectation,
when $V$ is advanced to $\Upt$.  However, no uniformity in scale $j$ or volume $\Lambda$
is implied in \refeq{Iex}, and both of these defects must be remedied.

Detailed estimates on $W_j$ and $I_j$ are provided in \cite{BS-rg-IE}.
In \cite{BS-rg-IE}, the context is the weakly self-avoiding walk, but
the estimates apply also to the $n$-component $|\varphi|^4$ model with
only superficial changes.
Recall that $\chicCov_j$ is defined in \refeq{chicCovdef}.

\begin{prop} \label{prop:IWbd}
    There exists $C>0$ such that
    for $V\in\Vcal$ and $B \in \Bcal_j$,
\begin{align}
  \label{e:Wbd}
  \|W_j(V,B)\|_{T_{\varphi,j}}
  &\leq C\chicCov_j \|V\|_\Vcal^2(1+ \|\varphi\|_{\Phi_j}^6)
  .
\end{align}
If, in addition, $g>0$, then
\begin{align}
\label{e:Vbd}
  \|e^{-V(B)}\|_{T_{\varphi,j}}
  &\leq
  C e^{C\|V\|_\Vcal (1+ \|\varphi\|_{\Phi_j}^2)},
  \\
\label{e:Ibd1}
  \|I_j(V,B)\|_{T_{\varphi,j}}
  &\leq C e^{C\|V\|_\Vcal(1+ \|\varphi\|_{\Phi_j}^2)},
  \\
\label{e:Ibd}
  \|I_j(V,B)-1\|_{T_{\varphi,j}}
  &\leq C\|V\|_\Vcal (1+\|\varphi\|_{\Phi_j}^6) e^{C\|V\|_\Vcal(1+ \|\varphi\|_{\Phi_j}^2)}
  .
\end{align}
\end{prop}

\begin{proof}
The inequality \refeq{Wbd} is proved in
\cite[Lemma~\ref{IE-lem:W-logwish}]{BS-rg-IE}.
The inequalities \refeq{Vbd}--\refeq{Ibd1} are proved in
\cite[Proposition~\ref{IE-prop:Iupper}]{BS-rg-IE}; the parameter $\epV$ appearing there
obeys $\epV \asymp \|V\|$ by definition and \refeq{relmon}--\refeq{marmon}.
The inequality \refeq{Ibd} follows by
writing $I-1=(e^{-V}-1)+e^{-V}W= -\int_0^1 Ve^{-tV}dt+e^{-V}W$, using the product
property of the $T_\varphi$ semi-norm, using \refeq{Wbd} and \refeq{Vbd},
and using the fact that $\|V\|_{T_\varphi} \le \|V\|_\Vcal (1+\|\varphi\|_{\Phi}^4)$ by
\cite[Proposition~\ref{norm-prop:T0K}]{BS-rg-norm}.
\end{proof}

\subsection{Perturbative flow of coupling constants}
\label{sec:pt2}

The polynomial $\Upt$ of \refeq{Vptdef} can be evaluated explicitly.
For the weakly self-avoiding walk, this is discussed in detail in \cite{BBS-rg-pt}.
We now extend that discussion to the $n$-component $|\varphi|^4$ model.
We first discuss the evaluation of $\Upt$, and then discuss a change of
variables that simplifies the system of equations for $\Upt$ by putting them
into triangular form.

\subsubsection{Explicit calculation of $\Upt$}

To evaluate \eqref{e:PdefF}, we use the equivalent formula (see \cite[Lemma~\ref{pt-lem:Palt}]{BBS-rg-pt})
\begin{equation}
    \label{e:Palt0}
   P(V,x)
   =
  \frac 12
  \LT_{x} F_{w+C} (e^{\Lcal_C} V_x,e^{\Lcal_C} V(\Lambda))
  -
  \frac 12
  e^{\Lcal_C}\LT_{x} F_{w} ( V_x, V(\Lambda))
  ,
\end{equation}
where $w=w_j$ and $C=C_{j+1}$.
The evaluation of $F$ is routine.
In fact, since $V_x$ is a polynomial in $\varphi$ of degree $4$,
for $x,y\in \Lambda$ (see \cite[Lemma~\ref{pt-lem:Fexpand}]{BBS-rg-pt}),
\begin{equation}
\label{e:Fexpand1Vpt}
    F_{w}(V_x,V_y) = \sum_{k=1}^4  \frac{1}{k!}
    \sum_{i_1,\ldots,i_k=1}^n
    \sum_{\substack{u_l,v_l \in \Lambda\\(l=1,\dots,k)}}  \left(\prod_{l=1}^k w_{u_l,v_l}\right)
    \frac{\partial^k V_x}{\partial \varphi_{u_1}^{i_1}\cdots \partial \varphi_{u_k}^{i_k}}
    \frac{\partial^k V_y}{\partial \varphi_{v_1}^{i_1}\cdots \partial \varphi_{v_k}^{i_k}}.
\end{equation}
Moreover, since $V_x$ only depends on the field at $x$ and its neighbours,
the terms in the above sum vanish unless $u_l$ is $x$ or its neighbour, for each $l$,
and similarly for each $v_l$.
From \eqref{e:Fexpand1Vpt}, it can be seen that
the coefficients of $P$ are polynomial in $n$.  In fact,
the degree of the polynomials in $n$ is bounded by $3$:
since $k \ge 1$, each $V$ is differentiated at least once
and hence has degree at most 3, so, in terms of Feynman diagrams,
the number of choices of components at each vertex is
$O(n^3)$.
Vertices must be paired componentwise, so there are $O(1)$ ways to do the pairing,
resulting in $O(n^3)$ overall.
As a consequence, the coefficients are uniquely determined by their values for $n=1,2,3,4$.
For fixed $n$, the
computation of \refeq{Palt0} is mechanical enough to be carried out on a computer
\cite{BBS-rg-ptsoft}.
The result of the computer calculation, together with an explicit and elementary
calculation of the Gaussian moments in the first term of \refeq{Vptdef},
leads to the explicit formulas given below in \eqref{e:gpt2a}--\eqref{e:zpta},
for $n=1,2,3,4$ and thus for all $n\in\N$.

To write down these formulas,
we use the following definitions.
To simplify the notation, we usually leave dependence on $j$ implicit.
Given $g,\nu \in \R$, let
\begin{equation}
    \eta' = (n+2)C_{0,0},
    \quad
   \nu^+ = \nu + \eta' g  ,
   \quad
   w_+=w+C,
   \lbeq{nuplusdef}
\end{equation}
and, given a function $f=f(\nu,w)$, let
\begin{equation}
    \label{e:delta-def}
    \delta[f (\nu ,w)]
=
    f (\nu^+ ,w_{+}) -  f (\nu  ,w )
.
\end{equation}
For a function $q:\Lambda \to \R$ that vanishes if $|x| > \half \diam{\Lambda}$,
we supplement the definitions \eqref{e:qprop1} with
\begin{equation} \label{e:wndef}
  (\nabla q)^2 = \frac 12 \sum_{e \in \Z^d: |e|_1 = 1}(\nabla^e q)^2, \quad
  q^{(n)} = \sum_{x\in\Lambda} q_{x}^n,\quad
  q^{(**)} = \sum_{x \in \Lambda} x_{1}^{2} q_{x}.
\end{equation}
The $q$ that we use are combinations of $w$ that are invariant under lattice rotations,
so that $x_1^2$ can be replaced by $x_i^2$ for any $i=1,\dots, d$ in \eqref{e:wndef}.
We set
\begin{alignat}{2}
  \lbeq{betadef}
  \beta &= (8+n) \delta[w^{(2)} ],
  \quad&
  \theta & = (2+n)\delta[(w^{3})^{(**)}]
  ,
  \\
  \xi' &=
  2(2+n)
  \big(
  \delta[w^{(3)}]- 3 w^{(2)}C_{0,0}
  \big)
  + \gamma \beta   \eta',
  \quad&
  \pi' &= (2+n)  \delta[(w\Delta w)^{(1)}]
  ,
  \\
  \label{e:sigzetadef}
  \sigma &=
  \half (2+n)\delta[(w\Delta w)^{(**)}],
  \quad&
  \zeta &=
  \half (2+n)\delta[((\nabla w)^2)^{(**)}]
  .
\end{alignat}
The dependence on $j$ in the above quantities has been left implicit.
We write
\begin{equation} \label{e:Uptcalc}
    U_{\pt,x} = \gpt \tau^2_x + \nupt \tau_x
    + \zpt \tau_{\Delta,x} + \ypt \tau_{\nabla\nabla,x}
    + \delta u_\pt.
\end{equation}
Then, recalling the definition of $\gamma$ from \eqref{e:gammadef},
the result of explicit calculation is
\begin{align}
  \gpt
  &
  =
  g
  - \beta g^{2}
  - 4g \delta[\nu w^{(1)}]
  ,
\label{e:gpt2a}
  \\
  \nu_\pt
  &=
  \nu
  +  \eta' (g + 4g\nu w^{(1)})
  - \xi' g^{2}
  - \gamma\beta g \nu  - \pi' g(z+y)
  - \delta[\nu^{2} w^{(1)}]
  ,
\label{e:nupta}
  \\
  y_\pt
  &=
  y +
  \sigma gz
  - \zeta gy
  - \half (2+n) g \delta [\nu (w^2)^{(**)}],
\label{e:ypta}
  \\
  z_\pt
  &=
  z
  -
  \theta g^{2}
  - \half \delta[\nu^{2} w^{(**)}]
  - 2 z \delta[\nu w^{(1)}]
  - (y_\pt -y)
\label{e:zpta}
  .
\end{align}
In \refeq{gpt2a}--\refeq{zpta}, $y=0$ in our application, but the formulas are valid
as stated also when $\Upt(V)$ is computed for a polynomial $V$ which contains a
term $y\tau_{\nabla\nabla}$.

The constant term $\delta u_\pt$ can also be calculated explicitly.
After simplification, the result of computer calculation \cite{BBS-rg-ptsoft} is
%
%
\begin{align} \lbeq{uptlong}
  \delta u_\pt &=    \kappa_g g + \kappa_\nu' \nu  - \kappa_z (y + z )
  - \kappa_{gg} g^2 - \kappa_{\nu\nu}' \nu^2 - \kappa_{zz}(y+z)^2
  + \kappa_{z\nu}' (y+z)\nu
\end{align}
where, with $C=C_{0,0}$ and $\Delta C = (\Delta C)_{0,0}$,
\begin{align}
 \kappa_g &= \tfrac{1}{4} n(n+2) C^2 ,
 \quad
 \kappa_\nu' = \half n C,
 \quad
 \kappa_z =  \half n  \Delta C ,
 \quad
 \kappa_{z\nu}' = \half n  \delta[(w\Delta w)^{(1)}],
 \nnb
 \kappa_{gg}
 &=
 \tfrac{1}{4}n(n+2) (\delta[w^{(4)}] - 4 Cw^{(3)} + 2 \Delta C (w^3)^{(**)} - 6 C^2 w^{(2)} + (n+2) C^2 \delta[w^{(2)}]),
 \nnb
 \kappa_{\nu\nu}'
 &= \tfrac{1}{4}n (\delta[w^{(2)}] - 2C w^{(1)}+\Delta C w^{(**)})
 , \quad
 \kappa_{zz} = \tfrac{1}{4} n  \delta[(\Delta w)^{(2)}]
 .
 \label{e:ugreeks}
\end{align}

We require bounds on the coefficients \eqref{e:betadef}--\eqref{e:sigzetadef}
and \refeq{ugreeks}.
The covariance estimate \refeq{scaling-estimate} can be used to show that
\begin{align} \lbeq{greekbds}
  \beta_j, \theta_j, \sigma_j , \zeta_j, \kappa_{\nu\nu}'
  &=
  O(\chicCov_j),
  \\
  \eta_j',\pi_j',\xi_j', \kappa_\nu', \kappa_{z\nu}'
  &=
  O(L^{-2j}\chicCov_j),
  \\
  \kappa_g,\kappa_z,\kappa_{gg},\kappa_{zz} &= O(L^{-4j} \chicCov_j);
  \lbeq{greekbds2}
\end{align}
this is proved in
\cite[Proposition~\ref{pt-prop:rg-pt-flow}]{BBS-rg-pt}
for \eqref{e:betadef}--\eqref{e:sigzetadef} and in
Lemma~\ref{lem:upt} for \eqref{e:ugreeks}.
Also, by \cite[Lemma~\ref{pt-lem:betalim}]{BBS-rg-pt},
\begin{equation} \label{e:betalim}
  \lim_{j\to\infty}\beta_j = \frac{8+n}{16 \pi^2} \log L
  \quad\quad \text{when $m^2 = 0$}
  .
\end{equation}
We summarise the above as follows.

\begin{prop} \label{prop:pt}
  For $U \in \Ucal$, the polynomial $\Upt$ of \refeq{Vptdef}
  is given by \eqref{e:Uptcalc}--\eqref{e:zpta} and \refeq{uptlong}.
  The coefficients of \eqref{e:Uptcalc}--\eqref{e:zpta} are bounded as in
  \refeq{greekbds}--\refeq{greekbds2}.
\end{prop}

The formulas \eqref{e:gpt2a}--\eqref{e:zpta} reduce to those in \cite{BBS-rg-pt}
for the weakly self-avoiding walk if we set $n=0$.
A simplification for weakly self-avoiding walk is that the
constant term vanishes as a result
of supersymmetry \cite[Lemma~\ref{pt-lem:ss2}]{BBS-rg-pt}.
This is consistent with the fact that $\delta u_\pt$ becomes zero
when $n$ is set equal to zero in \refeq{uptlong}, due to the explicit factor $n$
appearing in each coefficient of \refeq{ugreeks}.
The value of $\delta u_\pt$ plays a role
in our analysis only in Theorem~\ref{thm:pressure}.

\subsubsection{Change of variables}
\label{sec:tr}

A change of variables can be used to simplify the equations for $\Upt$, as we
discuss next.
First, the monomial $\tau$ is relevant, and
we absorb its growth from \refeq{relmon} into a rescaled coupling constant
\begin{equation} \label{e:munu}
  \mu_j = L^{2j}\nu_j,
\end{equation}
which is reminiscent of \refeq{Vnorm}.
We also define rescaled coefficients
\begin{equation} \label{e:wbardef}
  \gamma_j = L^{2(j+1)}\gamma_j' \quad (\gamma=\eta,\xi,\pi),
  \quad
  \bar{w}_j^{(1)} = L^{-2j}  w_j^{(1)},
  \quad
  \bar{w}_j^{(**)} = L^{-4j}w_j^{(**)}
  ,
\end{equation}
which are all shown in \cite[Lemma~\ref{pt-lem:wlims}]{BBS-rg-pt} to be uniformly bounded.

Summation by parts on the torus gives
$\sum_{x\in\Lambda} \tau_{\nabla\nabla,x} = \sum_{x\in\Lambda} \tau_{\Delta,x}$,
and hence
\begin{equation}
\label{e:zysbp}
    \zpt   \sum_{x\in\Lambda} \tau_{\Delta,x}
    + \ypt \sum_{x\in\Lambda} \tau_{\nabla\nabla,x}
    =
    (\zpt + \ypt) \sum_{x\in\Lambda} \tau_{\Delta,x}.
\end{equation}
Boundary terms do arise if the sum over $\Lambda$ is replaced by
a sum over a proper subset of $X \subset \Lambda$, and we do need to work
with $I(X)$ with such $X$.  Nevertheless, we are able to
make use of a version of \refeq{zysbp}
(our implementation occurs in
\cite[Section~\ref{step-sec:int-by-parts2}]{BS-rg-step}).
This suggests that $\zpt+\ypt$ is a natural variable, so we define
\begin{equation}
    z^{(0)} = y+z,
    \quad\quad
    \zpt^{(0)}= \ypt+\zpt .
\end{equation}
Then we define $V_{\pt}^{(0)}$ and $\Vpt^{(1)}$ by
\begin{align}
    V_{\pt}^{(0)}(V) &= g_\pt \tau^2 + \nu_\pt \tau + z^{(0)}_\pt \tau_\Delta,
    \\
    V_{\pt}^{(1)}(V) &=  V_{\pt}^{(0)}(V) +\delta u_\pt.
\end{align}
We also define $\Ucal^{(1)}$ to be the subspace of $\Ucal$ for which $y=0$,
with the norm induced by the norm on $\Ucal$ in \refeq{Vnorm}.
In particular, $\Vpt^{(1)} \in \Ucal^{(1)}$.

The equations \eqref{e:gpt2a}--\eqref{e:zpta} are coupled,
and it is useful to re-express the map
$V_{\pt}^{\smash{(0)}}$
in transformed coordinates, as follows.
We define maps $T = T_j:\R^3 \to \R^3$ by $T(g,z,\mu)= (\gch,\zch,\much)$ where
\begin{align}
  \label{e:gch-def}
  \gch  & = g + 4 g\mu \bar w^{(1)},
  \\
  \label{e:zch-def}
  \zch & = z + 2 z\mu \bar w^{(1)} +  \half \mu^2 \bar w^{(**)},
  \\
  \label{e:much-def}
  \much & = \mu + \mu^2 \bar w^{(1)}
  .
\end{align}
The map $T$ is identical to that in \cite[Section~\ref{pt-sec:ourflow}]{BBS-rg-pt}.
The transformation $T=T_j$ satisfies
\begin{equation} \label{e:TVV2}
  T_0(V) = V, \quad\quad T_j(V) = V + O(\|V\|^2),
\end{equation}
with error estimate uniform in $j$.
Since the $T$ are polynomials,
this implies that they are invertible in a neighbourhood of $0$ that is independent of $j$.

We define $\bar \phi = \bar \phi_j : \R^3 \to \R^3$
by $(\gbar_{+},\zbar_{+},\mubar_{+}) = \bar \phi_j(\gbar,\zbar,\mubar)$, with
\begin{align}
\label{e:gbar}
  \gbar_{j+1}
  &=
   \gbar_j
  -
  \beta_j  \gbar_j^{2}
  ,
  \\
\label{e:zbar}
  \zbar_{j+1}
  &=
  \zbar_j
  - \theta_j  \gbar_j^{2}
  ,
  \\
\label{e:mubar}
  \mubar_{j+1}
  &=
  L^2 \mubar_j ( 1- \gamma \beta_j \gbar_j)
  + \eta_j  \gbar_j
  -
  \xi_j  \gbar_j^{2}
   - \pi_j  \gbar_j  \zbar_j
   .
\end{align}
Similarly, we define $\bar\phi^{\delta u} = \bar\phi^{\delta u}_j: \R^3 \to \R$ by $\delta \bar u_+ = \bar\phi^{\delta u}(\gbar, \zbar, \mubar)$, with
\begin{align}
  \delta \bar u_+ &=
  u  + \kappa_g \gbar + \kappa_\nu \bar\nu   - \kappa_z\zbar
 - \kappa_{gg} \gbar^2
 - \kappa_{\mu\mu} \mubar^2
 - \kappa_{zz} \zbar^2
 + \kappa_{g\nu} \gbar\mubar
 + \kappa_{z\nu} \zbar\mubar
\lbeq{ubarlong}
,
\end{align}
with $\kappa_g,\kappa_z,\kappa_{gg},\kappa_{zz}$ as in \eqref{e:ugreeks}, and with
\begin{equation}  \lbeq{ugreeks2}
 \kappa_{\mu\mu}
 = \tfrac{1}{4} n L^{-4j} \delta[w^{(2)}],
 \quad
 \kappa_{g\mu} =  n(n+2) \bar w^{(1)} C^2  
 ,\quad
 \kappa_{z\mu} = L^{-2j} \kappa_{z\nu}'
 - n \bar w^{(1)} \Delta C 
.
\end{equation}
The next proposition shows that the transformation essentially
reduces the study of the map $V \mapsto (\delta u_\pt,V_\pt)$ to that of the simpler maps
$\bar\phi^{\delta u}, \bar\phi$.
For the part of the statement concerning $\bar\phi$,
the elementary proof is given in \cite[Proposition~\ref{pt-prop:transformation}]{BBS-rg-pt}.
The proof of the statement concerning $\bar\phi^{\delta u}$ is an analogous computation.

\begin{prop} \label{prop:T}
  The transformation $T$ and the maps $\bar\phi, \bar\phi^{\delta u}$ satisfy
  \begin{equation}
    \bar \phi=T_{+}\circ V_{\pt}^{\smash{(0)}}\circ T^{-1} + O(\|V\|^3),\quad
    \bar \phi^{\delta u} = \delta u_{\pt} \circ T^{-1} + O(\|V\|^3).
  \end{equation}
\end{prop}

The effect of the transformation $T$ is to \emph{triangularise} the
evolution equation to second order: the $\gbar$-equation does not
depend on $\zbar$ or $\mubar$, the $\zbar$-equation depends only on
$\gbar$, and the $\mubar$-equation depends both on $\gbar$ and
$\zbar$.
This second-order triangularisation is the natural coordinate system
to study the evolution of $V_j$.
In the transformed variables, the $\gbar\mubar$ term in \refeq{mubar}
is proportional to $\beta$, and
its coefficient $\gamma$ provides the power of the logarithm in Theorem~\ref{thm:suscept}.
Similarly, the $\mubar^2$ coefficient in \eqref{e:ubarlong} is
proportional to $\beta_j$, and this is important for the analysis of
the critical behaviour of the specific heat in
Theorem~\ref{thm:pressure}.

For $m^2>0$, the coefficients \eqref{e:betadef}--\eqref{e:sigzetadef} are essentially constant
for moderately large $j$ but then decay exponentially for $j\ge j_m$.  This exponential decay
effectively stops the sequence $\gbar_j$ obtained by iterating \eqref{e:gbar} from evolving further for $j >j_m$.
The sequence $\chicCov$ compensates for this, and it is shown in
\cite[Proposition~\ref{flow-prop:approximate-flow}]{BBS-rg-flow} that there is a unique
solution to the iteration of \refeq{gbar}--\refeq{mubar} with boundary conditions $\gbar_0=g_0$,
$\zbar_\infty=\mubar_\infty =0$ and that this solution obeys, for all real $p\in [1,\infty)$,
\begin{equation}
  \lbeq{chig}
  \chicCov_j\gbar_j^p = O(g_0/(1+g_0j))^p,
  \quad
  \zbar_j = O(\chicCov_j \gbar_j),
  \quad
  \mubar_j = O(\chicCov_j \gbar_j),
\end{equation}
with constants depending on $p$, but independent of $(m^2,g_0)$.

\subsection{Non-perturbative coordinate}
\label{sec:Kspace}

Our treatment of the perturbative flow of coupling constants
is in the spirit of Wilson's general
approach.  For a rigorous analysis, we must also understand the
non-perturbative coordinate $K$, and this poses substantial challenges.
A proper definition of $K$, and the analysis needed to control it,
is the topic of \cite{BS-rg-step}, which in turn relies on the main results of
\cite{BS-rg-IE}.
In our scheme, at scale $j$, $K$ lies in the space $\Kcal_j$ of maps from $\Pcal_j$ to
$\Ncal$, given in the following definition.
Our estimates require that the total number $p_\Ncal$ of derivatives
in the definition of $\Ncal$ in \refeq{Ncaldef} be a fixed integer $p_\Ncal \ge 10$.

The defining properties of $\Kcal_j$ are similar to those for $I$ given below \eqref{e:Idef},
with the important difference that while $I_j$ has the \emph{block} factorisation property,
$K_j$ has a weaker \emph{component} factorisation property.
Also, whereas $I_j(B) \in \Ncal(B^+)$ with $B^+$ the enlargement of $B$
obtained by adjoining all neighbouring blocks, now $K(X) \in \Ncal(X^\Box)$
with
\begin{equation}
    X^\Box = \bigcup_{Y \in\Scal_j: X\cap Y \not =\varnothing } Y,
\end{equation}
where $\Scal_j\subset\Pcal_j$ is the set of connected polymers
consisting of at most $2^d$
blocks.  Elements of $\Scal_j$ are called \emph{small sets}.
The \emph{small set neighbourhood} $X^\Box$ of $X$ is a greater enlargement than
adjoining all neighbouring blocks.
Recall that we write ${\rm Comp}_j( X)$ for the set of connected components
of a polymer $X\in\Pcal_j$.

\begin{defn} \label{def:Kspace}
Let $\Kspace_{j} = \Kspace_{j} (\Lambda_N)$ be the vector space of functions $K :
\Pcal_j  \to \Ncal$ with the properties:
\begin{itemize}
\item Field locality: $K(X) \in \Ncal (X^{\Box})$ for each connected $X\in\Pcal_j$,
\item Symmetry: $K$ is $O(n)$ invariant and $K$ is Euclidean invariant,
\item Component factorisation: $K (X) = \prod_{Y
\in {\rm Comp}_j( X)}K (Y)$ for all $X\in\Pcal_j$.
\end{itemize}
\end{defn}

We write $U \in \Ucal^{(1)}$ as $U = (\delta u,V)$, with $V \in \Vcal$.
In \cite{BS-rg-step}, a map $(V,K) \mapsto (U_+,K_+)$ satisfying \eqref{e:IcircKdu}
is defined, and we use this map from now on.
More precisely, \cite{BS-rg-IE,BS-rg-step} are written explicitly
for the $V$ of the weakly self-avoiding walk, but they apply {\it mutatis mutandis} to the
$V$ we have here for the $|\varphi|^4$ model.
The polynomial $V$ is an element of $\Vcal$, whereas $U_+$ includes
a constant monomial and hence lies in $\Ucal^{(1)}$.
The map $(V,K) \mapsto U_+$ is explicit and relatively simple.
Let $\LT_{Y,B}$ denote the operator defined by
$\LT_{Y,B} F = P_Y(B)$, where $P_Y$ is the polynomial determined by $P_Y(Y) = \Loc_Y F$.
As in \cite[Section~\ref{step-sec:Rconstruction}]{BS-rg-step}, the map $(V,K) \mapsto U_{+}$
is given by
\begin{equation}
\lbeq{Vplusdef}
  U_{+}(V,K) = \Vpt^{(1)}(V-Q) \quad\text{with}\quad
  Q(B)
  =
  \sum_{\substack{Y \in \Scal\\Y \supset B}}
  \LT_{Y,B} \left( \frac{K (Y)}{I(Y,V)} \right).
\end{equation}

When $K=0$,  $U_{+}(V,0)$ is
just the quadratic polynomial $\Vpt^{\smash{(1)}}(V)$ discussed in Section~\ref{sec:pt2}.
We also write $U_+=(\delta u_+,V_+)$ and then
$\delta u_+(V,0) = \delta u_\pt(V,0)$, $V_+(V,0) = \Vpt^{\smash{(0)}}(V)$.
The formula \refeq{Vplusdef} incorporates
the marginal and relevant parts
of the non-perturbative coordinate $K$ into the flow of coupling constants;
the fact that this goal is achieved is
shown by Theorem~\ref{thm:step-mr} below.  We express estimates on $U_+$
in terms of $R_+$ defined by
\begin{equation} \label{e:Rdef}
    R_+(V,K) = U_+(V,K)-U_+(V,0) \in \Ucal^{(1)}
    .
\end{equation}

The definition of the map $(V_j,K_j) \mapsto K_{+}$ is explicit
in \cite{BS-rg-step}, but it is too elaborate to write down.
This map captures the errors in the perturbative calculation,
and it suffices to estimate it rather than studying its explicit form.
To estimate $K_+$, we need appropriate norms on the space $\Kcal_j$ for each $j$,
and we use the
same norms $\Wcal_j$ defined in \cite[Section~\ref{log-sec:flow-norms}]{BBS-saw4-log}.
These norms have some dependence on the values of $g_0$, $m^2$, and $N$,
but all conclusions we reach by employing the norms are uniform in these parameters.
These dependencies are treated carefully in \cite{BBS-saw4-log}, but
they do not play an explicit role in this paper
and we suppress this dependence in the notation.

To state estimates for the map $(V,K) \mapsto (U_+,K_+)$,
we first define its domain.
Given $C_\DV>1$, $\DVa>0$, $\delta >0$, and
$\sgen=(\mgen^2, \ggen)\in [0,\delta) \times (0,\delta)$, let
\begin{equation} \label{e:domRG}
  \domRG_j(\sgen)
  =
  \{ (g_j, z_j, \nu_j): C_\DV^{-1} \ggen < g_j < C_\DV \ggen, \;
  |z_j|, |\mu_j| < C_\DV \ggen \}
  \times
  B_{\Wcal_j}(\DVa \chicCovgen_j\ggen^3)
  ,
\end{equation}
where $B_X(r)$ is the open ball of radius $r$ in the Banach space $X$,
and $\tilde\chicCov_j = \chicCov_j(\mgen^2)$ with $\chicCov_j$ defined by \refeq{chicCovdef}.
The space $\Wcal_j$ also depends on $\sgen$, but we suppress this dependence
in our notation.
The domain \eqref{e:domRG} permits small $g_j > 0$ that is bounded away from zero,
with $z_j,\mu_j = O(g_j)$,
and with $K_j$ bounded in a precise but
non-trivial fashion by $O(g_j^3)$.
The domain $\domRG_j(\sgen)$ is equipped with the norm of $\Vcal \times \Wcal_j$.

The main result of \cite{BS-rg-step}
is \cite[Theorem~\ref{step-thm:mr}]{BS-rg-step}.  It was formulated there for
the weakly self-avoiding walk, but it applies equally well to the $n$-component
$|\varphi|^4$ model.
The absence of the fermions needed for the weakly self-avoiding walk
is a minor simplification for $|\varphi|^4$.
(The basic estimate which deals with the fermions is
\cite[Proposition~\ref{norm-prop:EK}]{BS-rg-norm}.)
The main change for the $|\varphi|^4$ model is the occurrence
of $\delta u_+$, and this is addressed explicitly in
\cite[Remark~\ref{step-rk:uphi4}]{BS-rg-step}.
The use of observables in \cite{BS-rg-step} can be ignored by setting $\sigma=0$
there; the observables are needed only to study the critical two-point function,
to be studied for $|\varphi|^4$ elsewhere \cite{BST-phi4}.
The conclusions of \cite[Theorem~\ref{step-thm:mr}]{BS-rg-step}
are summarised in the following theorem.  It is possible to promote the statement of
infinite differentiability in Theorem~\ref{thm:step-mr} to an analyticity statement,
but since we have restricted attention to real variables here and do not need
the analyticity, we do not make a formal statement.
To address continuity of the map $(V,K) \mapsto (\delta u_+, V_+,K_+)$ in the
mass $m^2$, we recall $\Iint$ from \eqref{e:massint} and set
\begin{equation}
\lbeq{Itilint}
    \Igen_j = \Igen_j(\mgen^2) =
    \begin{cases}
    [\frac 12 \mgen^2, 2 \mgen^2] \cap \Iint_j & (\mgen^2 \neq 0)
    \\
    [0,L^{-2(j-1)}] \cap \Iint_j & (\mgen^2 =0).
    \end{cases}
\end{equation}

\begin{theorem} \label{thm:step-mr}
  Let $d =4$.
  Let $C_\DV$ and $L$ be sufficiently large, and let $p,q\in \N_0$.
  There exist $M>0$ (depending on $p,q$)  and $\kappa = O(L^{-1})$ such that
  for any $\DVa >M$ (defining \refeq{domRG})
  there exists $\delta >0$ (depending on $\DVa$) such that
  for $\ggen \in (0,\delta)$ and $\mgen^2 \in \Iint_+$, the maps
  $R_+, K_+$ are defined and infinitely differentiable
  from the domain $\domRG(\sgen) \times \Igen_+(\mgen^2)$ to
  $\Ucal^{(1)},\Wcal_{+}(\sgen_+)$ respectively,
  are continuous in $m^2 \in \Igen(\mgen^2)$, and satisfy
  the estimates
  \begin{align}
\label{e:Rmain-g}
    \|D_V^p D_K^q R_+\|_{L^{p,q}}
    & \le
    \begin{cases}
    M
    \tilde\chicCov\ggen^{3-p} & (p\ge 0,\, q=0)\\
    M
     \ggen^{1-p-q} & (p\ge 0,\, q = 1,2)\\
    \rlap{$0$}\hspace{3.3cm}  & (p\ge 0,\, q \ge  3),
    \end{cases}
\\
\lbeq{DVKbd}
    \|D_{V}^pD_{K}^{q}K_+\|_{L^{p,q}}
    &\le
    \begin{cases}
    M  \tilde\chicCov \ggen^{3-p}
    &
    (p \ge 0)
    \\
    \rlap{$\kappa$}\hspace{3.3cm}
    & (p=0,\, q=1)
    \\
    M  \ggen^{-p}
    (
    \tilde\chicCov
    \ggen^{10/4}
    )^{1-q}
    &
    (p \ge 0,\, q \ge 1)
    .
    \end{cases}
  \end{align}
\end{theorem}

Since $\kappa < 1$, the second bound of \refeq{DVKbd} provides the crucial contraction which
is our implementation of the concept that the perturbative coordinate has
accurately captured the marginal and relevant parts of the renormalisation group map.
Theorem~\ref{thm:step-mr} provides a kind of local existence theorem,
which permits the renormalisation
group map to be iterated as long as $V_j$ remains in the correct domain.
We describe in Section~\ref{sec:rgf} conditions under which there is global existence.

\subsection{Renormalisation group flow}
\label{sec:rgf}

We write $U_j=(\delta u_j,V_j)$, and
say that $(U_j,K_j)_{0\leq j \leq N}$ is a \emph{flow} of the
renormalisation group if
\begin{equation} \label{e:flowVK}
  (U_{j+1}, K_{j+1}) 
  = (U_{+}(V_j,K_j),K_{+}(V_j,K_j)) \quad \text{for all $0 \leq j < N$}.
\end{equation}
Here $(U_+,K_+)$ denotes the map of of Section~\ref{sec:Kspace}.
In particular, $U_+$ acts on
a polynomial in $\Vcal$ and produces one on $\Ucal^{(1)}$.
The following key theorem constructs a sequence with the desired properties.

For its statement,
the parameters $\ggen = \ggen_j$ in \eqref{e:domRG}
need to be chosen appropriately.
First, the sequence $\gbar_j= \gbar_j(m^2,g_0)$ is defined as the solution
of the recursion \refeq{gbar}, and it obeys the estimate \eqref{e:chig}.
Then, as in \cite{BBS-saw4-log}, we define $\ggen_j$ by
\begin{equation}
\label{e:ggendef}
  \ggen_j(m^2,g_0) =
  \gbar_j(0,g_0) \1_{j \le j_m} + \gbar_{j_m}(0,g_0) \1_{j > j_m}.
\end{equation}
The sequences $\ggen_j$ and $\gbar_j$ are almost the same,
in fact,
 $\ggen_j = \gbar_j + O(\gbar_j^2)$ by \cite[Lemma~\ref{log-lem:gbarmcomp}]{BBS-saw4-log},
but $\ggen_j$ is more convenient for aspects of the analysis.
In addition, we make a specific (somewhat arbitrary) choice of the parameter $\DVa$
in $\domRG$ of \eqref{e:domRG} in \cite{BBS-saw4-log}.
It plays no direct role in this paper.

\begin{theorem} \label{thm:flow-flow}
  Let $d=4$.  Let $\delta>0$ be sufficiently small.
  There is
  an infinite sequence
  of functions $U_j=(\delta u_j^c,V_j)$, $V_j=(g_j^c,\mu_j^c,z_j^c)$ of
  $(m^2,g_0) \in [0,\delta)^2$,
  independent of $N$, such that:
  \smallskip
  \\
  (i) for $N \in \N$,
  initial conditions $V_0 = (g_0,\mu_0^c,z_0^c)$ with $g_0 \in (0,\delta)$, $K_0=\1_{\varnothing}$,
  and mass $m^2$, the flow \eqref{e:flowVK}
  exists for all $j < N$,
  and if $m^2 \in [\delta L^{-2(N-1)},\delta)$, also for $j=N$.
  Its $U$-component is given by the sequence $U_j$,
  and
  $(V_j,K_j) \in \domRG_j(m^2,\ggen_j)$.  
  In particular, then
  \begin{align}
    \lbeq{VVbar1}
    \|K_j\|_{\Wcal_j}
    &\leq O(\chicCov_j \gbar_j^3)
    \quad ( j \le N)
  \end{align}
  and $\gch_j, g_j = O(\gbar_j)$. In addition, $\zch_j, \much_j, z_j,\mu_j = O(\chicCov_j \gbar_j)$.
  \smallskip

  \noindent
  (ii) $z_0^c, \mu_0^c$ are continuous in $(m^2,g_0) \in [0,\delta)^2$ and differentiable in
  $g_0 \in (0,\delta)$ with uniformly bounded $g_0$-derivative.
\end{theorem}

A version of
Theorem~\ref{thm:flow-flow}
is proved in \cite[Proposition~\ref{log-prop:KjNbd}]{BBS-saw4-log}
for the weakly self-avoiding walk.
The proof holds without modification for the $|\varphi|^4$ model---the
perturbative flow differs in an unimportant $n$-dependent manner, and the estimates
on the non-perturbative component have an identical form for both
models---so we do not repeat the proof.

The sequence $U_j=(\delta u_j,V_j)$ is defined for all $j\in \N$, and is independent of
$N$.  (In \cite{BBS-saw4-log} this independence is established in a setting where
$\delta u$ is absent, but it applies also to $\delta u$ for the same reason.)
Of course $U_j$ is only applicable to $\Lambda$ as long as $j \le N$, but for
any fixed $j$ it remains constant in the limit $N\to\infty$.
Since $(V_j,K_j) \in \domRG(m^2,\ggen_j)$ for all $j<N$, the bounds \refeq{Rmain-g}
hold for $L^{4j}R_+^{\delta u}$, as stated in Theorem~\ref{thm:step-mr}.
Let
\begin{equation}
  \lbeq{usum}
  u_j = \sum_{i=1}^j \delta u_i.
\end{equation}
By \refeq{IcircKnew}, Theorem~\ref{thm:flow-flow} then implies that,
if $(m^2,g_0) \in [\delta L^{-2(N-1)},\delta) \times (0,\delta)$, then,
with $(V_j,K_j)$ as given in the theorem and $I_j=I_j(V_j)$,
\begin{equation}
\lbeq{Zju}
  Z_j = e^{-u_j|\Lambda|} (I_j \circ K_j)(\Lambda)
  \quad\quad
  (j \le N).
\end{equation}
In particular,
since $\Pcal_N$ consists only of the two polymers $\varnothing,\Lambda$
(see Section~\ref{sec:IK}),
\begin{equation}
  \lbeq{ZNu}
  Z_N = e^{-u_N|\Lambda|} (I_N +  K_N)(\Lambda).
\end{equation}
We use the identity \refeq{ZNu}, in conjunction with the estimates provided by
Theorem~\ref{thm:step-mr} to control $u_N, V_N, K_N$, to prove
Theorems~\ref{thm:suscept}--\ref{thm:clt}
in Section~\ref{sec:pf}.

The control of $K_j$ is in terms of the $\Wcal_j$ norm.
The precise details of the definition of the $\Wcal_j$ norm are important for the
proofs of Theorems~\ref{thm:step-mr}--\ref{thm:flow-flow}, but  not
for our current discussion.
To have some idea of what the norm estimates accomplish, we
recall from \cite[\eqref{step-e:T0dom}]{BS-rg-step} that there is a constant $C$ such that
\begin{equation}
\label{e:Tphidom}
  \|F(X)\|_{T_{\varphi,j}} \leq \|F\|_{\Wcal_j} e^{C\|\varphi\|_{\Phi_{j}(X^\Box)}^2}
  \quad
  (F \in \Kcal_j, \; X\in\Scal_j).
\end{equation}
In particular, \eqref{e:VVbar1} and \eqref{e:Tphidom} provide a uniform bound
\begin{equation} \label{e:Kbd}
  \|K_N(\Lambda)\|_{T_{\varphi,N}} \leq C\chicCov_N g_N^3 e^{C\|\varphi\|_{\Phi_N}^2}.
\end{equation}
In fact, a more general estimate than \refeq{Tphidom}
holds for arbitrary connected polymers $X$,
with a factor $\gbar_j^{a(|X|_j-2^d)_+}$ on the right-hand side, where $a>0$,
$|X|_j$ is the number of scale-$j$ blocks in $X$, and $x_+=\max\{x,0\}$.
This exponential decay in the size of $X$ provides
a rigorous statement that long-range interactions are small.

\begin{figure}
  \begin{center} \label{fig:phaseport}
    \input{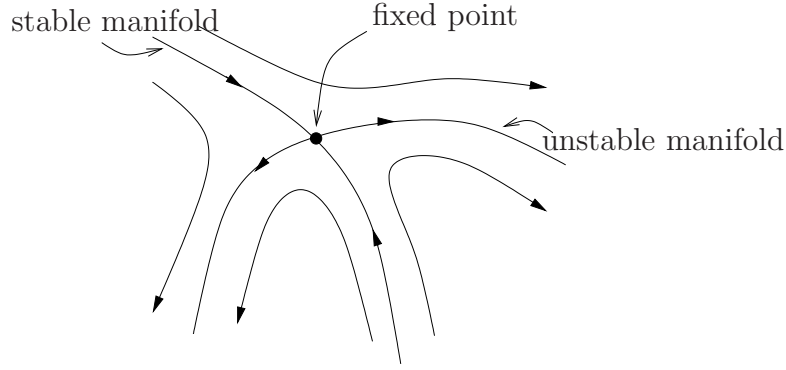}
  \end{center}
  \caption{Schematic phase portrait of the renormalisation group
    flow. In the situation of Theorem~\ref{thm:flow-flow},
    the part of the stable manifold near the fixed point $(V=(0,0,0),K=0)$ restricted to $K=0$ consists of the
    points $V = (g_0,\mu_0^c(g_0), z_0^c(g_0))$, $g_0 \in [0,\delta]$.}
\end{figure}

The proof of Theorem~\ref{thm:flow-flow}
is based on an interpretation of $(V,K) \mapsto (V_+,K_+)$ as a dynamical system.
Note that $\delta u$ plays no role in the dynamical system.
This is possible since $\delta u$ does not appear on the right-hand side of \refeq{flowVK}.
For the dynamical system, it is useful use the transformed variables $\Vch = T(V_j)$,
for which the perturbed flow is triangular to second order. We define
variants $\Rch_{+}^{\smash{(0)}}, \Kch_{+}$ of the maps $R_+,K_+$, which act on $\Vch$ rather than $V$.
Namely, we define
\begin{equation} \label{e:RchKch}
  \Rch_+^{(0)}(\Vch,K) = T_{+}(V_+(T^{-1}(\Vch),K)) - \bar\phi(\Vch),
  \quad
  \Kch_+(\Vch,K) = K_+(T^{-1}(\Vch),K),
\end{equation}
with $\bar \phi$ given by \refeq{gbar}--\refeq{mubar}.
By \cite[Corollary~\ref{log-cor:step-mr-tr}]{BBS-saw4-log},
the maps $\Rch_+^{\smash{(0)}},\Kch_+$ satisfy the estimates for $R_+^{\smash{(0)}},K_+$
in Theorem~\ref{thm:step-mr}, with the domain $\domRG$ replaced by $T(\domRG)$
(where we extend the transformation $T$ of Section~\ref{sec:tr}
to act as the identity on $K$).
The evolution equation for the sequence $(\Vch_j, K_j) = (T_j(V_j),K_j)$ can be written
(with subscripts $j$ omitted and $j+1$ written as $+$) as
\begin{equation}
  \label{e:VchK-rec}
  (\Vch_{+}, K_{+})
  =
  (\bar \phi(\Vch), 0) + (\Rch_{+}^{(0)}(\Vch, K), \Kch_{+}(\Vch,K))
  \quad
  (0 \leq j < N).
\end{equation}
The limitation on $j$ corresponds to the fact that the map ends at the scale $N$ of
the torus $\Lambda = \Lambda_N$.

However, there is a natural way to
pass to an inductive limit, as $N\to\infty$, to obtain maps defined for all $j \in \N_0$.
This is discussed at length in \cite[Section~\ref{step-sec:rg-iv}]{BS-rg-step},
and that discussion applies without change to the $|\varphi|^4$ model.
For the inductive limit, as discussed in \cite{BBS-saw4-log},
we obtain a time-dependent dynamical system $\Phi = (\Phi_j)$, with
\begin{equation}
\lbeq{Phi-flow}
    \Phi_{j}: (\Vch_j,K_j) \mapsto (\Vch_{j+1},K_{j+1}).
\end{equation}
The renormalisation group map $(V_+,K_+)$ is not defined for $V=0$,
but it is natural, by the estimates of Theorem~\ref{thm:step-mr}, to extend it so that
$(V_+(0,0),K_+(0,0))=(0,0)$.
In particular, $(V,K)=(0,0)$ can be regarded as a fixed point for the dynamical system $\Phi$.
This fixed point is
\emph{non-hyperbolic}: the $g$- and $z$-directions neither contract nor expand,
as \refeq{gbar}--\refeq{zbar} indicate.
Such a class of non-hyperbolic dynamical systems is studied in \cite{BBS-rg-flow},
from which, together with Theorem~\ref{thm:step-mr}, Theorem~\ref{thm:flow-flow}
can be deduced.
To be able to iterate the map over all scales, the iteration must begin
at the correct \emph{critical} value of $V_0$, in the domain of attraction
of the fixed point.  The domain of attraction corresponds to the stable manifold,
depicted schematically in Figure~\ref{fig:phaseport}.
This is where the critical value $\nu_c$ plays its role.

\section{Proof of main results}
\label{sec:pf}

We now prove our main results, Theorems~\ref{thm:suscept}--\ref{thm:clt}.
Throughout this section, we write
\begin{equation}
  (f,g) = (f,g)_{\Lambda} = \sum_{i=1}^n \sum_{x\in\Lambda} f_x^i g_x^i,
  \quad (f,g: \Lambda \to \R^n)
\end{equation}
to denote the inner product on $L^2(\Lambda,\R^n)$.

In the proofs of Theorems~\ref{thm:suscept}--\ref{thm:pressure}, we study limits
in which first $N\to\infty$ and then $\nu \downarrow \nu_c$.
Since $m^2>0$ corresponds to $\nu>\nu_c$,
and $\delta L^{-2N} \leq m^2$ for any fixed $\delta,m^2>0$ if $N$ is large enough,
we can fix the parameter $\delta > 0$ discussed below \eqref{e:scaling-estimate} arbitrarily.

\subsection{Proof of Theorem~\ref{thm:suscept}: susceptibility}
\label{sec:pfsuscept}

A result exactly analogous to Theorem~\ref{thm:suscept} is proved in
\cite[Theorems~\ref{log-thm:suscept}--\ref{log-thm:nuc}]{BBS-saw4-log}
for the weakly self-avoiding walk.  We now show how that proof can be modified
slightly so as to apply also
to the $n$-component $|\varphi|^4$ model.
We adapt the definitions of $\bubble$, ${\sf b}$, and $\gamma$ in
\cite[\eqref{log-e:freebubble}, \eqref{log-e:gamconv}]{BBS-saw4-log}, and define now
\begin{equation}
\lbeq{bubgam}
  \bubble_{m^2} = (n+8) B_{m^2}, 
  \quad
  {\sf b} = \frac{n+8}{16 \pi^2},
  \quad
  \gamma = \frac{n+2}{n+8}.
\end{equation}
Then \refeq{freebubble} gives
\begin{equation}
\lbeq{newbubble}
    \bubble_{m^2} \sim {\sf b}(\log m^{-2})^{-1},
\end{equation}
which is the source of the logarithmic correction for the susceptibility.

As in Section~\ref{sec:method}, we denote by $\Ex_C$ the Gaussian expectation
with covariance $C=(-\Delta+m^2)^{-1}$, and write $Z_0=e^{-V_0}$ as in \eqref{e:V0Z0}.
From \eqref{e:ZNdef}, we recall the definition
\begin{equation}
\label{e:ZNdef-bis}
  Z_N(\varphi) = \Ex_{C}\theta Z_0 = \Ex_C Z_0(\varphi +\zeta).
\end{equation}
We write the map $V_+$
of Section~\ref{sec:Kspace} as $V_+=(\delta u_+,V_+^{(0)})$.
The $\delta u$ component is handled explicitly in the following,
but its role is limited since it effectively cancels in ratios.

Let $V_0=(g_0,\nu_0^c,z_0^c)$ as in Theorem~\ref{thm:flow-flow},
and let $K_0 = \1_\varnothing$.  We determine
a finite sequence $(V_j,K_j)_{0\leq j \leq N}$  by
the recursion $(V_{j+1},K_{j+1}) = (V_+^{(0)}(V_j,K_j),K_+(V_j,K_j))$.
We write $I_N = I_N(V_N)$ with $I_N$ as in Section~\ref{sec:pt}.
By \eqref{e:ZNu},
\begin{equation}
  \lbeq{ZIKN}
  Z_N = e^{-u_N|\Lambda_N|}(I_N \circ K_N)(\Lambda)
  = e^{-u_N|\Lambda_N|}(I_N(\Lambda) + K_N(\Lambda))
  .
\end{equation}
The identity \refeq{ZIKN} continues to hold under slight ($N$-dependent) variation of the
initial conditions, since it is a finite recursion and the domains 
are open sets, so we can differentiate \refeq{ZIKN} with respect to the initial conditions.
In \refeq{ZIKN}, $I_N$ is the leading term and $K_N$ a remainder.
For example, for $\varphi=0$, $I_N(\Lambda;0) = 1$ by \eqref{e:Idef},
and, by \refeq{Kbd},
\begin{equation}
\lbeq{KNg3}
  |K_N(\Lambda; 0)| \leq \|K(\Lambda)\|_{T_{0,N}} \leq O(\chicCov_N g_N^3)
  ,
\end{equation}
so that
\begin{equation} \label{e:ZN0}
  Z_N(0)
  = e^{-u_N|\Lambda_N|} (1+O(\chicCov_N g_N^3)).
\end{equation}
Both bounds \refeq{KNg3}--\refeq{ZN0} hold uniformly in
$m^2 \in [\delta L^{-2(N-1)},\delta)$.

For the proofs of Theorems~\ref{thm:suscept} and \ref{thm:clt},
we use the Laplace transform $\Sigma_N(h)$, defined for $h : \Lambda_N \to \R^n$ by
\begin{equation}
\lbeq{Sigdef}
  \Sigma_N(h) = \Ex_{C}(Z_0(\varphi)e^{(\varphi,h)})
  .
\end{equation}
Completion of the square in the Gaussian expectation gives
\begin{equation}
\lbeq{LTC}
  \Sigma_N(h)
  = e^{\frac 12(h,Ch)} \Ex_{C}(Z_0(\varphi + Ch))
  = e^{\frac 12(h,Ch)} Z_N(Ch).
\end{equation}

\begin{proof}[Proof of Theorem~\ref{thm:suscept}]
The susceptibility $\chi$ is defined as a function of $(g,\nu)$,
but it is useful to work with variables $(m^2,g_0,\nu_0,z_0)$ instead.
According to \refeq{ExF}, if the two sets of variables are such that \refeq{g0gnu0nu}
is satisfied, then the susceptibility on $\Lambda_N$ is given by
\begin{equation}
\lbeq{chichihat}
    \chi_N(g,\nu) = (1+z_0) \hat\chi_N(m^2,g_0,\nu_0,z_0),
\end{equation}
with
\begin{equation}
\lbeq{chiSig}
    \hat\chi_N(m^2,g_0,\nu_0,z_0)
    = \frac{1}{|\Lambda_N|} \frac{D^2 \Sigma_N(0;\1,\1)}{Z_N(0)},
\end{equation}
where $\Sigma_N$ is defined by \refeq{Sigdef}, and we use the $n$-component constant
test function
$\1=(1,0,\dots,0)$,
as in \refeq{sigN}.
By \refeq{LTC},
\begin{equation}
    D^2 \Sigma_N(0;\1,\1)
    = \frac{1}{m^2} |\Lambda_N| Z_N(0) + \frac{1}{m^4} D^2 Z_N(0;\1,\1),
\end{equation}
and hence
\begin{equation}
    \hat\chi_N(m^2,g_0,\nu_0,z_0) =
    \frac{1}{m^2}
    +
    \frac{1}{m^4} \frac{1}{|\Lambda_N|} \frac{D^2 Z_N(0;\1,\1)}{Z_N(0)}.
\end{equation}

For the moment, we regard the variables $(m^2,g_0,z_0,\nu_0)$ as primary, and fix them
equal to $(m^2,g_0,z_0^c,\nu_0^c)$, with $z_0^c=z_0^c(m^2,g_0)$ and $\nu_0^c=\nu_0^c(m^2,g_0)$
as in Theorem~\ref{thm:flow-flow}.
By \eqref{e:ZIKN},
\begin{align}
\label{e:chibarm-bisIK}
  D^2 Z_N(0;\1,\1)
  =
  e^{-u_N|\Lambda_N|}
  \left( D^2I_N(0; \1,\1)
  + D^2K_N(0; \1,\1) \right).
\end{align}
By \eqref{e:Idef},
$I_N(\Lambda) = e^{-V_N(\Lambda)}(1+W_N(\Lambda))$, so
\begin{equation}
\lbeq{D2Icalc}
   D^2 I_N(\Lambda; 0; \1, \1)
  =  D^2 e^{-V_N}(\Lambda; 0; \1, \1)
  + D^2W_N(\Lambda; 0; \1, \1)
\end{equation}
since cross-terms cancel when $\varphi=0$
because $W_N$ is a polynomial in $\varphi$ with no monomials
of degree below two.
The first term on the right-hand side of \eqref{e:D2Icalc} can be evaluated directly,
yielding
\begin{equation}
\label{e:D2eVN}
   D^2 e^{-V_N}(\Lambda;0; \1, \1)
   = - \nu_N |\Lambda_N|
  ,
\end{equation}
where we have used the facts that the quartic term $\tau^2$ does not contribute to
\eqref{e:D2eVN}, and $\Delta 1 = 0$. This gives
\begin{equation}
  \label{e:chibarm-bis2}
  \hat\chi_N
  =
  \frac{1}{m^{2}} +
  \frac{A_N}{B_N}
\end{equation}
with
\begin{align}
\lbeq{Aid}
  A_N&= - \frac{\nu_N}{m^4}
  + \frac{1}{m^4}\frac{1}{|\Lambda|}  D^2W_N(0; \1,\1)
  + \frac{1}{m^4}\frac{1}{|\Lambda|}  D^2K_N(0; \1,\1) ,
  \\
  B_N & = e^{u_N|\Lambda_N|}Z_N(0)
.
\end{align}

Since $z_0=z_0^c(m^2,g_0)$, $\nu_0 = \nu_0^c(m^2,g_0)$, it follows
as in \cite[Section~\ref{log-sec:pfmr}]{BBS-saw4-log} that
$A_N$
tends to $0$ as $N\to\infty$.
For $\nu_N$, this
follows from the fact that $\nu_N=O(L^{-2N}\chicCov_N \gbar_N)$
since $(V_N,K_N) \in \domRG_N$ by Theorem~\ref{thm:flow-flow},
and the other two terms are
smaller by factors $g_N$ and $g_N^2$ respectively.  In fact,
by \refeq{Wbd} and \refeq{Kbd} (cf.
\cite[\eqref{log-e:WNNbd}, \eqref{log-e:KNbd2}]{BBS-saw4-log}),
\begin{alignat}{2}
  \label{e:WNKNbd}
  \|W_{N} \|_{T_{0,N}} &\leq O(\chicCov_N g_N^2),
  &\qquad
  \|K_{N}\|_{T_{0,N}} &\leq O(\chicCov_N g_N^3)
  ,
\end{alignat}
and these estimates give rise to bounds of order $L^{-2N}\chicCov_N g_N^2$ and
$L^{-2N}\chicCov_N g_N^3$ for the last two terms of \refeq{Aid} as in
\cite[\eqref{log-e:WNbd}]{BBS-saw4-log}.
More simply, \eqref{e:ZN0} implies that $B_N \to 1$.
Therefore, with \refeq{ZN0}, \refeq{chibarm-bis2} gives
\begin{align}
\label{e:chi-m}
  &\hat\chi \left( m^2,g_0,\nu_0^c(m^2,g_0),z_0^c(m^2,g_0) \right)
  = \frac{1}{m^2}
  .
\end{align}

So far we have considered the six variables $\{g,\nu,m^2,g_0,\nu_0,z_0\}$.
For \eqref{e:chichihat}, we assumed that they satisfy the two equations in \eqref{e:g0gnu0nu},
while for \eqref{e:chi-m}, we used
\begin{equation} \label{e:nu0cz0c}
   z_0 = z_0^c(m^2,g_0), \quad \nu_0 = \nu_0^c(m^2,g_0),
\end{equation}
with $z_0^c,\nu_0^c$ the functions of Theorem~\ref{thm:flow-flow}.
In \cite[Proposition~\ref{log-prop:changevariables}]{BBS-saw4-log}, it is shown
that given $(g,\nu_c+ \varepsilon) \in [0,\delta)^2$ for $\delta>0$ small,
it is possible to choose
\begin{equation}
\lbeq{tildemap}
    (m^2,g_0,z_0,\nu_0) =
    (\tilde m^2(g,\varepsilon), \tilde g_0(g,\varepsilon),
    \tilde z_0(g,\varepsilon), \tilde \nu_0(g,\varepsilon))
\end{equation}
so that \eqref{e:g0gnu0nu} and \eqref{e:nu0cz0c} both hold,
with the right-hand side of \refeq{tildemap} right continuous as $\varepsilon \downarrow 0$,
moreover with
$\tilde m^2(g,\varepsilon) \downarrow 0$ as $\varepsilon \downarrow 0$.
The construction of \eqref{e:tildemap} involves elementary calculus,
and uses \eqref{e:chi-m} for the identification
of the critical point, $\tilde m(g,0) = 0$. We do not repeat it here.

This leads to the identity
\begin{equation}
  \lbeq{chim2}
  \chi(g,\nu) = \frac{1+\tilde z_0}{\tilde m^2}
  .
\end{equation}
We now study $\ddp{}{\nu}\chi(g,\nu)$ similarly, by studying the derivative
$\ddp{}{\nu_0}\hat \chi(m^2,g_0,\nu_0,z_0)$, evaluated
at $z_0=z_0^c(m^2,g_0)$, $\nu_0=\nu_0^c(m^2,g_0)$.
It is convenient to use primes to denote derivatives with respect to $\nu_0$,
evaluated at $(m^2,g_0,\nu_0^c,z_0^c)$.
From \refeq{chibarm-bis2}, we obtain
\begin{equation}
\lbeq{chiNprime}
  \hat\chi_N'
  =
  \frac{A_N'}{B_N}  - \frac{A_NB_N'}{B_N^2}.
\end{equation}

Then, by \refeq{ZN0}, $B_N\to 1$, and by definition,
\begin{equation}
  \label{e:chiprimebarm-bis2}
  A_N'
  =
  \frac{1}{m^4}
  \left(
    - \ddp{\nu_N}{\nu_0}
  + \frac{1}{|\Lambda|} \ddp{}{\nu_0}  D^2W_N(0; \1,\1)
  + \frac{1}{|\Lambda|} \ddp{}{\nu_0} D^2K_N(0; \1,\1)
  \right)
  .
\end{equation}
By \cite[Lemma~\ref{log-lem:gzmuprime}]{BBS-saw4-log}
(with the new interpretation of $\gamma$ in \refeq{nupta}), as $j \to \infty$,
\begin{equation} \label{e:muprime}
  \nu_j' 
  \sim (1+O(g_0))
  \left(\frac{g_j}{g_0}\right)^\gamma
  ,
\end{equation}
and by \cite[Lemma~\ref{log-lem:ginfty}]{BBS-saw4-log}, as $m^2\downarrow 0$,
\begin{equation} \label{e:ginfty}
  g_\infty \sim \frac{1}{\bubble_{m^2}}
  .
\end{equation}
The combination of \refeq{muprime}--\refeq{ginfty} gives
\begin{equation}
\lbeq{nuNprimelim}
    \lim_{N\to\infty} \nu_N'
    \sim
    (1+O(g_0))
    \left(\frac{1}{g_0\bubble_{m^2}}\right)^\gamma.
\end{equation}
It is important to compute derivatives exactly at the
critical $\nu_0^c(m^2,g_0),z_0^c(m^2,g_0)$,
as these remain bounded in the infinite
volume limit as in \refeq{nuNprimelim}.
As argued below \cite[\eqref{log-e:WNbd}]{BBS-saw4-log}, the $W$ and $K$ terms in \refeq{chiprimebarm-bis2} are
respectively $O(\chicCov_N g_N \nu_N')$ and $O(\chicCov_N g_N^2 \nu_N')$,
and hence are relatively small.
Thus, as $m^2 \downarrow 0$,
\begin{equation}
  \lim_{N\to\infty}
  A_N' \sim
  - (1+O(g_0))
  \frac{1}{m^4}
  \left(\frac{1}{g_0\bubble_{m^2}}\right)^\gamma.
\end{equation}

By \eqref{e:ZN0}, and by the bound on the derivative of $K_N$ of
\cite[\eqref{log-e:KNbd2}]{BBS-saw4-log}
(discussed further in Lemma~\ref{lem:der2} below),
\begin{equation}
  B_N'
  = \ddp{}{\nu_0}(1+K_N(\Lambda;0))
  = O(\chicCov_N g_N^2L^{2N} \nu_N').
\end{equation}
Therefore, since
$L^{2N}\nu_N=O(\chicCov_Ng_N)$ as noted above,
\begin{equation}
    A_NB_N' =  m^{-4}O(\nu_N L^{2N}\chicCov_N g_N^2 (g_0\bubble_{m^2})^{-1})
    =
    m^{-4}O(\chicCov_N g_N^{3} (g_0\bubble_{m^2})^{-1}),
\end{equation}
and this contribution to $\hat\chi_N'$ can be absorbed into the error term of
the leading contribution due to the $A_N'$ term.
The convergence of $\hat\chi_N'$ to its limiting value
can be seen to be uniform on compact
subsets of $m^2 \in (0,\delta)$.
Therefore the limit and derivative can be interchanged.
As in the proof of \cite[Theorem~\ref{log-thm:suscept}]{BBS-saw4-log},
for $\hat g_0 \in (0,\delta)$,
\begin{align}
  \label{e:chiprime-m}
  \hat\chi' \left(m^2,g_0,\nu_0^c(m^2,g_0),z_0^c(m^2,g_0) \right)
  &\sim - \frac{1}{m^4} \frac{1+O(\hat g_0)}{(\hat g_0\bubble_{m^2})^{\gamma}}
  \quad \text{as $(m^2,g_0) \to (0,\hat g_0)$}.
\end{align}
With \eqref{e:tildemap},
  the chain rule, \refeq{newbubble}, and \refeq{chim2} imply, and with $\varepsilon = \nu_c-\nu > 0$,
\begin{align} \label{e:chideriv}
  \ddp{}{\nu} \chi(g,\nu)
  &=
  \ddp{}{\nu} \lim_{N\to\infty} \chi_N(g,\nu)
  \nnb
  &= (1+\tilde z_0(g,\varepsilon))^2 \lim_{N\to\infty} \ddp{\hat\chi_N}{\nu_0}(\tilde m^2(g,\varepsilon),\tilde g_0(g,\varepsilon),\tilde\nu_0(g,\varepsilon),\tilde z_0(g,\varepsilon))
  \nnb
  & \sim
  - (1+O(g))\chi^2(g,\nu)
  (g \bubble_{\tilde m^2(g,\varepsilon)})^{-\gamma}
  \nnb
  &\sim
  - \frac{1+O(g)}{(g {\sf b})^{\gamma}}
  \chi^2(g,\nu) (\log \chi(g,\nu))^{-\gamma}
  .
\end{align}
It is now an exercise in calculus, carried out in \cite{BBS-saw4-log},
to conclude the desired asymptotic formula \refeq{suscept4} for the susceptibility.

The proof of \eqref{e:Anuc} is exactly as in \cite{BBS-saw4-log}. In particular,
the asymptotic formula for $\nu_c$ is proved as in \cite[Theorem~\ref{log-thm:nuc}]{BBS-saw4-log}.
This concludes the proof.
\end{proof}

By \refeq{suscept4} and \refeq{chim2},
with $\gamma$ as in \refeq{bubgam}, the map $\tilde m$ introduced in \eqref{e:tildemap}
satisfies, as $\varepsilon \downarrow 0$,
\begin{equation} \label{e:masymp}
  \tilde m^2(g,\varepsilon) \sim
  c_g \varepsilon (\log \varepsilon^{-1})^{-\gamma}
  \qquad
  \text{with} \quad c_g = (1+\tilde{z}_0(g,0))/A(g).
\end{equation}

\subsection{Proof of Theorem~\ref{thm:pressure}: pressure and its derivatives}
\label{sec:pfpressure}

\subsubsection{Proof of Theorem~\ref{thm:pressure}(i): pressure}

We now prove Theorem~\ref{thm:pressure}(i).
The proof uses the following lemma concerning the coupling constants $u_j$.
In its statement, we extend the renormalisation group flow to $g_0=0$
by continuity, to discuss $u_j$ at $g_0=0$.
The proof of Lemma~\ref{lem:uinfty} is deferred to Lemma~\ref{lem:uinfty-app}.

\begin{lemma} \label{lem:uinfty}
  For $(m^2,g_0) \in [0,\delta)^2$,
  the limit
  $u_\infty = \lim_{j\to\infty}u_j$ exists, is continuous in $(m^2,g_0) \in [0,\delta)^2$,
  and obeys
  \begin{equation}
    \lbeq{uinfty}
    u_\infty = \lim_{j\to\infty}u_j = u_j + O(L^{-4j}\chicCov_j\gbar_j).
  \end{equation}
  In particular, since $u_0=0$, $u_\infty = O(g_0)$.
\end{lemma}

\begin{proof}[Proof of Theorem~\ref{thm:pressure}(i)]
  Let $\delta>0$ be sufficiently small, and let $\nu=\nu_c
  +\varepsilon$ with $\varepsilon \in (0,\delta)$.
  Let $p_N(g,\nu)=|\Lambda_N|^{-1} \log Z_{g,\nu,N}$, as in \refeq{pNdef}.
  We prove that the limit
  $p(g,\nu)=\lim_{N\to\infty}|\Lambda_N|^{-1}p_N(g,\nu)$ of
  \eqref{e:pressuredef} exists, and is given by
  \begin{equation}
    p(g,\nu)
    =
    p(0,1/\chi) (1+ O(g))
    ,
  \end{equation}
  with the constant in $O(g)$ uniform in $\varepsilon \in (0,\delta)$.
  We also prove that $\lim_{\nu \downarrow \nu_c} p(g,\nu) =
  p(0,0)(1+ O(g))$.

  By definition, the partition function is
  given by the integral
  \begin{equation} \label{e:Zpress}
    Z_{g,\nu,N} = \int e^{-V_{g,\nu}(\varphi)} d\varphi.
  \end{equation}
  Given
  $(g,\nu)$, we choose
  $(m^2,g_0,\nu_0,z_0)=(\tilde m^2, \tilde g_0, \tilde \nu_0, \tilde z_0)$
  as in \eqref{e:tildemap}.  Since $\varepsilon >0$, we have $m^2>0$.
  Then, with \eqref{e:Vsplit} and the change of variables used to obtain \refeq{ExF},
  \begin{align}
  \lbeq{Znew}
    Z_{g,\nu,N}
    &= (1+ z_0)^{n|\Lambda_N|/2}
    Z_{0,m^2,N}
    \Ex_C Z_0
    = (1+  z_0)^{n|\Lambda_N|/2}
    e^{|\Lambda_N| p_N(0,m^2)}
    Z_N(0),
  \end{align}
  where $Z_{0,m^2,N}$  is the normalisation
  of $\Ex_C$, and $p_N(0,m^2)$ is the free pressure.
  By \eqref{e:ZN0} and \refeq{uinfty},
  \begin{equation}
    |\Lambda_N|^{-1} \log Z_N(0) =
    - u_N + O(|\Lambda_N|^{-1} \chicCov_N g_N^3)
    =
    -
    u_\infty  + O(L^{-4N}\chicCov_N g_N)
    ,
  \end{equation}
  with the error uniform as long as $N$ is large enough that
  $m^2 \in [\delta L^{-2(N-1)},\delta)$.
  Setting $q(m^2,g_0) = \frac{n}{2}\log(1+ z_0^c(m^2,g_0)) -  u_\infty(m^2, g_0)$, this gives
  \begin{equation}
  \lbeq{pN-2}
    p_N(g,\nu) = p_N(0,m^2) + q(m^2,g_0) + O(L^{-4N}\chicCov_N g_N).
  \end{equation}
  Therefore
  \begin{equation}
  \lbeq{plim}
    p(g,\nu)= \lim_{N \to\infty}p_N(g,\nu)=p(0,m^2) + q(m^2,g_0)
  \end{equation}
  exists.

  By Theorem~\ref{thm:flow-flow} and Lemma~\ref{lem:uinfty}, $q = O(g_0) = O(g)$.
  Also, since $z_0$ and $u_\infty$ are continuous in $(m^2,g_0) \in [0,\delta)^2$,
  and
  since $\tilde m^2, \tilde g_0$
  are continuous as $\varepsilon \downarrow 0$,
  $q=q(\tilde m^2, \tilde g_0)$
  is continuous as $\nu \downarrow \nu_c$.  Therefore,
  with $p(0,0)=\lim_{m^2 \downarrow 0}p(0,m^2)$ (as discussed at
  \refeq{pressure0}), since  $\mgen^2 \downarrow 0$,
  as $\varepsilon \downarrow 0$, it follows from \refeq{plim} that
  \begin{equation}
    \lim_{\nu \downarrow \nu_c}p(g,\nu)  = p(0,0) +O(g) = p(0,0)(1+O(g)).
  \end{equation}
  Since $m^{2}=(1+z_0)\chi^{-1}$ by \refeq{chim2}, and since $z_0 = O(g)$,
  \refeq{plim} and the regularity of the free pressure imply that
  \begin{equation}
    p(g,\nu)
    =p(0,\chi^{-1}) + O(g) = p(0,(1+z_0)\chi^{-1})(1+O(g)),
  \end{equation}
  and the proof is complete.
\end{proof}

\subsubsection{Derivatives of the pressure}

In this section, we prepare for the proof of Theorem~\ref{thm:pressure}(ii--iii).
By \refeq{Znew} and \refeq{ZN0}, the finite volume pressure is given by
\begin{equation}
\lbeq{pNformula}
  p_N(g,\nu) =
  \frac{n}{2}\log(1+z_0) + p_N(0,m^2)
  -  u_N + \frac{1}{|\Lambda|} \log (1+K_N(\Lambda; 0)),
\end{equation}
when the right-hand side is defined via
$(m^2,g_0,\nu_0,z_0) = (\tilde m^2,\tilde g_0,\tilde \nu_0, \tilde z_0)$.
In fact,
the identity \eqref{e:pNformula} continues to hold as long as
 \eqref{e:g0gnu0nu} holds and the right-hand side is well-defined.
The latter is the case in an $N$-dependent neighbourhood
of $(\tilde m^2,\tilde g_0,\tilde \nu_0,\tilde z_0)$, by continuity.

As in the proof of Theorem~\ref{thm:suscept},
we temporarily consider $(m^2,g_0)$ rather than $(g,\nu)$ as the primary variables, and
consider derivatives with respect to $\nu_0$ instead of $\nu$. We denote
$\nu_0$-derivatives evaluated at $(m^2,g_0,\nu_0^c,z_0^c)$
with primes, e.g., $u_N'' = \ddp{^2}{\nu_0^2} u_N$.
Since $\nu_0 = \nu(1+z_0) - m^2$, by \eqref{e:g0gnu0nu}, by assumption,
differentiation of \refeq{pNformula}, with \refeq{ZN0} to bound $K_N$, gives
\begin{align} \label{e:pNderiv1}
  \ddp{}{\nu} p_N(g,\nu)
  &=
  -
  (1+z_0) u_N'
  + |\Lambda|^{-1} O(K_N'(\Lambda;0)),
  \\
  \label{e:pNderiv2}
  \ddp{^2}{\nu^2} p_N(g,\nu)
  &=
  -
  (1+z_0)^2 u_N''
  + |\Lambda|^{-1} O\big(|K_N'(\Lambda;0)|^2 +|K_N''(\Lambda;0)|\big).
\end{align}
Thus we require estimates on the $\nu_0$-derivatives of $u_N$ and $K_N(\Lambda;0)$.
These are provided by the following two lemmas.
We defer part of the proof of Lemma~\ref{lem:der2}, and all of the proof of
Lemma~\ref{lem:u2p-bis}, to Lemmas~\ref{lem:gzmuprime2} and \ref{lem:u2p}, respectively.
The lemmas are most conveniently expressed in terms of the transformed
variables $(\gch_j,\much_j,\zch_j)$, defined in \eqref{e:gch-def}--\eqref{e:much-def},
instead of $(g_j,\mu_j,z_j)$.

\begin{lemma}
\label{lem:der2}
Let $(m^2,g_0)\in [0,\delta) \times (0,\delta)$, and let $(z_0,\nu_0)=(z_0^c,\nu_0^c)$.
There exists a function $c(m^2,g_0) = 1+O(g_0)$, continuous on $[0,\delta)^2$,
such that, for all $j \in \N_0$,
\begin{equation}
  \label{e:muprime2}
  \much_j' = L^{2j} \left(\frac{\gch_j}{g_0}\right)^{\gamma} (c(m^2,g_0)+O(\chicCov_j\gch_j)),
\end{equation}
Also, for $N \in \N$, uniformly in $(m^2,g_0) \in [\delta L^{-2(N-1)},\delta) \times (0,\delta)$,
\begin{equation} \label{e:Kprime12}
  K_N'(\Lambda,0) = O(\chicCov_j\much_j'\gch_j^2),
  \quad K_N''(\Lambda,0) = O(\chicCov_j(\much_j')^2 \gch_j).
\end{equation}
\end{lemma}

\begin{proof}
The proof for $\much_j'$ is given in
\cite[Lemma~\ref{log-lem:gzmuprime}]{BBS-saw4-log}
(stated there for the weakly self-avoiding walk, but as discussed in Section~\ref{sec:method},
the same proof applies to the $n$-component $|\varphi|^4$ model).
It is also shown there that $\|K_N'\|_{\Wcal_j} = O(\chicCov_j\much_j' \gch_j^2)$
which, with \eqref{e:Tphidef}, \eqref{e:Tphidom}, implies
$|K_N'(\Lambda;0)| \leq \|K_N'(\Lambda)\|_{T_{0,N}} \leq
\|K_N'\|_{\Wcal_j} = O(\chicCov_j\much_j' \gch_j^2)$.
Analogous considerations can be used to prove the bound on $K_N''$, and we defer its
proof to Lemma~\ref{lem:gzmuprime2}.
\end{proof}

The $\nu_0$-derivatives of $u_N$ of \eqref{e:usum}
are estimated in the following lemma, whose proof
 can be found in Lemma~\ref{lem:u2p}.
In the proof, not only $\much_j'$ arises, but also all of the
other first and second derivatives of $\gch_j,\zch_j,\much_j$,
as well as a non-perturbative error due to $K_j$.
The leading contribution is seen to be due only to $\mu_j'$,
with all other terms accounted for in the error terms.

\begin{lemma} \label{lem:u2p-bis}
Let $(m^2,g_0)\in(0,\delta)^2$, and let $(z_0,\nu_0)=(z_0^c,\nu_0^c)$.
There exist $u_\infty'$ continuous in $(m^2,g_0) \in [0,\delta)^2$, and
$u_\infty''$ continuous in $(m^2,g_0) \in (0,\delta)^2$ such that
\begin{align}
  \label{e:uNprime1lim-bis}
  u'_\infty &= \lim_{N\to\infty} u_N' = \frac{n}{2} \sum_{j=1}^\infty \nuch_j' C_{j;0,0} + O(g),
  \\
  \label{e:uNprime2lim-bis}
  u''_\infty &= \lim_{N\to\infty} u_N''
  = -\frac{n}{2(8+n)} \sum_{j=0}^\infty  \beta_j (\nuch_j')^2 + O(1).
\end{align}
The convergence $u_N'\to u_\infty'$ is uniform in $(m^2,g_0) \in [0,\delta)^2$,
and the convergence $u_N'' \to u_\infty''$ is uniform on compact subsets of $(m^2,g_0) \in (0,\delta)^2$.
\end{lemma}

Given Lemmas~\ref{lem:der2}--\ref{lem:u2p-bis}, we now show that $p$ is
twice differentiable in the interval $\nu\in (\nu_c,\nu_c+\delta)$, with
\begin{equation} \label{e:pderiv12}
  \ddp{}{\nu} p(g,\nu)
  =
  -
  (1+z_0) u_\infty'
  ,\qquad
  \ddp{^2}{\nu^2} p(g,\nu)
  =
  -
  (1+z_0)^2 u_\infty''
  .
\end{equation}
In Theorem~\ref{thm:pressure}(i), it is shown that
$\lim_{N\to\infty} p_N(g,\nu) = p(g,\nu)$, for $\nu\in (\nu_c,\nu_c+\delta)$.
We now argue that
the derivatives $\ddp{^s}{\nu^s} p_N(g,\nu)$
converge compactly (uniformly on compact subsets)
on $\nu \in (\nu_c,\nu_c+\delta)$  to a limiting function, for $s=1,2$.
The compact convergence of the derivatives implies that $p(g,\nu)$
is twice differentiable, with derivatives given by the limits of the finite
volume derivatives, i.e.,
for $s=1,2$,
\begin{equation} \label{e:pNinfderiv}
  \ddp{^s}{\nu^s} p(g,\nu) = \lim_{N\to\infty} \ddp{^s}{\nu^s} p_N(g,\nu).
\end{equation}

To establish the compact convergence, it suffices to show that each term on the
right-hand sides of \eqref{e:pNderiv1}--\eqref{e:pNderiv2} converges compactly in
$m^2 \in (0,\delta)$.
For the terms on its right-hand sides involving derivatives of $K_N$, we use
\eqref{e:Kprime12}.
These bounds hold
uniformly on $[\delta L^{-2N},\delta)$, and thus uniformly on compact subsets of $m^2 \in (0,\delta)$,
for sufficiently large $N$ (depending on the subset).
By \eqref{e:muprime2}, both members of \eqref{e:Kprime12} 
are bounded by $O(\chicCov_N \gbar_N)$, and thus
converge to $0$ as $N\to\infty$, compactly on $m^2 \in (0,\delta)$.
For $u_N'$ and $u_N''$, compact convergence as functions of $m^2$
is asserted by Lemma~\ref{lem:u2p-bis}.
To translate this into compact convergence in $\nu \in (\nu_c,\nu_c+\delta)$, let
$I \subset (0,\delta)$ be a compact $\nu$-interval, and let $J$ be the closure of its image
under $\mgen^2$.  It is not possible that $0 \in J$.  In fact,
since $(m^2,g_0) \mapsto \nu = (\nu_0^c+m^2)/(1+z_0^c)$
is continuous with $\nu \downarrow \nu_c$ as $m^2\downarrow 0$, if $0$ were in $J$
then $0$ would also have to be a limit point of $I$, which is impossible.
Thus compact convergence on $m^2$-intervals implies compact convergence on $\nu$-intervals.
With \eqref{e:pNderiv1}--\eqref{e:pNderiv2} and \eqref{e:pNinfderiv},
this implies that \eqref{e:pderiv12}--\eqref{e:pNinfderiv} hold
for $\nu \in (\nu_c,\nu_c+\delta)$.

\subsubsection{Proof of Theorem~\ref{thm:pressure}(ii)--(iii): derivatives}
\label{sec:pfspecificheat}

By \eqref{e:pderiv12},
to prove Theorem~\ref{thm:pressure}(ii)--(iii),
it remains to analyse the asymptotic behaviour of
$(1+z_0) u_\infty'$ and $(1+z_0)^2u_\infty''$ as $\varepsilon = \nu-\nu_c \downarrow 0$.
We do this now.

By \eqref{e:VchK-rec}, \eqref{e:TVV2}, and Theorems~\ref{thm:step-mr}--\ref{thm:flow-flow},
the sequence $\gch_j$ satisfies $\gch_{j+1}=\gch_j - \beta_j \gch_j^2 + r_j$
with $r_j = O(\chicCov_j \gch_j^3)$.
As a consequence,
for any continuously differentiable function $\psi: (0,\infty) \to \R$ and any $k \ge j$,
\begin{equation}
  \label{e:gbarsumbisx}
  \sum_{l=j}^{k} (\beta_l \gch_l^2 - r_l) \psi(\gch_l)
  = \int_{\gch_{k+1}}^{\gch_{j}} \psi(t) \; dt
  + O\left(\int_{\gch_{k+1}}^{\gch_{j}} t^2 |\psi'(t)| \; dt
  \right).
\end{equation}
The formula \eqref{e:gbarsumbisx} was proved in
\cite[\eqref{flow-e:gbarsumbis}]{BBS-rg-flow} for the special case $r_j=0$,
but the same proof applies also when $r_j = O(\chicCov_j \gch_j^3)$.
Also, by \cite[Lemma~\ref{flow-lem:elementary-recursion}(ii)]{BBS-rg-flow},
for every $p>1$,
\begin{equation}
  \label{e:gbarpsum}
  \sum_{l=j}^k \chicCov_l \gch_l^p \leq O(\chi_j \gch_j^{p-1})
  ,
\end{equation}
with a constant depending on $p$, but independent of $j$ and $k$.

\begin{proof}[Proof of Theorem~\ref{thm:pressure}(ii)]
By \eqref{e:muprime2} and \eqref{e:uNprime1lim-bis},
with $c=c(m^2,g_0)$ the function of \eqref{e:muprime2},
\begin{align} \label{e:uinfprimepf}
  u_\infty'
  &= n c \sum_{j=0}^\infty \left(\frac{\gch_j}{g_0}\right)^\gamma C_{j;0,0} + O(g)
  \nnb
  &= n c \sum_{j=0}^\infty  C_{j;0,0} + O(1) \sum_{j=0}^N \left(\left(\frac{\gch_j}{g_0}\right)^\gamma-1\right) C_{j;0,0} + O(g)
  .
\end{align}
From \eqref{e:gbarsumbisx} with $\psi(t) = \gamma t^{-1+\gamma}$,
it follows that
\begin{equation}
  g_0^\gamma - \gch_j^\gamma
  = \int_{\gch_j}^{g_0} \psi(t) \, dt
  = \sum_{l=0}^{j-1} O(\chicCov_l\gch_l^{1+\gamma}).
\end{equation}
Thus, with \eqref{e:scaling-estimate} to bound $C_{j;0,0} = O(L^{-2j})$,
\begin{align} \label{e:sumuinfprimepf}
  \sum_{j=0}^\infty \left(\gch_j^\gamma-g_0^\gamma\right) C_{j;0,0}
  &=\sum_{j=0}^\infty \sum_{l=0}^{j-1} O(\gch_l^{1+\gamma}) C_{j;0,0}
  =\sum_{l=0}^{\infty} O(\gch_l^{1+\gamma}) \sum_{j=l+1}^{\infty} C_{j;0,0}
  \nnb
  &=\sum_{l=0}^{\infty} O(\gch_l^{1+\gamma}) O(L^{-2l})
  = O(g_0^{1+\gamma}).
\end{align}
Since $z_0 = O(g)$ and $c= 1+O(g)$, \eqref{e:pderiv12}, \eqref{e:uinfprimepf}, \eqref{e:sumuinfprimepf} imply
\begin{equation}
  \ddp{}{\nu} p(g,\nu) = - (1+z_0)cnC_{0,0} +O(g) = nC_{0,0} + O(g).
\end{equation}
The covariance on the right-hand side is given by
$C_{0,0} = C_{m^2}(0) = (-\Delta_{\Zd}+m^2)^{-1}_{0,0}$
with $m^2 = (1+  z_0)/\chi$ by \eqref{e:chim2}.
To compare it with the covariance $C_{1/\chi}(0)$ appearing in the statement of the theorem,
we note that $\ddp{}{m^2}C_{m^2}(0) = -B_{m^2} = O(\log m^{-2})$ and that $C_{0}(0) = \lim_{m^2\downarrow 0} C_{m^2}(0) >0$ exists,
so that $C_{m^2}(0)$ is in particular uniformly bounded from below, for $m^2 \geq 0$ sufficiently small.
This shows $C_{m^2}(0) = C_{1/\chi}(0) + O(B_{m^2} m^2 g) = C_{1/\chi}(0) (1+ O(g))$,
and the proof is complete.
Since $u_\infty'$ and $z_0$ are continuous in $(m^2,g_0) \in [0,\delta)^2$,
the statement about the limit $\nu\downarrow \nu_c$ of $p(g,\nu)$ follows from the right-continuity
as $\varepsilon \downarrow 0$ of \eqref{e:tildemap}.
\end{proof}

\begin{proof}[Proof of Theorem~\ref{thm:pressure}(iii)]
By definition, the specific heat $c_H$ is the second derivative in \eqref{e:pderiv12}.
The estimates \refeq{uNprime2lim-bis} for $u_\infty''$ and \refeq{muprime2} for $\nu_j'$
imply
\begin{equation} \label{e:sumg2gamma1}
  c_H
  =
  (1+z_0)^2 c g_0^{-2\gamma}
  \frac{n}{2(n+8)}
  \sum_{j=0}^\infty \beta_j \gch_j^{2\gamma} + O(1),
\end{equation}
where $c =c(m^2,g_0)$ is the function from \eqref{e:muprime2}.
Since $2\gamma>0$, by \eqref{e:gbarsumbisx}--\eqref{e:gbarpsum},
\begin{align}
  \sum_{j=0}^N \beta_j \gch_j^{2\gamma}
  &= \int_{\gch_N}^{g_0} t^{2\gamma-2} \; dt
  + \int_{\gch_{N}}^{g_{0}} O(t^{2\gamma-1}) \; dt
  + \sum_{j=0}^N O(\chi_j \gch_j^{1+2\gamma})
  \nnb & = \int_{\gch_N}^{g_0} t^{2\gamma-2} \; dt + O(g_0^{2\gamma})
  \nnb &
  =
  \begin{cases}
    (1-2\gamma)^{-1}(\gch_N^{2\gamma-1}-g_0^{2\gamma-1}) +O(g_0^{2\gamma}) & (2\gamma<1)
    \\
    \log g_0 / \gch_N +O(g_0) & (2\gamma =1)
    \\
    (2\gamma-1)^{-1}(g_0^{2\gamma-1}-\gch_N^{2\gamma-1}) +O(g_0^{2\gamma}) & (2\gamma>1).
  \end{cases}
\end{align}
We now apply
$\gch_\infty \sim 1/\bubble_{m^2} \sim 1/({\sf b} \log m^{-2}) = 16\pi^2/((n+8)\log m^{-2})$,
by \refeq{ginfty} and \refeq{newbubble},
and use $2\gamma - 1 = \frac{n-4}{n+8}$ to conclude that,
as $m^2\downarrow 0$,
\begin{equation} \label{e:sumg2gamma}
  c_H
  \sim
  (1+z_0)^2 c g_0^{-2\gamma}
  \frac{n}{2(n+8)}
  \begin{cases}
  \frac{n+8}{4-n}\left(\frac{n+8}{16\pi^2} \log m^{-2} \right)^{(4-n)/(n+8)}
  & (n<4)
  \\
  \log \log m^{-2} & (n=4)
  \\
  \frac{n+8}{n-4}g_0^{(n-4)/(n+8)}(1 + O(g_0)) & (n>4)
  .
  \end{cases}
\end{equation}
To conclude \eqref{e:sumg2gamma},
we have used the fact that the $O(1)$ term in \eqref{e:sumg2gamma1} is negligible in the asymptotics of \eqref{e:sumg2gamma} for $n \leq 4$,
and that, for $n > 4$, it can be included in the $O(g_0)$ error term in
\eqref{e:sumg2gamma}, since $O(g_0^{\smash{-2\gamma}})O(g_0) = O(1)$.
Since $m^2 = \tilde m^2(g,\varepsilon) \sim c_g \varepsilon (\log \varepsilon^{-1})^{-\gamma}$ as
$\varepsilon \downarrow 0$ by \eqref{e:masymp},
it follows that $\log m^{-2} \sim \log \varepsilon^{-1}$ and
$\log \log m^{-2} \sim \log \log \varepsilon^{-1}$,
giving the $\varepsilon$-dependence claimed in \eqref{e:cHasy}.
\end{proof}

Since $z_0=O(g)$, $c=1+O(g)$, and $g_0=g(1+O(g))$, we see from \refeq{sumg2gamma}
that, as $g \downarrow 0$, the amplitude $D(g,n)$ has the asymptotic formula
stated in \eqref{e:Dgn}.

\subsection{Proof of Theorem~\ref{thm:clt}: scaling limits}
\label{sec:pfclt}

We first prove the following purely analytic estimate.
The proof shows that
for Theorem~\ref{thm:clt} it is in fact sufficient if $h\in C^{p_\Phi+d}=C^8$,
rather than $h \in C^\infty$.  We have made no attempt to optimise the regularity
assumption.

\begin{lemma}
Let $h \in C^{p_\Phi+d}(\T^d,\R^n)$ and $m_N^2>0$.  Let $C^{(N)}=(-\Delta_{\Lambda_N}+m_N^2)^{-1}$.
Then
\begin{equation}  \label{e:Cfbd}
  \|C^{(N)}h_N\|_{\Phi_N} \leq O(L^N m_N^{-2}) \|h\|_{C^{p_\Phi + d}(\T^d,\R^n)}
  .
\end{equation}
\end{lemma}

\begin{proof}  Let $f: \Lambda_N \to \R$.
  By the lattice Sobolev inequality \cite[Lemma~\ref{norm-lem:sobolev2}]{BS-rg-norm},
  \begin{align}
    \|\nabla^\alpha (C^{(N)} f)\|_\infty
    &\leq
    O( L^{-Nd}) \max_{|\beta|_\infty \leq 1}
    \|L^{N|\beta|_1}
    \nabla^{\alpha+\beta}(C^{(N)}f)\|_2
    \nnb & =
    O( L^{-Nd}) \max_{|\beta|_\infty \leq 1}
    \|C^{(N)}(L^{N|\beta|_1}
    \nabla^{\alpha+\beta}f)\|_2
    .
  \end{align}
  Since $\|C^{(N)} \tilde f\|_{2} \leq O(m_N^{-2}) \|\tilde f\|_2$
  and $\|\tilde f\|_2 \leq |\Lambda|^{1/2}\|\tilde f\|_\infty = L^{Nd/2} \|\tilde f\|_\infty$,
  \begin{align}
    L^{N|\alpha|_1}\|\nabla^\alpha (C^{(N)} f)\|_\infty
    &\leq O(m_N^{-2}) L^{-Nd/2}
    \max_{|\beta|_\infty \leq 1} L^{N(|\alpha|_1+|\beta|_1)} \|\nabla^{\alpha+\beta}f\|_2
    \nnb
    &\leq O(m_N^{-2})  \max_{|\beta|_\infty \leq 1}
    L^{N(|\alpha|_1+|\beta|_1)}
    \|\nabla^{\alpha+\beta} f\|_\infty
    .
  \end{align}
Moreover, by estimating discrete differences by derivatives, it follows that
for $h \in C^{s}(\T^d,\R^n)$ and $h_N(x) = h(L^{-N}x)$,
\begin{equation}
  L^{N|(\alpha|_1 +|\beta|_1} \|\nabla^{\alpha+\beta} h_N\|_\infty
  \leq O(1) \|h\|_{C^{s}(\T^d,\R^n)}
  \quad\quad
  (|\alpha|_1+|\beta|_1 \leq s)
  .
\end{equation}
Thus, by \eqref{e:Phinormdef} with $\ell_N^{-1} \propto L^{N}$,
\begin{equation}  \label{e:Cfbd-pf}
  \|C^{(N)}h_N\|_{\Phi_N} \leq O(L^N m_N^{-2}) \|h\|_{C^{p_\Phi+d}(\T^d,\R^n)}
  ,
\end{equation}
and this proves \refeq{Cfbd}.
\end{proof}

The coupling constant $g$ does not play a role in this section, so we fix it
(small) and drop it from the notation.
The following lemma verifies that, under the hypotheses of Theorem~\ref{thm:clt},
the mass $m_N^2=\tilde m^2(\varepsilon_N)$ defined as in \eqref{e:tildemap} obeys
the condition needed for the covariance decomposition estimates and hence for our
application of the renormalisation group method.

\begin{lemma}
\label{lem:mLlim}
Let $\rho=1+\tilde z_0(0)$, let $m^2 \in (0,\infty]$, and suppose that
$\lim_{N\to \infty} \chi^{(N)}L^{-2N} = \rho m^{-2}$
(with $\infty^{-2}=0$).  Then
$\lim_{N\to\infty} m_N^{-2}L^{-2N} =m^{-2}$.
In particular, $m_N^{-2}L^{-2N}$ is bounded uniformly in $N$.
\end{lemma}

\begin{proof}
Let $z_{0,N} = \tilde z_0(\varepsilon_N)$.
By \refeq{chim2},
\begin{equation}
    \chi^{(N)}=\chi(\nu_c+\varepsilon_N) = \frac{1+z_{0,N}}{m_N^2},
\end{equation}
and hence, since $z_{0,N}=O(g)$,
\begin{equation}
\lbeq{mLlim}
    m_N^{-2}L^{-2N} = \chi^{(N)}L^{-2N}(1+z_{0,N})^{-1}
    \sim \rho m^{-2} (1+z_{0,N})^{-1}.
\end{equation}
This proves the desired result if $m^{-2}=0$.  On the other hand, if $m^{-2}>0$ then
by \refeq{mLlim} $m_N^2 \to 0$,
and since $\lim_{m_N^2 \to 0}1+z_{0,N} = 1+\tilde z_0(0)=\rho$, the
right-hand side converges to $m^{-2}$.
\end{proof}

\begin{proof}[Proof of Theorem~\ref{thm:clt}]
Fix $h \in C^\infty(\T^d,\R^n)$, and define $h_N:\Lambda_N \to \R^n$
by $h_N(x) = h(L^{-N}x)$.
Let $\rho=1+\tilde z_0(0)$.
Suppose that $\chi^{(N)}L^{-2N} \to \rho m^{-2}$, with $m^{2}=\infty$ allowed and $\infty^{-2} = 0$,
and set
\begin{equation}
  \bar C = \begin{cases}
    (-m^{-2}\Delta_{\T^d}+1)^{-1} & (m^2>0)\\
    1 & (m^2 = \infty),
  \end{cases}
\end{equation}
with the usual convention that $\bar C$ acts component-wise on vector-valued functions.
To prove \eqref{e:clt}, it suffices to show that
\begin{alignat}{2}
  \label{e:U1}
  \lim_{N\to\infty}
  \log \langle e^{(\varphi,h_N)/\sigma_N} \rangle_{\nu_c+\varepsilon_N,N}
  &=
  \half (h,\bar C h)_{\T^d}
  ,
\end{alignat}
where, for $f,g : \T^d \to \R^n$,
\begin{equation}
  (f,g)_{\T^d} = \sum_{i=1}^n \int_{\T^d} f^i(x)g^i(x) \; dx.
\end{equation}

Let $m_N^2=\tilde m^2(\varepsilon_N)$ and $z_{0,N}=\tilde z_0(\varepsilon_N)$.
Let  $C^{(N)}=(-\Delta_{\Lambda_N}+m_N^2)^{-1}$,
and define $Z_0 = Z_0^{(N)}$ and $Z_N = Z_N^{(N)}$ by \eqref{e:V0Z0} and \eqref{e:ZNdef}.
It follows from \refeq{ExF} and \refeq{Sigdef}--\refeq{LTC} that
\begin{align}
\label{e:UZ}
  \log \langle e^{(\varphi,h_N)/\sigma_N} \rangle_{\nu_c+\varepsilon_N,N}
  &= \frac 12 \frac{1+z_{0,N}}{\sigma_N^2}(h_N,C^{(N)}h_N)
  \nnb & \qquad
  +
  \log \left( \frac{Z_N((1+z_{0,N})^{1/2}\sigma_N^{-1}C^{(N)}h_N)}{Z_N(0)} \right)
  .
\end{align}
We start with the first term on the right-hand side of
\refeq{UZ}.
We claim that
\begin{equation}
\lbeq{Riemann}
  \lim_{N\to\infty} \frac{1+z_{0,N}}{\sigma_N^2}
  \sum_{x,y \in \Lambda_N} h_N( x)
  C^{(N)}_{xy}
  h_{N}(y)
  =
  (h, \bar C h)_{\T^d}
  .
\end{equation}
Once this is established, it then suffices to prove that the ratio
inside the logarithm on the right-hand side of \refeq{UZ} has limit
$1$ as $N \to \infty$, since this gives \refeq{U1}.

To prove \refeq{Riemann},
we set $\T^d_N = \T^d \cap L^{-n}\Zd$
and
$\lambda(k) = 4\sum_{j=1}^d \sin^2(\pi k_j)$ for $k\in \T^d$,
as in Section~\ref{sec:mr}.
For $l \in \Lambda_N$, $\lambda(L^{-N}l) \sim 4\pi^2|l|^2$ as $N \to \infty$.
We write the rescaled covariance in terms of its Fourier transform, as
\begin{align}
    C^{(N)}_{L^{N}x,L^{N}y}
    &=
    \frac{1}{L^{dN}}
    \sum_{k\in \T^d_N} \frac{1}{\lambda(k)+ m_N^2}
    e^{2\pi i k\cdot L^{N}(x-y)}
    =
    \frac{1}{L^{dN}}
    \sum_{l\in \Lambda_N} \frac{1}{\lambda(L^{-N}l)+ m_N^2}
    e^{2\pi i l\cdot (x-y)}
    .
\end{align}
Since  $\chi^{(N)} = (1+z_{0,N})m_N^{-2}$ by \refeq{chim2},
by rewriting the convolution in terms of the Fourier transform
we obtain
(with $|\hat h(k)|^2 = \sum_{i=1}^d |\hat h^i(k)|^2$)
\begin{align}
  \frac{1+z_{0,N}}{\sigma_N^2}
     \sum_{x,y \in \Lambda_N}
  h_{N}(x)
  C^{(N)}_{xy}
  h_N(y)
  & =
  \frac{m_N^2}{ L^{dN}} \sum_{u,v \in \T_N^d} h(u) C^{(N)}_{L^Nu,L^Nv}
  h(v)
  \nnb
  & =
  \frac{m_N^2}{L^{dN}} \sum_{l \in \Lambda_N} \frac{|\hat h(l)|^2}{\lambda(L^{-N} l)+ m_N^2}
   \nnb
  & \sim
  \frac{1}{L^{dN}}
  \sum_{l \in \Lambda_N}
  \frac{|\hat h(l)|^2}
  {m_N^{-2}L^{-2N}4\pi^2|l|^2 + 1}
  .
\lbeq{R1pf}
\end{align}
Therefore, by Lemma~\ref{lem:mLlim}, \eqref{e:Riemann} holds as claimed.

Thus it suffices to show that the ratio of two $Z_N$ in \refeq{UZ} has limit 1.
By \refeq{ZN0}, this will follow once we show that
\begin{equation}
\lbeq{IKclt1}
    \lim_{N\to\infty}
    I_N(\Lambda_N; \sigma_N^{-1} C^{(N)}h_N)
    =
    1,\qquad
    \lim_{N\to\infty}
     K_N(\Lambda_N; \sigma_N^{-1} C^{(N)}h_N)
    =
    0.
\end{equation}
  By \refeq{Cfbd} and \refeq{chim2},
\begin{align}
    \|\sigma_N^{-1} C^{(N)}h_N\|_{\Phi_N}
    &
    = O(\sigma_N^{-1}L^N m_N^{-2})
    = O(\sqrt{\chi^{(N)}L^{-2N}} (\chi^{(N)}m_N^2)^{-1})
    \nnb &
    = O(\sqrt{\chi^{(N)}L^{-2N}} )
    =
    \begin{cases}
    o(1) & (\chi^{(N)}L^{-2N} \to 0)
    \\
    O(1) & (\chi^{(N)}L^{-2N} =O(1)).
    \end{cases}
\end{align}
  Since $\|V_N\|\to 0$ by Theorem~\ref{thm:flow-flow},
  the bounds \eqref{e:Ibd} and \eqref{e:Kbd} imply \refeq{IKclt1},
  and the proof of \eqref{e:clt} 
  is complete. We now prove (i)--(ii).

\smallskip \noindent
  (i)
  We fix $\varepsilon>0$ and a sequence $\varepsilon_N \sim \varepsilon$.
  It suffices to show that $\chi^{(N)}\to \chi(\nu_c+\varepsilon)$, and this follows
  immediately from the continuity of $\chi$ for $\nu>\nu_c$ (differentiability was
  established in \refeq{chideriv}).

\smallskip \noindent
  (ii)
We fix $m^2>0$ and a sequence $\varepsilon_N \sim \alpha m^2 L^{-2N} (\log L^N)^{\gamma}$
with $\alpha>0$ to be determined, and set
$m_N^2 = \tilde m^2(\varepsilon_N)$.
  It suffices to prove that there exists $\alpha=\alpha(g)>0$ such that
$\chi^{(N)}L^{-2N} \to (1+\tilde z_0(0))m^{-2}$, or equivalently
(by \refeq{chim2} and the continuity $\tilde z_0(\varepsilon_N) \to \tilde z_0(0)$),
that $m_N^2 L^{2N} \to m^2$.
By \eqref{e:masymp}, as $\varepsilon \downarrow 0$,
\begin{equation}
  \tilde m^2(\varepsilon) \sim
  c_g \varepsilon (\log \varepsilon^{-1})^{-\gamma}
  .
\end{equation}
Therefore, as $N \to \infty$,
\begin{equation}
\lbeq{m2L2N}
  m_N^2
  \sim c_g\alpha m^2 L^{-2N} (\log L^N)^{\gamma} [-\log (\alpha m^2 L^{-2N} (\log L^N)^{\gamma})]^{-\gamma}
  \sim c_g\alpha 2^{-\gamma} m^2 L^{-2N},
\end{equation}
and the proof is completed by taking $\alpha = 2^\gamma c_g^{-1}$.
\end{proof}

\appendix
\normalsize

\section{Bounds on renormalisation group map}

We now prove
Lemma~\ref{lem:uinfty}, the second bound of \eqref{e:Kprime12},
and Lemma~\ref{lem:u2p-bis}.
These are restated here as
Lemmas~\ref{lem:uinfty-app}, \ref{lem:gzmuprime2}, and \ref{lem:u2p}, respectively.
(Lemma~\ref{lem:gzmuprime2} does more, in preparation for the proof of Lemma~\ref{lem:u2p}.)
This involves a detailed analysis of the sequence $u_j$, as well as estimates
on second derivatives of the renormalisation group flow with respect to the
initial condition $\nu_0$.

From \eqref{e:usum}, we recall that the sequence $u_j$ is defined by
\begin{equation} \label{e:usumpf}
  u_j = \sum_{i=0}^{j-1} \delta u_{i+1}.
\end{equation}
The coupling constants $\delta u_{j+1}$ are given by
\begin{equation} \label{e:uerror}
  \delta u_{j+1} = \delta u_{\pt}(V_j) + R^{\delta u}_{j+1}(V_j,K_j),
\end{equation}
with $(V_j,K_j)$ the renormalisation group flow of Theorem~\ref{thm:flow-flow},
$\delta u_\pt$ defined in \eqref{e:uptlong},
and $R^{\delta u}_+$ the $\delta u$ component of \eqref{e:Rdef}.

\subsection{The coupling constant $u$: proof of Lemma~\ref{lem:uinfty}}
\label{sec:u}

We begin with the following lemma concerning $\delta u_\pt$ of \refeq{uptlong}.

\begin{lemma}
  \label{lem:upt}
  The coefficients in \eqref{e:uptlong} are continuous in $m^2 \in [0,\delta)$
  and are uniformly bounded by $O(L^{-dj}\chicCov_j)$.
\end{lemma}

\begin{proof}
  Except for the coefficient of $g^2$, the claim follows from \eqref{e:greekbds}--\refeq{greekbds2}
  and
  the facts that $C_{0,0} = O(L^{-2j})$, $\Delta C_{0,0} = O(L^{-4j})$ by
  \refeq{scaling-estimate} (the $\delta[(\Delta w)^{(2)}]$ term can
  be handled similarly to $\zeta$).

  We fix any $k>0$ and set $M_j = (1+m^2L^{2j})^{-k}$.
  With a $k$-dependent constant, $M_j=O(\chicCov_j)$.
  The remaining bound to be established is
  \begin{equation} \label{e:ug2bd}
      \delta[w^{(4)}] - 4 C_{0,0}w^{(3)} + 2 \Delta C_{0,0} (w^3)^{(**)} - 6 C_{0,0}^2 w^{(2)}
      = O(M_jL^{-4j}).
  \end{equation}
  The left-hand side of \eqref{e:ug2bd} is equal to
  \begin{equation} \label{e:ug2bd2}
    4 \sum_x w_x^3 (C_x-C_0-\frac{1}{2} x_1^2 \Delta C_0) + 6 \sum_x w_x^2 (C_x^2-C_0^2) + 4\sum_x w_x C_x^3 + \sum_x C_x^4.
  \end{equation}
  Since $\nabla^e\nabla^{-e}C_0 = \nabla^e\nabla^{e} C_{0} + O(\|\nabla^3 C\|_\infty)$,
  and by invariance under lattice rotations,
  $x_1^2\Delta C_0$ can be replaced by
  $\sum_{i,j=1}^d x_i x_j \nabla^{e_i}\nabla^{e_j}C_0 +  O(\|\nabla^3 C\|_\infty)$.
  By a discrete Taylor approximation
  (e.g., as in the proof of \cite[Lemma~\ref{loc-lem:taylor-theorem}]{BS-rg-loc}),
  \begin{align}
  \sum_x w_x^3 (C_x-C_0-\frac{1}{2} x_1^2 \Delta C_0)
  &= \sum_x w_x^3 O(|x|^3 \|\nabla^3 C\|_\infty).
\end{align}
  Therefore, using \eqref{e:scaling-estimate} to estimate the $C_n$
(e.g., $\|\nabla^3 C\|_\infty = M_jL^{-5j}$), and
since $C_n$ is supported in a cube with $O(L^{4n})$ points, we obtain
  \begin{align}
  \sum_x w_x^3 (C_x-C_0-\frac{1}{2} x_1^2 \Delta C_0)
  &= O(M_jL^{-5j}) \sum_{j \geq i\geq l\geq m} \sum_{x} C_{i;x}C_{l;x}C_{m;x} |x|^3
  \nnb
  &= O(M_jL^{-5j}) \sum_{j \geq i \geq l} L^{-2i} L^{-2l} L^{5l}
  \nnb
  &= O(M_jL^{-5j}) \sum_{j \geq i} L^{i}
  = O(M_jL^{-4j}).
\end{align}
Similarly,
\begin{equation}
  \sum_x w_x^2 (C_x^2-C_0^2) = O(M_j L^{-6j}) \sum_x w_x^2 |x|^2
  = O(M_jL^{-6j}) \sum_{k\leq j} L^{2k} = O(M_jL^{-4j}).
\end{equation}
The last two terms in \eqref{e:ug2bd2} are $O(M_jL^{2j} L^{-6j}) = O(M_jL^{-4j})$ and $O(M_jL^{4j}L^{-8j}) = O(M_jL^{-4j})$ as claimed.
This completes the proof.
\end{proof}

The proof of Lemma~\ref{lem:uinfty} uses the following definition
and proposition from \cite{BBS-saw4-log}.  The proof of the proposition is
given in \cite[Proposition~\ref{log-prop:Fcont}]{BBS-saw4-log}.

\begin{defn} \label{def:flowcont}
  (i)
  A map $(V,K,m^2) \mapsto F(V,K,m^2)$ acting on a subset
  of $\Vcal \times \Kspace_j \times [0,\delta)$ with values in a Banach space $E$
  is a \emph{continuous function of the renormalisation group coordinates
  at scale-$j$},
  if its domain includes $\domRG_j(\sgen_j) \times \Igen_{j+1}(\mgen^2)$
  for all $\sgen_0 \in [0,\delta) \times (0,\delta)$, and if its
  restriction to the domain $\domRG_j(\sgen_j) \times \Igen_{j+1}(\mgen^2)$
  is continuous as a map $F: \domRG_j(\sgen_j) \times \Igen_{j+1}(\mgen^2) \to E$,
  for all $\sgen_0 \in [0,\delta) \times (0,\delta)$.
  We also say that $F$ is a $C^0$ map of the renormalisation group coordinates.
  \smallskip
  \\
  \noindent
  (ii) For $k \in \N$, a map $F$ is a
  \emph{$C^k$ map of the renormalisation group coordinates at scale-$j$},
  if it is a $C^0$ map  of the renormalisation group coordinates,
  its restrictions to the domains  $\domRG_j(\sgen_j) \times \Igen_{j+1}(\mgen^2)$
  are $k$-times continuously Fr\'echet differentiable in $(V,K)$,
  and every Fr\'echet derivative in $(V,K)$,
  when applied as a multilinear map to directions $\dot{V}\in\Vcal^{p}$
  and $\dot{K}\in\Wcal^{q}$, is jointly continuous in all arguments,
  $m^{2}, V,K, \dot{V}, \dot{K}$.
\end{defn}

\begin{prop} \label{prop:Fcont}
  Let $j < N(\volume)$, $k \in \N_0$,
  and let $F$ be a $C^k$ map of the renormalisation group coordinates at scale-$j$.
  Then, for every $p \leq k$,
  all $s_0 \in [0,\delta) \times (0,\delta)$, the derivative $D_{V_0}^p F(s_0)$ exists, and
  \begin{equation} \label{e:Fcont}
    \text{$s_0 \mapsto D_{V_0}^p F(s_0)$ is a continuous map $[0,\delta) \times (0,\delta) \to L^p(\Vcal, E)$,}
  \end{equation}
  where $L^p(\Vcal,E)$ is the space of $p$-linear maps from $\Vcal$ to $E$ with the operator norm.
\end{prop}

The following lemma is a restatement of Lemma~\ref{lem:uinfty}.

\begin{lemma} \label{lem:uinfty-app}
  For $(m^2,g_0) \in [0,\delta)^2$,
  the limit
  $u_\infty = \lim_{j\to\infty}u_j$ exists, is continuous in $(m^2,g_0) \in [0,\delta)^2$,
  and obeys
  \begin{align}
    \lbeq{uinfty-app}
    u_\infty &= \lim_{j\to\infty}u_j = u_j + O(L^{-4j}\chicCov_j\gbar_j).
  \end{align}
  In particular, since $u_0=0$, $u_\infty = O(g_0)$.
\end{lemma}

\begin{proof}
The first term on the right-hand side of \eqref{e:uerror} is $O(L^{-dj}\chicCov_j\gbar_j)$,
by Lemma~\ref{lem:upt} and Theorem~\ref{thm:flow-flow}.
The second term is $O(L^{-dj}\chicCov_j\gbar_j^3)$, by Theorems~\ref{thm:step-mr}--\ref{thm:flow-flow}.
Thus $\delta u_{j+1}= O(L^{-4j}\chicCov_j \gbar_j)$.

By Theorems~\ref{thm:step-mr}--\ref{thm:flow-flow}, $U_+=(\delta u_+,V_+)$ is a continuous function of the
renormalisation group coordinates at scale $j$.
Thus, by Proposition~\ref{prop:Fcont}, $(m^2,g_0) \mapsto \delta u_{j+1}$ is a continuous
function on $[0,\delta) \times (0,\delta)$. Since $\delta u_{j+1} = O(L^{-4j}\chicCov_j \gbar_j) = O(g_0) \to 0$
as $g_0\downarrow 0$, it follows that $\delta u_{j+1}$ is continuous on $[0,\delta)^2$.

The existence of the limit $u_\infty$ and \eqref{e:uinfty-app} follow immediately from
$\delta u_{j+1}= O(L^{-4j}\chicCov_j \gbar_j)$.
Since $\delta u_{j+1} = O(L^{-4j})$, the sum \eqref{e:usumpf} converges
uniformly on $(m^2,g_0) \in [0,\delta)^2$ as $j\to\infty$,
so
$u_\infty$ is also continuous on $[0,\delta)^2$.  This completes the proof.
\end{proof}

\subsection{Derivatives of flow} 

For a function $f = f(m^2,g_0,\nu_0,z_0)$, we recall the notation
\begin{align}
  f' &= \ddp{}{\nu_0} f(m^2,g_0, \nu_0^c(m^2,g_0), z_0^c(m^2,g_0)),
  \\
  f'' &= \ddp{^2}{\nu_0^2} f(m^2,g_0, \nu_0^c(m^2,g_0), z_0^c(m^2,g_0)),
\end{align}
with $(z_0^c,\nu_0^c)$ as in Theorem~\ref{thm:flow-flow}.
As in \cite[Lemma~\ref{log-lem:gzmuprime}]{BBS-saw4-log},
\begin{gather}
  \label{e:muprime1}
  \much_j' = L^{2j}\left(\frac{\gch_j}{g_0}\right)^{\gamma} (c(m^2, g_0) + O(\vartheta_j \gch_j)),
  \intertext{where $c(m^2,g_0) = 1+O(g_0)$, and}
  \label{e:gzprime1}
  \gch_j', \zch_j' = O(\chicCov_j \much_j'\gch_j^2),
  \quad \|K_j'\|_{\Wcal_j} = O(\chicCov_j \much_j'\gch_j^2).
\end{gather}
The following lemma gives similar bounds for second derivatives, via an
extension of the proof of \cite[Lemma~\ref{log-lem:gzmuprime}]{BBS-saw4-log}.

\begin{lemma} \label{lem:gzmuprime2}
  Let $(m^2,g_0)\in [0,\delta) \times (0,\delta)$, let $(z_0,\nu_0)=(z_0^c,\mu_0^c)$.
  Then
  \begin{equation} \label{e:gzmuprime2pf1}
    \much_j'', \gch_j'', \zch_j'' = O(\chicCov_j(\much_j')^2 \gch_j), \quad \|K_j''\|_{\Wcal_j} = O(\chicCov_j(\much')^2\gch_j).
  \end{equation}
\end{lemma}

\begin{proof}
The proof is by induction, with the induction hypothesis that there exist constants
$M_1,M_2$ such that
\begin{equation} \label{e:induct1}
  |\much_j''|, |\gch_j''|, |\zch_j''| \leq M_1 \chicCov_j(\much_j')^2 \gch_j, \quad
  \|K_j''\|_{\Wcal_j} \leq M_2 \chicCov_j(\much_j')^2 \gch_j.
\end{equation}
This case $j=0$ is trivial since the left-hand sides are $0$.
The advancement of the induction uses the fact that, by \eqref{e:muprime1},
\begin{equation}
  \frac{\gch_{j}}{\gch_{j+1}} = 1+O(\gch_j),
  \quad
  \frac{\much_j'}{\much_{j+1}'} =
  L^{-2}(1+O(\gch_j)).
\end{equation}
Also, assuming $\Omega \leq L$, we have $\chicCov_j/\chicCov_{j+1} \leq L$.
Assuming also that $L \geq 4$, we therefore have
\begin{align} \label{e:gmuprime2}
  \chicCov_j \gch_j (\much'_j)^2
  \leq \frac{2L}{L^4} \chicCov_{j+1} (\much_{j+1}')^2 \gch_{j+1}
  &\leq
  \frac{1}{2L^2}\chicCov_{j+1} \gch_{j+1} (\much_{j+1}')^2
  \leq
  \frac{1}{2} \chicCov_{j+1} (\much_{j+1}')^2 \gch_{j+1}
  .
\end{align}

As in \eqref{e:VchK-rec}--\eqref{e:Phi-flow}, we write
the recursion relation for $(\Vch_j,K_j)$  as
\begin{equation} \label{e:VchK-rec-bis}
  \Vch_{j+1} = \bar\phi_j(\Vch_j) + \Rch_{j+1}^{(0)}(\Vch_j,K_j), \quad K_{j+1}=\Kch_{j+1}(V_j,K_j).
\end{equation}
With $F$ equal to either $\Rch_{j+1}^{(0)}$ or $\Kch_{j+1}$, the chain rule gives
\begin{align}
  F''(\Vch_j,K_j)
  &=
  D_{\Vch}F(\Vch_j,K_j)\Vch_j'' + D_KF(\Vch_j,K_j)K_j''
  + D_{\Vch}^2F(\Vch_j,K_j)\Vch_j'\Vch_j'
  \nnb &\quad
  + D_K^2F(\Vch_j,K_j)K_j'K_j'
  + 2D_{\Vch}D_KF(\Vch_j,K_j)\Vch_j'K_j
\end{align}
(here $D_{\Vch}D_KF(\Vch,K)AB$ denotes the second derivative of $F$ with derivative in the variable
$\Vch$ taken in direction $A$ and derivative in $K$ taken in direction $B$).
We use $\|\cdot\|$ to denote either the norm $\|\cdot\|_{\Vcal}$ on $\Vcal \cong \R^3$
or the norm $\|\cdot\|_{\Wcal_j}$.
By the versions of \eqref{e:Rmain-g}--\eqref{e:DVKbd} for $\Rch_+,\Kch_+$ discussed
below \eqref{e:RchKch}, and by \eqref{e:induct1},
\begin{align}
  \|D_VF(\Vch_j,K_j)\Vch_j''\| &\leq O(\chicCov_j \gch_j^2)M_1(\much_j')^2\gch_j
  \\
  \|D_V^2F(\Vch_j,K_j)\Vch_j'\Vch_j'\| &\leq O(\chicCov_j \gch_j) (\much_j')^2
  \\
  \|D_VD_KF(\Vch_j,K_j)\Vch_j'K_j'\| &\leq
  O(\gch_j^{-1}) \much_j' (\chicCov_j \much_j' \gch_j^2 )
  \\
  \|D_K^2F(\Vch_j,K_j)K_j'K_j'\| &\leq
  O(\chicCov_j^{-1} \gch_j^{-10/4}) (\chicCov_j \much_j' \gch_j^2 )^2
  \leq O(\chicCov_j (\much_j')^2 \gch_j^{3/2} )
  \\
  \|D_K\Rch_{j+1}^{(0)}(\Vch_j,K_j)K_j''\| &\leq O(M_2)\chicCov_j(\much_j')^2 \gch_j
  \\
  \|D_K\Kch_{j+1}(\Vch_j,K_j)K_j''\| &\leq M_2\chicCov_j (\much_j')^2 \gch_j
  .
\end{align}
This implies, for $M_2 \gg 1$,
\begin{equation} \label{e:phiprime}
  \|(\Rch_{j+1}^{(0)})''(\Vch_j,K_j)\| \leq O(M_2) \chicCov_j (\much_j')^2 \gch_j,
  \quad
  \|\Kch_{j+1}''(\Vch_j,K_j)\| \leq 2M_2 \chicCov_j (\much_j')^2 \gch_j.
\end{equation}%
The second bound and \eqref{e:gmuprime2} immediately advance the induction for $K_j''$.
For $\gch_{j+1}''$, we use \eqref{e:VchK-rec-bis}.
The second derivative of the first term of \eqref{e:VchK-rec-bis} can be bounded using
the recursion \refeq{gbar} for $\gbar$,
Proposition~\ref{prop:pt} to estimate the coefficients,
and \refeq{gzprime1} and the induction hypothesis \refeq{induct1} to estimate
the first and second derivatives.  With \eqref{e:phiprime}, this gives
\begin{align}
  |\gch_{j+1}''|
  &\leq ((1+O(g_j))M_1+O(M_2))\chicCov_j (\much_j')^2 \gch_j
  .
\end{align}
Therefore,
\begin{align}
  |\gch_{j+1}''|
  &\leq \frac{1}{2}(M_1+O(M_2)) \chicCov_{j+1} (\much_{j+1}')^2 \gch_{j+1}
  \leq M_1 \chicCov_{j+1} (\much_{j+1}')^2 \gch_{j+1},
\end{align}
by \eqref{e:gmuprime2} for the second inequality,
and using $M_1 \gg M_2$ in the last inequality.
The estimates for $\zch_j'', \much_j''$ are analogous,
with the difference that for $\much_j''$ there is an additional factor $L^{2}$
(which is bounded analogously, using the second rather than the third inequality in \eqref{e:gmuprime2}).
This completes the proof.
\end{proof}

\subsection{Derivatives of $u$: proof of Lemma~\ref{lem:u2p-bis}}

The following lemma is a restatement of Lemma~\ref{lem:u2p-bis}.

\begin{lemma} \label{lem:u2p}
Let $(m^2,g_0)\in(0,\delta)^2$, and let $(z_0,\nu_0)=(z_0^c,\nu_0^c)$.
There exist $u_\infty'$ continuous in $(m^2,g_0) \in [0,\delta)^2$ and
$u_\infty''$ continuous in $(m^2,g_0) \in (0,\delta)^2$ such that
\begin{align}
  \label{e:uNprime1lim}
  u'_\infty &= \lim_{N\to\infty} u_N' = \frac{n}{2
} \sum_{j=1}^\infty \nuch_j' C_{j;0,0} + O(g),
  \\
  \label{e:uNprime2lim}
  u''_\infty &= \lim_{N\to\infty} u_N''
  = -\frac{n}{2(8+n)} \sum_{j=0}^\infty  \beta_j (\nuch_j')^2 + O(1).
\end{align}
The convergence $u_N'\to u_\infty'$ is uniform in $(m^2,g_0) \in [0,\delta)^2$,
and the convergence $u_N'' \to u_\infty''$ is uniform on compact subsets of $(m^2,g_0) \in (0,\delta)^2$.
\end{lemma}

In the proof of Lemma~\ref{lem:u2p},
we use the transformed variables $(\Vch,K) = (T(V),K)$.
As in \eqref{e:RchKch},
the corresponding version of \eqref{e:uerror} is
\begin{equation} \label{e:uerrortr}
  \delta u_{j+1}
  = \bar\phi^{\delta u}_{j}(\Vch_j) + \Rch^{\delta u}_{j+1}(\Vch_j,K_j),
\end{equation}
where the map $\bar\phi^{\delta u}$ is given by \eqref{e:ubarlong}, and where
$\Rch^{\delta u}_+$ is the transformed version of $R^{\delta u}_+$, defined by
$\Rch^{\delta u}_+(\Vch,K) = \delta u_+(T^{-1}(\Vch),K) - \bar\phi^{\delta u}(\Vch)$.
As noted around \eqref{e:RchKch},
the estimates stated for $R_+$ in Theorem~\ref{thm:step-mr} hold \emph{mutatis mutandis}
for $\Rch_+$. In particular,
\begin{equation}
\label{e:DRch}
    \|D_V^p D_K^q \Rch_+^{\delta u}\|_{L^{p,q}}
    \le
    \begin{cases}
    O(\tilde\chicCov\ggen^{3-p}) & (p\ge 0,\, q=0)\\
    O(\ggen^{1-p-q}) & (p\ge 0,\, q = 1,2).
    \end{cases}
\end{equation}
We recall that the norm \eqref{e:Vnorm} appearing on the left-hand side of \eqref{e:DRch}
scales the $\delta u$ component
by a factor $L^{4j}$. Thus, when estimating absolute values of derivatives
of $\Rch^{\delta u}$, we obtain an additional factor $O(L^{-4j})$.

\begin{proof}
We first note that,
by Lemma~\ref{lem:upt}, Theorems~\ref{thm:step-mr}--\ref{thm:flow-flow}, and \eqref{e:uerror},
$\delta u_+$ is a $C^2$ function
of the renormalisation group coordinates at scale-$j$.
By Proposition~\ref{prop:Fcont},
each of $\delta u_+,\delta u_+', \delta u_+''$
is therefore
continuous on $[0,\delta) \times (0,\delta)$ (in particular, the derivatives exist).

We now prove the convergence and bounds for $u_j'$.
By \eqref{e:ubarlong} and Lemma~\ref{lem:upt},
and by \eqref{e:muprime1}--\eqref{e:gzprime1},
\begin{equation} \label{e:uerror1pt}
  (\bar\phi^{\delta u}_{j+1})' = O(L^{-4j} \chicCov_j \much_j').
\end{equation}
Similarly, by \eqref{e:muprime1}--\eqref{e:gzprime1}, \eqref{e:DRch}, and the chain rule,
\begin{equation} \label{e:uerror1R}
  (\Rch_{j+1}^{\delta u})'= O(L^{-4j} \chicCov_j \much_j' \gch_j^2).
\end{equation}
The latter bound is obtained when the derivative acts in the $\mu_j$ or $K_j$ direction,
with the derivatives in the $g_j,z_j$ directions smaller by a factor $O(\gch_j^2)$.
By \eqref{e:muprime}, it follows in particular that
$\delta u_{j+1}' = O(L^{-4j}\much_j') = O(L^{-2j})$.
Thus $\delta u_{j+1}'$ is summable, uniformly in $(m^2,g_0)\in[0,\delta)^2$.
Since $\delta u_{j+1}'$ is continuous in $(m^2,g_0) \in [0,\delta)^2$,
as noted in the first paragraph of the proof, this implies that
$u'_\infty$ is also continuous on $[0,\delta)^2$, as claimed.
By \eqref{e:muprime1}--\eqref{e:gzprime1}, and since the coefficients of $\bar\phi^{\delta u}$
are uniformly bounded,
the dominant contribution in \eqref{e:ubarlong} is given by
$n\nuch_j' C_{j;0,0}$, and
its sum over $j$ yields the main term of \eqref{e:uNprime1lim}.
The other terms in \eqref{e:ubarlong} as well as $(R_{j+1}^{\delta u})'$ are bounded by
$O(\chicCov_j L^{-4j} \gch_j \much_j') = O(L^{-2j}\gch_j)$, whose sum is $O(g)$ as claimed.

We now consider $u_N''$.
By \eqref{e:muprime1}--\eqref{e:gzmuprime2pf1},
the dominant contribution in \eqref{e:ubarlong} for $(\bar\phi^{\delta u})''$ is
given by the term proportional to
\begin{equation} \label{e:mu2psq}
  (\nuch_j^2)'' = 2(\nuch_j')^2 + 2 \nuch_j''\nuch_j = 2(\nuch')^2 (1+O(\chicCov_j\gch_j^2)),
\end{equation}
with the other terms bounded by $O(L^{-4j} \chicCov_j (\much_j')^2 \gch_j^2)$.
To see the latter, observe that differentiating every monomial in \eqref{e:ubarlong} gives either
one factor from $\gch_j,\zch_j,\much_j$ multiplied with one of $\gch_j'',\zch_j'',\much_j''$,
which is $O(L^{-4j} \chicCov_j (\much_j')^2 \gch_j^2)$,
or two factors of $\gch_j',\zch_j',\much_j'$ of which the largest is $(\much')^2$,
i.e., \eqref{e:mu2psq},
with all other combinations bounded by $O(L^{-4j} \chicCov_j (\much_j')^2 \gch_j^2)$.
From \eqref{e:muprime1}--\eqref{e:gzmuprime2pf1} and \eqref{e:DRch},
it similarly follows that
\begin{equation} \label{e:uerror2}
  (\Rch^{\delta u}_{j+1})'' = O(L^{-4j} \chicCov_j (\much_j')^2 \gch_j^2),
\end{equation}
which is obtained when both derivatives act in the $\much_j$ direction,
or if one acts in $\much_j$ direction and one in the $K_j$ direction.
With \eqref{e:ubarlong}--\eqref{e:ugreeks2}
for the coefficient of the $\nuch^2$ term,
since $\delta_j[w^{(2)}] = \beta_j/(8+n)$,
it follows that
\begin{equation} \label{e:uNpp}
  u_N'' = - \frac{n}{2(8+n)} \sum_{j=0}^{N-1}  \beta_j (\nuch_j')^2 +
  \sum_{j=0}^{N-1} O(\chicCov_j \gch_j (\nuch_j')^2).
\end{equation}
The second term  on right-hand side is bounded by
\begin{equation}
  \sum_{j=0}^\infty \vartheta_j \gch_j (\nuch_j')^2
  = O(1)  \sum_{j=0}^\infty \vartheta_j \frac{\gbar_j^{1+2\gamma}}{g_0^{2\gamma}} = O(1),
\end{equation}
by \eqref{e:gbarpsum}.

We finally show that $u_N \to u_\infty$
uniformly in $m^2 \in [\varepsilon, \delta)$, for any $\varepsilon \in (0,\delta)$.
It suffices to show that this holds for the restriction of both sums in  \eqref{e:uNpp}
to $j \geq j_{\varepsilon} = \lfloor \log_L \varepsilon \rfloor$.
Then the summands are uniformly bounded by $O(\chicCov_j) = O(2^{-(j-j_{\varepsilon})})$,
from which the claim is immediate.
Thus $u_N'' \to u_\infty''$ compactly on $(m^2,g_0) \in (0,\delta)^2$, and
since $u_j''$ is continuous, it follows that
$u_\infty''$ is also continuous on $(0,\delta)^2$, as claimed.
This completes the proof.
\end{proof}

\section*{Acknowledgements}
This work was supported in part by NSERC of Canada.
This material is also based upon work supported by the National
Science Foundation under agreement No.\ DMS-1128155.
We thank Alexandre Tomberg for useful discussions.

\bibliographystyle{plain}

\end{document}